\documentclass[12pt]{article}
\usepackage{graphicx}
\usepackage{eepic}
\usepackage{epic}
\usepackage{cite}

\newcommand{\beq}{\begin{equation}}
\newcommand{\eeq}{\end{equation}}
\def\bea{\begin{eqnarray}}
\def\eea{\end{eqnarray}}
\def\nn{\nonumber}
\def\bec{\begin{center}}
\def\eec{\end{center}}

\begin{document}

\newcommand{\EWtitle}{  
                   Large electroweak penguin   
                         contribution \\ in $B\rightarrow K\pi$ and 
                                               $\pi\pi$ decay modes }
\newcommand{\EWauthor}{
Satoshi Mishima$^{a,}$\footnote{E-mail: mishima@tuhep.phys.tohoku.ac.jp} 
and 
Tadashi Yoshikawa$^{b,}$\footnote{E-mail: tadashi@eken.phys.nagoya-u.ac.jp}
}
\newcommand{\EWaddress}{\small
${}^a$ Department of Physics, Tohoku University, Sendai 980-8578, Japan

${}^b$ Department of Physics, Nagoya University, Nagoya 464-8602, Japan
}


\begin{titlepage}
\begin{flushright}
TU-725 ~ \\
hep-ph/0408090\\
August 6, 2004
\end{flushright}
\vspace{.1in}
\begin{center}
{\large{\bf \EWtitle}}
\bigskip \medskip \\
\EWauthor \\
\mbox{} \\
\medskip
{\it \EWaddress} \\
\vspace{.5in}
\bigskip \end{center} \setcounter{page}{0}

\begin{abstract}
We discuss about 
a possibility of large electroweak penguin contribution in 
$B\rightarrow K \pi$ and $\pi \pi$ from recent experimental data.  
The experimental data may be suggesting that there are some
discrepancies between the data and theoretical estimation in the 
branching ratios of them. In $B\rightarrow
K\pi$ decays, to explain it, 
a large electroweak penguin contribution and large
strong phase differences seem to be needed. 
The contributions should appear also in $B\rightarrow \pi\pi$.   
We show, as an example, a solution to solve 
the discrepancies in both $B\rightarrow K\pi$ and $B\rightarrow
\pi\pi$.  However the magnitude of the parameters and the strong
phase estimated from experimental data 
are quite large compared with the theoretical estimations.  
It may be suggesting some new physics effects are including in these 
processes. We will have to discuss about the dependence of the new
physics. To explain both modes at once, we may need large electroweak penguin
contribution with new weak phases and some SU(3) breaking effects by
new physics in both QCD and electroweak penguin type processes.    
\end{abstract}

\vspace*{6cm}

\end{titlepage}

\section{Introduction}
One of the main targets at the $B$ factories is to determine the all
CP angles in the unitarity triangles of the Cabbibo-Kobayashi-Maskawa (CKM)
matrix~\cite{CKM}.  $\phi_1$~\cite{SM} as one of the angles has already been
measured and established the CP violation in the $B$ meson system
by Belle~\cite{BF1} and BaBar~\cite{BF2} collaborations.
The next step is to determine the remaining
angles and to confirm the unitarity of the CKM matrix.  
The good decay modes for measuring $\phi_2$ and $\phi_3$ are
$B^0 \rightarrow \pi^+ \pi^-$ and $B^\pm \rightarrow D K^\pm $ respectively
but their methods have some difficulties to extract cleanly the angles. 
To avoid the difficulties, isospin relation~\cite{ISO} 
and SU(3) relation including 
$B\rightarrow K\pi$ modes~\cite{GHLR1,GHLR,GR,KLY,FMB,BF,NEU,GNK,GPY,FM} 
are being considered as a method to extract the weak phases. 
However there seems to be anomalies in the recent experimental data of 
$B\rightarrow K\pi$, $\pi \pi$~\cite{TOMURA,NBELLE,NBABAR}. 
To explain the discrepancies in 
$B\rightarrow K\pi $ modes, a large electroweak (EW) penguin contribution 
will be requested~\cite{Y2}\footnote{In the previous work\cite{Y2}, one of
authors showed several relations for $B\rightarrow K\pi $. Some of
these relations was not including a cross term. 
In this work, it is corrected but the discussion is almost same and
it do not make any large changes.}~\cite{GROROS,BFRS,CL}. In addition, the
magnitude of the
branching ratios for $B\rightarrow \pi \pi $ do not also 
agree with the theoretical estimations~\cite{BFRS,CL}. In other words, 
without solving the problems, we can not extract any informations from 
these $B$ decays so clear.        

$B\rightarrow K\pi $ modes have already been
measured~\cite{TOMURA,NBELLE,NBABAR} (and see 
also the web page by Heavy Flavor Averaging Group~\cite{HFAG}) and they will 
be useful informations to understand the CP violation through 
the Kobayashi-Maskawa (KM) phases. 
If we can directly solve these modes, it is very elegant way to 
determine the parameters and the weak phases. 
To extract the weak phases through this mode, there are several 
approaches: diagram decomposition~\cite{GHLR1,GHLR,GR,KLY,FMB,BF,NEU,FM}, 
QCD factorization~\cite{BBNS}, PQCD~\cite{YLS,S-U} and so on. The
contributions including the weak phase $\phi_3$ come from tree type  
diagrams which have a CKM suppression factor and they are usually dealt with 
small parameter compared with gluonic penguin diagram. 
If we can deal the contributions except for gluonic penguin with 
the small parameters, then, there are several relations among 
the averaged branching ratios of $B\rightarrow K\pi$ modes:  
$Br(K^+ \pi^- )/2 Br(K^0 \pi^0 ) \approx 2 Br(K^+ \pi^0)/Br(K^0
\pi^+)$~\cite{BBNS,Y2} et al. However, the recent data do not
seem to satisfy them so well. To satisfy the data, we find that 
the role of a color-favored EW penguin 
may be important. The color favored 
EW penguin diagram is included in $B^0\rightarrow K^0 \pi^0$ and 
$B^0\rightarrow K^+ \pi^0$ and the data of their branching ratio are 
slightly larger than half of that for $B^0\rightarrow K^+ \pi^-$, where
the $1/2$ 
comes from the difference between $\pi^0$ and $\pi^+$ in final state. 
Thus we need to know the EW penguin 
contributions in $B \rightarrow K \pi$ decay modes to extract the weak
phases.
The role of the EW penguin was pointed out and their magnitude was
estimated in several works~\cite{BF,NEU,GNK,GPY,FM,BBNS,YLS,S-U}.  
They said that the ratio between 
gluonic and EW penguins is about $0.14$ as the central value,
but the experimental data may suggest that the magnitude seems to be slightly 
larger than the estimation~\cite{Y2,GROROS,BFRS,CL}. 
Furthermore, one of the most difficult points to
explain them is that we will need quite large strong phase difference of EW
penguin diagram compared with the other strong phases. To produce the
such large strong phase is difficult in the SM. 
If there is quite large deviation in the contribution 
from the EW penguin, it may suggest a possibility of new physics in 
these modes. 

Under the flavor SU(3) symmetry, these discussions in $B\rightarrow
K\pi$ have to relate to $B\rightarrow \pi \pi$ modes, too. If
the contributions from EW penguin are so large, they have to appear also
in $B\rightarrow \pi \pi$. However it is slightly difficult to explain 
the discrepancies between theoretical estimations and experimental
data of the branching ratios by only the EW penguin contribution,
because in these modes the leading contribution comes from tree type 
diagram and the EW penguin is the sub-leading contribution.  
It seems to need other contributions such as the SU(3) breaking
effects. Perhaps, they are induced from some new physics effects. 
In the usual sense, new physics contributions should be through in some
loop effects such as the penguin-type diagram so that there should not be
any discrepancies in tree-type diagrams. We put the new physics
contributions with weak phase into both gluonic and EW penguins to 
find the allowed regions for each parameters. 

This paper is organized as follows. In Section 2, we review the
diagram decomposition approach in $B\rightarrow K\pi$ and
$\pi\pi$. In Section 3, the large EW penguin contribution in
$B\rightarrow K\pi $ decay modes is discussed. We find that we do
not only need the large magnitude but also the large strong phase
differences. In Section 4 and 5, the SU(3) breaking effect and the
large EW penguin contributions in $B\rightarrow \pi\pi $ is
discussed. And Section 6 shows what are needed to explain both modes at
once if the new physics contributions were exiting in these modes. 
Section 7 summarizes our discussions.

\section{$B\rightarrow K\pi $ and $\pi \pi $ under flavor SU(3) symmetry}
Using the diagram decomposition
approach~\cite{GHLR1,GHLR,GR,KLY,FMB,BF,NEU,FM},
the decay amplitudes of $B \rightarrow K\pi$ and $\pi \pi $ are 
\bea
A^{0+}_K &\equiv & A(B^+\rightarrow K^0 \pi^+) \nn \\
         &=& 
         \left[ A V_{ub}^*V_{us} 
             + \sum_{i=u,c,t}( P_i + EP_i - \frac{1}{3}P_{EWi}^C 
                                  + \frac{2}{3}EP^C_{EWi} )
                       V_{ib}^*V_{is} 
                                        \right] , \\[5mm]
A^{00}_K  &\equiv & A(B^0\rightarrow K^0\pi^0) \nn \\
         &=& 
      - \frac{1}{\sqrt{2}}
         \left[ C  V_{ub}^*V_{us} 
             - \sum_{i=u,c,t}( P_i + EP_i - P_{EWi} - \frac{1}{3}P^C_{EWi} 
                                                    - \frac{1}{3}EP^C_{EWi})
                       V_{ib}^*V_{is} 
                                        \right] , \\[5mm]
A^{+-}_K  &\equiv & A(B^0\rightarrow K^+\pi^-) \nn \\
          &=&
            - \left[ T  V_{ub}^*V_{us} 
             + \sum_{i=u,c,t}( P_i + EP_i + \frac{2}{3}P^C_{EWi}
                                   - \frac{1}{3}EP^C_{EWi} )
                       V_{ib}^*V_{is} 
                                        \right] , \\[5mm]
A^{+0}_K  &\equiv & A(B^+\rightarrow K^+ \pi^0) \nn \\
          &=& 
            - \frac{1}{\sqrt{2}}
                      \left[ (T + C + A)  V_{ub}^*V_{us} \right.\nn \\
          &\ & ~~~~~~~~~~~~~~~~
             \left.
        + \sum_{i=u,c,t}( P_i  + EP_i + P_{EWi} + \frac{2}{3}P^C_{EWi}
                                   + \frac{2}{3}EP^C_{EWi} )
                       V_{ib}^*V_{is} 
                                        \right] ,  \\[1.5cm]
A^{00}_\pi  &\equiv & A(B^0\rightarrow \pi^0\pi^0) \nn \\
         &=& 
       \frac{1}{\sqrt{2}}
         \left[ (- C + E ) V_{ub}^*V_{ud} 
             + \sum_{i=u,c,t}( P_i + EP_i - P_{EWi} - \frac{1}{3}P^C_{EWi} 
                                                    - \frac{1}{3}EP^C_{EWi})
                       V_{ib}^*V_{id} 
                                        \right] , \\[5mm]
A^{+-}_\pi  &\equiv & A(B^0\rightarrow \pi^+\pi^-) \nn \\
          &=&
            - \left[ (T + E) V_{ub}^*V_{ud} 
             + \sum_{i=u,c,t}( P_i + EP_i + \frac{2}{3}P^C_{EWi}
                                   - \frac{1}{3}EP^C_{EWi} )
                       V_{ib}^*V_{id} 
                                        \right] , \\[5mm]
A^{+0}_\pi  &\equiv & A(B^+\rightarrow \pi^+ \pi^0) \nn \\
          &=& 
            - \frac{1}{\sqrt{2}}
                      \left[ (T + C )  V_{ub}^*V_{ud} 
        + \sum_{i=u,c,t}( P_{EWi} + P^C_{EWi}
                                   )
                       V_{ib}^*V_{id} 
                                        \right] , 
\eea 
where $T$ is a color-favored tree amplitude, $C$ is a color-suppressed 
tree, $A$($E$) is an annihilation (exchange), $P_i$ $(i=u,c,t)$ 
is a gluonic penguin, $EP_i$ is a penguin exchange, $P_{EWi}$ is a
color-favored EW penguin, $P_{EWi}^C$ is a color-suppressed 
EW penguin and $EP_{EWi}^C$ is a color-suppressed EW 
penguin exchange. In the following study, for simplicity, we neglect 
the $u$- and $c$- EW penguins because of their smallness, Here
we redefine the each terms as follows:
\bea
T + P_u +EP_u - P_c - EP_c &\rightarrow & T\;, \\
C - P_u -EP_u + P_c + EP_c &\rightarrow & C\;, \\
A + P_u +EP_u - P_c - EP_c &\rightarrow & A\;, \\
E                          &\rightarrow & E\;, \\
P_t + EP_t  - P_c - EP_c - \frac{1}{3}P_{EW}^C + \frac{2}{3}EP_{EW}^C 
                           &\rightarrow & P\;, \\ 
P_{EW}+EP_{EW}^C  &\rightarrow & P_{EW}\;, \\
P_{EW}^C-EP_{EW}^C  &\rightarrow & P_{EW}^C\;. 
\label{redP}
\eea 
One can reduce the number of complex parameters up to 7. 
By using the unitarity relation of the CKM matrix, the amplitudes are
written as
\bea
A^{0+}_K &=& 
         \left[ P V_{tb}^*V_{ts} + A V_{ub}^*V_{us} 
                                        \right] , \\
A^{00}_K &=& 
     \frac{1}{\sqrt{2}}
         \left[ (P - P_{EW} )V_{tb}^*V_{ts} 
             - C
                       V_{ub}^*V_{us} 
                                        \right]  , \\
A^{+-}_K &=&
            - \left[ (P + P_{EW}^C)  V_{tb}^*V_{ts} 
             + T
                       V_{ub}^*V_{us} 
                                        \right]  , \\
A^{+0}_K &=& 
            - \frac{1}{\sqrt{2}}
                      \left[ (P + P_{EW} + P_{EW}^C)  V_{tb}^*V_{ts} 
             + ( T + C + A )
                       V_{ub}^*V_{us} 
                                        \right]  , \\[1cm] 
A^{00}_\pi &=& 
     \frac{1}{\sqrt{2}}
         \left[ (P - P_{EW} )V_{tb}^*V_{td} 
             - ( C - E ) 
                       V_{ub}^*V_{ud} 
                                        \right]  , \\
A^{+-}_\pi &=&
            - \left[ (P + P_{EW}^C)  V_{tb}^*V_{td} 
             + (T + E)
                       V_{ub}^*V_{ud} 
                                        \right]  , \\
A^{+0}_\pi &=& 
            - \frac{1}{\sqrt{2}}
                      \left[ (P_{EW} + P_{EW}^C)  V_{tb}^*V_{td} 
             + ( T + C )
                       V_{ub}^*V_{ud} 
                                        \right]  . 
\eea 
By this diagram decomposition~\cite{GHLR}, 
one can easily find the isospin relation among the amplitudes, 
\bea
\sqrt{2}A^{+0}_K + A^{0+}_K &=& \sqrt{2} A^{00}_K + A^{+-}_K\;, 
\label{isospinkpi}
\\
\sqrt{2}A^{+0}_\pi &=& \sqrt{2} A^{00}_\pi + A^{+-}_\pi\;.
\label{isospin}
\eea 
The largest contribution in $B\rightarrow K\pi $ is the gluonic penguin 
and that in $B\rightarrow \pi\pi $ is the color-favored tree so that by
factoring out them the amplitudes are rewritten as follows: 
\bea
A^{0+}_K &=& - P |V_{tb}^*V_{ts}|
         \left[ 1 - r_A e^{i\delta^A}e^{i\phi_3}  
                                        \right]  , \\
A^{00}_K &=& 
     - \frac{1}{\sqrt{2}} P |V_{tb}^*V_{ts}|
         \left[ 1 - r_{EW}e^{i\delta^{EW}}  
             + r_C e^{i\delta^C}e^{i\phi_3}
                                        \right]  , \\
A^{+-}_K &=&
            P |V_{tb}^*V_{ts}| 
            \left[ 1 + r_{EW}^C  e^{i\delta^{EWC}} 
             - r_T e^{i\delta^T}e^{i\phi_3}
                                        \right]  , \\
A^{+0}_K &=&  \frac{1}{\sqrt{2}}
            P |V_{tb}^*V_{ts}|
                      \left[ 1 + r_{EW} e^{i\delta^{EW}} 
                               + r_{EW}^C e^{i\delta^{EWC}} 
             - ( r_Te^{i\delta^T} + r_C e^{i\delta^C} + r_A e^{i\delta^A} )
                       e^{i\phi_3} 
                                        \right]  , 
\eea
\bea
A^{00}_\pi &=& 
     \frac{1}{\sqrt{2}} T |V_{ub}^*V_{ud}|
         \left[ (\tilde{r}_P e^{-i\delta^{T}} 
              - \tilde{r}_{EW} e^{i(\delta^{EW}-\delta^{T})})e^{-i\phi_1}  
             - (\tilde{r}_C e^{i(\delta^C-\delta^T)} 
                - \tilde{r}_E e^{i(\delta^E-\delta^T)} )e^{i\phi_3}
                                        \right]  , \nn \\
           & &                                         \\
A^{+-}_\pi &=&
            - T |V_{ub}^*V_{ud}| 
            \left[ (\tilde{r}_P e^{-i\delta^{T}} + \tilde{r}_{EW}^C
         e^{i(\delta^{EWC} - \delta^T) })e^{-i\phi_1}   
            + (1 + \tilde{r}_E e^{i(\delta^E-\delta^T)}) e^{i\phi_3}
                                        \right]  , \\
A^{+0}_\pi &=&  -\frac{1}{\sqrt{2}}
           T|V_{ub}^*V_{ud}|
                      \left[ (\tilde{r}_{EW} e^{i(\delta^{EW}-\delta^T)} 
                               + \tilde{r}_{EW}^C
         e^{i(\delta^{EWC}-\delta^T)}
                ) e^{-i\phi_1}  
         + (1 + \tilde{r}_C e^{i(\delta^C-\delta^T)} )
                       e^{i\phi_3} 
                                        \right], \nn \\
            & &  
\eea 
where $\phi_1$ and 
$\phi_3$ are the weak phases in $V_{tb}^*V_{td}$ and $V_{ub}^*V_{us}$
respectively, 
$\delta^X$'s are the strong phase differences between 
each diagram and gluonic penguin, and 
\bea 
r_A &=& \frac{ |A V_{ub}^*V_{us}| }{ |P V_{tb}^*V_{ts}|}\;, ~~~~
r_T\, =\, \frac{ |T V_{ub}^*V_{us}| }{ |P V_{tb}^*V_{ts}|}\;, ~~~~ 
r_C\, =\, \frac{ |C V_{ub}^*V_{us}| }{ |P V_{tb}^*V_{ts}|}\;, \\
& & r_{EW}\, =\, \frac{ |P_{EW}|}{ |P|}\;, ~~~~
r_{EW}^C\, =\, \frac{ |P_{EW}^C|}{ |P|}\;, 
\eea
\bea
\tilde{r}_C &=& \frac{|C|}{|T|}\, =\, \frac{r_C}{r_T}\;, ~~~ 
\tilde{r}_E\, =\, \frac{|E|}{|T|}\;, \\
\tilde{r}_P &=& \frac{|P V_{tb}^*V_{td}|}{|T V_{ub}^*V_{ud}|}\, =\,
\frac{1}{r_T} \frac{|V_{td}V_{us}|}{|V_{ud}V_{ts}|}\;, \\
\tilde{r}_{EW} &=& \frac{|P_{EW} V_{tb}^*V_{td}|}{|T V_{ub}^*V_{ud}|}\,
                =\, r_{EW} \tilde{r}_P\;, ~~~ 
\tilde{r}_{EW}^C\, =\, \frac{|P_{EW}^C V_{tb}^*V_{td}|}{|T
V_{ub}^*V_{ud}|}\, =\, r_{EW}^C \tilde{r}_P\;.
\label{eq:tilderew}
\eea 
We assume the hierarchy of the ratios as
$ 1 > r_T, r_{EW} > r_C, r_{EW}^C > r_A $ and 
$ 1 > \tilde{r}_P > \tilde{r}_{EW}, 
\tilde{r}_C > \tilde{r}_{EW}^C, \tilde{r}_E $~\cite{GHLR}. 
$r_T$ can be estimated from the the ratio of $Br(B^+ \rightarrow \pi^0
\pi^+)$ to $Br(B^+ \rightarrow K^0
\pi^+)$~\cite{LU,CONV1,CONV2,AALI,YOSHI},   
which are almost pure gluonic penguin and pure tree process,
respectively. Under the naive factorization approach,
the ratio should be proportional to
$r_T^2$:
\bea
\frac{2Br(B\rightarrow \pi^0\pi^+)}{Br(B\rightarrow K^0\pi^+) }\,
=\, \frac{|T|^2}{|P|^2}
   \frac{f_\pi^2 |V_{ub}^*V_{ud}|^2}{f_K^2 |V_{tb}^*V_{ts}|^2}\,  
\sim\, r_T^2 \frac{f_\pi^2}{\lambda^2 f_K^2 }\;, 
\eea
where the difference between the tree diagrams in $B\rightarrow \pi
\pi$ and $B\rightarrow K \pi$ decays comes from the decay constants, 
$T_\pi \sim T_K \frac{f_\pi}{ f_K } $, and $\lambda $ is the Cabbibo
angle, so that $r_T \sim 0.2 $ with $10\%$ error from recent
experimental data. $r_C$ and $r_{EW}^C$ must be suppressed by color
factor from $r_T$ and $r_{EW}$. 
Comparing the Wilson coefficients, which correspond to the diagrams 
under the factorization method,
we can assume that  $r_C \sim 0.1\, r_{T}$ and $r_{EW}^C \sim 0.1\,
r_{EW}$~\cite{LU,BBNS}. Here we do not put any assumption for the
magnitude of  
$r_{EW}$.  $r_A$ could be negligible because it should have $B$ meson 
decay constant and it works as a suppression factor $f_B/M_B$. 
While, by the similar way one obtains $\tilde{r}_P \sim 0.3$,
$\tilde{r}_C = 0.1$.
Indeed,
the estimations for each parameters in the PQCD approach are 
\bea 
r_T\, =\,  0.21,~~~ r_{EW}\, =\, 0.14 ~~~~ &:& O(0.1)\\
r_C\, =\,  0.018, ~~~ r^C_{EW}\, =\, 0.012, ~~~
r_A\, =\, 0.0048 &:& O(0.01)
\eea 
According to this assumption, 
we neglect the $r^2$ terms including $r_C, r_A$ and $r_{EW}^C$ in 
$B\rightarrow K\pi $. 
In $B\rightarrow \pi\pi $ we will neglect $\tilde{r}_{EW}^C$ 
and $\tilde{r}_E$ for simplicity, but keep $\tilde{r}_{EW}$ to discuss
its magnitude in the both modes.        
Consequently, the averaged branching ratios are 
\bea
\bar{B}_K^{0+}&\propto & \frac{1}{2}\left[|A^{0+}|^2 + |A^{0-}|^2\right] \nn \\
       &=& |P|^2 |V_{tb}^*V_{ts}|^2 
             \left[ 1 - 2 r_A \cos\delta^A\cos\phi_3  \right],
\label{B0+} \\[5mm]
2 \bar{B}_K^{00}&\propto &   
                           \left[|A^{00}|^2 + |\bar{A}^{00}|^2\right]
      \nn \\ 
      &=& |P|^2 |V_{tb}^*V_{ts}|^2 
                \left[ 1 + r_{EW}^2 
                       - 2 r_{EW} \cos\delta^{EW} 
                       + 2 r_C \cos\delta^C\cos\phi_3
                \right], \\[5mm]
\bar{B}_K^{+-}&\propto & \frac{1}{2}\left[|A^{+-}|^2 + |A^{-+}|^2\right]
       \nn \\
       &=& |P|^2 |V_{tb}^*V_{ts}|^2 
                \left[ 1 + r_T^2 
                       + 2 r_{EW}^C \cos\delta^{EWC} 
                       - 2 r_T \cos\delta^T\cos\phi_3
                \right], 
                           \\[5mm]
2 \bar{B}_K^{+0}&\propto & \left[|A^{+0}|^2 + |A^{-0}|^2\right]
       \nn \\
       &=& |P|^2 |V_{tb}^*V_{ts}|^2 
                \left[ 1 +  r_{EW}^2 + r_T^2 
                       + 2 r_{EW} \cos\delta^{EW} 
                       + 2 r_{EW}^C \cos\delta^{EWC} \right.\nn \\
       & & ~~~~~~~~~~
                       - ( 2 r_T \cos\delta^T + 2 r_C \cos\delta^C 
                         + 2 r_A \cos\delta^A )\cos\phi_3 \nn \\
       & & ~~~~~~~~~~  \left.
                - 2 r_{EW} r_T \cos(\delta^{EW}-\delta^{T})\cos\phi_3
                \right], \label{B+0}
                                              \\[3mm]
2 \bar{B}_\pi^{00}&\propto &   
                           \left[|A^{00}_\pi|^2 + |\bar{A}^{00}_\pi|^2\right]
      \nn \\ 
      &=& |T|^2 |V_{ub}^*V_{ud}|^2 
                \left[ \tilde{r}_C^2 +\tilde{r}_P^2 + \tilde{r}_{EW}^2 
                       - 2 \tilde{r}_P \tilde{r}_{EW} \cos\delta^{EW}
        \right. \nn \\ 
        & & ~~~~~~~~~~  \left.
                       - 2 \tilde{r}_C ( \tilde{r}_P \cos\delta^C 
                         - \tilde{r}_{EW}\cos(\delta^{EW}-\delta^C) )
                       \cos(\phi_1+\phi_3 )
                \right], \\[5mm]
\bar{B}_\pi^{+-}&\propto & 
         \frac{1}{2}\left[|A_\pi^{+-}|^2 + |A_\pi^{-+}|^2\right]
       \nn \\
       &=& |T|^2 |V_{ub}^*V_{ud}|^2 
                \left[ 1 + \tilde{r}_P^2 
                       + 2 \tilde{r}_{P} \cos\delta^{T} \cos(\phi_1+\phi_3)
                \right], \\[5mm]
2 \bar{B}_\pi^{+0}&\propto & \left[|A_\pi^{+0}|^2 + |A_\pi^{-0}|^2\right]
       \nn \\
       &=& |T|^2 |V_{ub}^*V_{ud}|^2 
                \left[ 1 +  \tilde{r}_C^2 + \tilde{r}_{EW}^2  
                       + 2 \tilde{r}_{C} \cos(\delta^{C}-\delta^T) 
                                                          \right.  \nn
       \\
       & & ~~~~~~~~~  \left.
                  +2 \tilde{r}_{EW} 
              \{ \cos(\delta^{EW}-\delta^T) 
                + \tilde{r}_C \cos(\delta^{EW}-\delta^C)\}  
                \cos(\phi_1+\phi_3)
                \right].  
\eea

\begin{table}
\begin{center}
\begin{tabular}{|c|c|c|c|c|}\hline
  & CLEO
  & Belle & BaBar & Average \\
\hline
$Br(B^0 \rightarrow K^+ \pi^-) \times 10^{6} $
             & 18.0 ${}^{+2.3+1.2}_{-2.1-0.9}$
                          & 18.5 $\pm$ 1.0
                          $\pm$ 0.7
                          & 17.9 $\pm$ 0.9 $\pm$ 0.7
                          & 18.2 $\pm$ 0.8  \\[2mm]
$Br(B^0 \rightarrow K^0 \pi^0) \times 10^{6} $
             & 12.8 ${}^{+4.0+1.7}_{-3.3-1.4}$
                          & 11.7 $\pm$ 2.3
                          ${}^{+1.2}_{-1.3}$
                          & 11.4 $\pm$ 1.7 $\pm$ 0.8
                          & 11.7 $\pm$ 1.4  \\[2mm]
$Br(B^+ \rightarrow K^+ \pi^0) \times 10^{6} $
             & 12.9 ${}^{+2.4+1.2}_{-2.2-1.1}$
                          & 12.0 $\pm$ 1.3
                          ${}^{+1.3}_{-0.9}$
                          & 12.8 ${}^{+1.2}_{-1.1}$ $\pm$ 1.0
                          & 12.5 $\pm$ 1.1  \\[2mm]
$Br(B^+ \rightarrow K^0 \pi^+) \times 10^{6} $
             & 18.8 ${}^{+3.7+2.1}_{-3.3-1.8}$
                          & 22.0 $\pm$ 1.7
                          $\pm$ 1.1
                          & 22.3 $\pm$ 1.9 $\pm$ 1.1
                          & 21.8 $\pm$ 1.4  \\[2mm]
\hline 
\hline 
$Br(B^0 \rightarrow \pi^+ \pi^-) \times 10^{6} $
             & 4.5 ${}^{+1.4+0.5}_{-1.2-0.4}$
                          & 4.4 $\pm$ 0.6
                          $\pm$ 0.3
                          & 4.7 $\pm$ 0.6 $\pm$ 0.2
                          & 4.6 $\pm$ 0.4  \\[2mm]
$Br(B^0 \rightarrow \pi^0 \pi^0) \times 10^{6} $
             & - 
                          & 1.7 $\pm$ 0.6
                          $\pm$ 0.2
                          & 2.1 $\pm$ 0.6 $\pm$ 0.3
                          & 1.9 $\pm$ 0.5  \\[2mm]
$Br(B^+ \rightarrow \pi^+ \pi^0) \times 10^{6} $
             & 4.6 ${}^{+1.8+0.6}_{-1.6-0.7}$
                          & 5.0 $\pm$ 1.2
                          $\pm$ 0.5
                          & 5.5 ${}^{+1.0}_{-0.9}$ $\pm$ 0.6
                          & 5.2 $\pm$ 0.8  \\[2mm]
\hline
\end{tabular}
\caption{The experimental data~\cite{TOMURA,NBELLE,NBABAR} 
         and the average~\cite{HFAG}. }
\end{center}
\end{table}

When we compare the theoretical predictions with the experimental data, 
the ratios among the branching ratios are very useful to reduce 
uncertainties in hadronic parts.
One can take several ratios among the branching ratios. 
{}From the averaged values of the recent experimental data in Table.~1,  
\bea
\frac{\bar{B}_K^{+-}}{2\bar{B}_K^{00}}\, =\, 0.78 \pm 0.10\;, ~~& &~
\frac{2 \bar{B}_K^{+0}}{\bar{B}_K^{0+}}\, =\, 1.15 \pm 0.12\;, 
\label{Bpm00-Bp00p}\\[5mm]
\frac{\tau^+}{\tau^0}\frac{\bar{B}_K^{+-}}{\bar{B}_K^{0+}}\,
  =\, 0.90 \pm 0.07\;, 
                                                    ~~& &~
\frac{\tau^0}{\tau^+}\frac{\bar{B}_K^{+0}}{\bar{B}_K^{00}}\, 
  =\, 0.98 \pm 0.15\;,
\label{datakpi2} 
                                                     \\[5mm]
\frac{\tau^+}{\tau^0}\frac{2 \bar{B}_K^{00}}{\bar{B}_K^{0+}}\, 
   =\, 1.17 \pm 0.16\;, 
                                                     ~~& &~
\frac{\tau^0}{\tau^+}\frac{2 \bar{B}_K^{+0}}{\bar{B}_K^{+-}}\, 
  =\, 1.26 \pm 0.13\;,
\label{data+0}   
\eea 
\bea
\frac{\tau^0}{\tau^+}\frac{2\bar{B}_\pi^{+0}}{\bar{B}_\pi^{+-}}\, 
                                   =\, 2.08 \pm 0.37\;, ~~& &~
\frac{2 \bar{B}_\pi^{00}}{\bar{B}_\pi^{+-}}\, =\, 0.83 \pm 0.23\;,
\label{Bpp}
\eea 
where $\frac{\tau^+}{\tau^0}$ is a lifetime ratio 
between the charged and the neutral $B$ mesons and 
$\tau(B^\pm)/\tau(B^0) = 1.086 \pm 0.017$~\cite{PDG}. 

\section{EW penguin contribution in $B\rightarrow K\pi $} 
\begin{figure}[thb]
\begin{center}
\begin{minipage}[l]{3.0in}
\setlength{\unitlength}{0.080450pt}
\begin{picture}(2699,2069)(0,0)
\footnotesize
\thicklines \path(370,249)(411,249)
\thicklines \path(2576,249)(2535,249)
\put(329,249){\makebox(0,0)[r]{ 0}}
\thicklines \path(370,539)(411,539)
\thicklines \path(2576,539)(2535,539)
\put(329,539){\makebox(0,0)[r]{ 0.2}}
\thicklines \path(370,829)(411,829)
\thicklines \path(2576,829)(2535,829)
\put(329,829){\makebox(0,0)[r]{ 0.4}}
\thicklines \path(370,1119)(411,1119)
\thicklines \path(2576,1119)(2535,1119)
\put(329,1119){\makebox(0,0)[r]{ 0.6}}
\thicklines \path(370,1408)(411,1408)
\thicklines \path(2576,1408)(2535,1408)
\put(329,1408){\makebox(0,0)[r]{ 0.8}}
\thicklines \path(370,1698)(411,1698)
\thicklines \path(2576,1698)(2535,1698)
\put(329,1698){\makebox(0,0)[r]{ 1}}
\thicklines \path(370,1988)(411,1988)
\thicklines \path(2576,1988)(2535,1988)
\put(329,1988){\makebox(0,0)[r]{ 1.2}}
\thicklines \path(370,249)(370,290)
\thicklines \path(370,1988)(370,1947)
\put(370,166){\makebox(0,0){-1}}
\thicklines \path(922,249)(922,290)
\thicklines \path(922,1988)(922,1947)
\put(922,166){\makebox(0,0){-0.5}}
\thicklines \path(1473,249)(1473,290)
\thicklines \path(1473,1988)(1473,1947)
\put(1473,166){\makebox(0,0){ 0}}
\thicklines \path(2025,249)(2025,290)
\thicklines \path(2025,1988)(2025,1947)
\put(2025,166){\makebox(0,0){ 0.5}}
\thicklines \path(2576,249)(2576,290)
\thicklines \path(2576,1988)(2576,1947)
\put(2576,166){\makebox(0,0){ 1}}
\thicklines \path(370,249)(2576,249)(2576,1988)(370,1988)(370,249)
\put(-32,968){\makebox(0,0)[l]{{$r_{EW}$ }}}
\put(1473,42){\makebox(0,0){ $ \cos\delta^{EW} $}}
\thinlines \path(1826,1713)(1848,1713)(1848,1713)(1859,1705)(1870,1698)(1870,1691)(1881,1684)(1881,1676)(1881,1669)(1892,1662)(1892,1655)(1892,1647)(1892,1640)(1903,1633)(1903,1626)(1903,1618)(1903,1611)(1903,1604)(1903,1597)(1914,1589)(1914,1582)(1914,1575)(1914,1568)(1914,1560)(1914,1553)(1914,1546)(1914,1539)(1914,1532)(1914,1524)(1914,1517)(1925,1510)(1925,1503)(1925,1495)(1925,1488)(1925,1481)(1925,1474)(1925,1466)(1925,1459)(1925,1452)(1925,1445)(1925,1437)(1925,1430)(1925,1423)(1925,1416)(1925,1408)(1925,1401)(1925,1394)(1925,1387)(1925,1379)(1925,1372)(1925,1365)(1925,1358)
\thinlines \path(1925,1358)(1925,1350)(1925,1343)(1925,1336)(1925,1329)(1925,1321)(1925,1314)(1925,1307)(1925,1300)(1925,1292)(1925,1285)(1925,1278)(1914,1271)(1914,1263)(1914,1256)(1914,1249)(1914,1242)(1914,1234)(1914,1227)(1914,1220)(1914,1213)(1914,1205)(1914,1198)(1914,1191)(1914,1184)(1914,1176)(1914,1169)(1903,1162)(1903,1155)(1903,1147)(1903,1140)(1903,1133)(1903,1126)(1903,1119)(1903,1111)(1903,1104)(1903,1097)(1903,1090)(1903,1082)(1892,1075)(1892,1068)(1892,1061)(1892,1053)(1892,1046)(1892,1039)(1892,1032)(1892,1024)(1892,1017)(1892,1010)(1892,1003)(1881,995)
\thinlines \path(1881,995)(1881,988)(1881,981)(1881,974)(1881,966)(1881,959)(1881,952)(1881,945)(1881,937)(1881,930)(1881,923)(1870,916)(1870,908)(1870,901)(1870,894)(1870,887)(1870,879)(1870,872)(1870,865)(1870,858)(1870,850)(1870,843)(1870,836)(1870,829)(1870,821)(1870,814)(1870,807)(1870,800)(1870,792)(1870,785)(1870,778)(1870,771)(1870,763)(1870,756)(1870,749)(1870,742)(1870,734)(1870,727)(1870,720)(1870,713)(1870,705)(1870,698)(1870,691)(1870,684)(1870,677)(1870,669)(1881,662)(1881,655)(1881,648)(1881,640)(1881,633)
\thinlines \path(1881,633)(1892,626)(1892,619)(1892,611)(1892,604)(1903,597)(1903,590)(1903,582)(1914,575)(1914,568)(1925,561)(1925,553)(1925,546)(1936,539)(1947,532)(1947,524)(1958,517)(1969,510)(1969,503)(1980,495)(1991,488)(1991,481)(1991,474)(1991,466)(1980,459)(1980,452)(1969,445)(1947,437)(1925,430)(1903,423)(1870,416)(1826,408)(1771,401)(1705,394)(1616,387)(1495,379)(1330,372)(1065,365)(500,365)(400,365)(370,365)
\thinlines \path(1826,1713)(1826,1713)(1804,1705)(1804,1698)(1793,1691)(1782,1684)(1771,1676)(1771,1669)(1760,1662)(1760,1655)(1749,1647)(1749,1640)(1738,1633)(1738,1626)(1727,1618)(1727,1611)(1716,1604)(1716,1597)(1716,1589)(1705,1582)(1705,1575)(1694,1568)(1694,1560)(1683,1553)(1683,1546)(1683,1539)(1672,1532)(1672,1524)(1661,1517)(1661,1510)(1661,1503)(1649,1495)(1649,1488)(1638,1481)(1638,1474)(1627,1466)(1627,1459)(1627,1452)(1616,1445)(1616,1437)(1605,1430)(1605,1423)(1605,1416)(1594,1408)(1594,1401)(1583,1394)(1583,1387)(1572,1379)(1572,1372)(1572,1365)(1561,1358)
\thinlines \path(1561,1358)(1561,1350)(1550,1343)(1550,1336)(1539,1329)(1539,1321)(1528,1314)(1528,1307)(1517,1300)(1517,1292)(1506,1285)(1506,1278)(1506,1271)(1495,1263)(1495,1256)(1484,1249)(1484,1242)(1473,1234)(1473,1227)(1462,1220)(1462,1213)(1451,1205)(1440,1198)(1440,1191)(1429,1184)(1429,1176)(1418,1169)(1418,1162)(1407,1155)(1407,1147)(1396,1140)(1396,1133)(1385,1126)(1374,1119)(1374,1111)(1363,1104)(1363,1097)(1352,1090)(1341,1082)(1341,1075)(1330,1068)(1330,1061)(1319,1053)(1308,1046)(1308,1039)(1297,1032)(1285,1024)(1285,1017)(1274,1010)(1263,1003)(1263,995)
\thinlines \path(1263,995)(1252,988)(1241,981)(1230,974)(1230,966)(1219,959)(1208,952)(1197,945)(1197,937)(1186,930)(1175,923)(1164,916)(1164,908)(1153,901)(1142,894)(1131,887)(1120,879)(1109,872)(1098,865)(1098,858)(1087,850)(1076,843)(1065,836)(1054,829)(1043,821)(1032,814)(1021,807)(1010,800)(999,792)(988,785)(977,778)(955,771)(944,763)(933,756)(922,749)(910,742)(888,734)(877,727)(866,720)(844,713)(833,705)(822,698)(800,691)(789,684)(767,677)(756,669)(734,662)(712,655)(701,648)(679,640)(657,633)
\thinlines \path(657,633)(635,626)(613,619)(591,611)(569,604)(546,597)(524,590)(491,582)(469,575)(436,568)(414,561)(381,553)(370,546)(370,539)(370,532)(370,524)(370,517)(370,510)(370,503)(370,495)(370,488)(370,481)(370,474)(370,466)(370,459)(370,452)(370,445)(370,437)(370,430)(370,423)(370,416)(370,408)(370,401)(370,394)(370,387)(370,379)(370,372)(370,365)
\put(1683,1176){\makebox(0,0){$\times$}}
\end{picture}
\end{minipage}
    \hspace*{8mm}
\begin{minipage}[r]{3.0in}
\setlength{\unitlength}{0.080450pt}
\begin{picture}(2699,2069)(0,0)
\footnotesize
\thicklines \path(370,249)(411,249)
\thicklines \path(2576,249)(2535,249)
\put(329,249){\makebox(0,0)[r]{ 0}}
\thicklines \path(370,539)(411,539)
\thicklines \path(2576,539)(2535,539)
\put(329,539){\makebox(0,0)[r]{ 0.2}}
\thicklines \path(370,829)(411,829)
\thicklines \path(2576,829)(2535,829)
\put(329,829){\makebox(0,0)[r]{ 0.4}}
\thicklines \path(370,1119)(411,1119)
\thicklines \path(2576,1119)(2535,1119)
\put(329,1119){\makebox(0,0)[r]{ 0.6}}
\thicklines \path(370,1408)(411,1408)
\thicklines \path(2576,1408)(2535,1408)
\put(329,1408){\makebox(0,0)[r]{ 0.8}}
\thicklines \path(370,1698)(411,1698)
\thicklines \path(2576,1698)(2535,1698)
\put(329,1698){\makebox(0,0)[r]{ 1}}
\thicklines \path(370,1988)(411,1988)
\thicklines \path(2576,1988)(2535,1988)
\put(329,1988){\makebox(0,0)[r]{ 1.2}}
\thicklines \path(370,249)(370,290)
\thicklines \path(370,1988)(370,1947)
\put(370,166){\makebox(0,0){-1}}
\thicklines \path(922,249)(922,290)
\thicklines \path(922,1988)(922,1947)
\put(922,166){\makebox(0,0){-0.5}}
\thicklines \path(1473,249)(1473,290)
\thicklines \path(1473,1988)(1473,1947)
\put(1473,166){\makebox(0,0){ 0}}
\thicklines \path(2025,249)(2025,290)
\thicklines \path(2025,1988)(2025,1947)
\put(2025,166){\makebox(0,0){ 0.5}}
\thicklines \path(2576,249)(2576,290)
\thicklines \path(2576,1988)(2576,1947)
\put(2576,166){\makebox(0,0){ 1}}
\thicklines \path(370,249)(2576,249)(2576,1988)(370,1988)(370,249)
\put(-32,968){\makebox(0,0)[l]{{ $r_{EW}$ }}}
\put(1473,42){\makebox(0,0){ $ r_{T} \cos(\delta^{EW}+\delta^{T})\cos \phi_{3} $}}
\thinlines \path(701,1713)(751,1713)(751,1713)(767,1705)(789,1698)(800,1691)(817,1684)(828,1676)(839,1669)(850,1662)(861,1655)(872,1647)(877,1640)(888,1633)(899,1626)(905,1618)(916,1611)(922,1604)(933,1597)(944,1589)(949,1582)(960,1575)(966,1568)(977,1560)(982,1553)(988,1546)(999,1539)(1004,1532)(1015,1524)(1021,1517)(1026,1510)(1037,1503)(1043,1495)(1048,1488)(1059,1481)(1065,1474)(1070,1466)(1076,1459)(1087,1452)(1092,1445)(1098,1437)(1109,1430)(1115,1423)(1120,1416)(1126,1408)(1137,1401)(1142,1394)(1148,1387)(1153,1379)(1164,1372)(1170,1365)(1175,1358)
\thinlines \path(1175,1358)(1186,1350)(1192,1343)(1197,1336)(1203,1329)(1214,1321)(1219,1314)(1225,1307)(1230,1300)(1241,1292)(1247,1285)(1252,1278)(1263,1271)(1269,1263)(1274,1256)(1285,1249)(1291,1242)(1297,1234)(1308,1227)(1313,1220)(1319,1213)(1330,1205)(1335,1198)(1346,1191)(1352,1184)(1357,1176)(1368,1169)(1374,1162)(1385,1155)(1390,1147)(1401,1140)(1407,1133)(1418,1126)(1423,1119)(1434,1111)(1445,1104)(1451,1097)(1462,1090)(1467,1082)(1479,1075)(1490,1068)(1501,1061)(1506,1053)(1517,1046)(1528,1039)(1539,1032)(1545,1024)(1556,1017)(1567,1010)(1578,1003)(1589,995)
\thinlines \path(1589,995)(1600,988)(1611,981)(1622,974)(1633,966)(1644,959)(1655,952)(1672,945)(1683,937)(1694,930)(1710,923)(1721,916)(1732,908)(1749,901)(1760,894)(1776,887)(1787,879)(1804,872)(1820,865)(1831,858)(1848,850)(1865,843)(1881,836)(1898,829)(1914,821)(1931,814)(1947,807)(1964,800)(1986,792)(2002,785)(2019,778)(2041,771)(2063,763)(2080,756)(2102,749)(2124,742)(2146,734)(2168,727)(2190,720)(2212,713)(2240,705)(2262,698)(2289,691)(2311,684)(2339,677)(2366,669)(2394,662)(2422,655)(2455,648)(2482,640)(2515,633)
\thinlines \path(2515,633)(2548,626)(2576,620)
\thinlines \path(701,1713)(701,1713)(690,1705)(684,1698)(679,1691)(679,1684)(673,1676)(673,1669)(673,1662)(673,1655)(673,1647)(673,1640)(673,1633)(679,1626)(679,1618)(679,1611)(679,1604)(684,1597)(684,1589)(684,1582)(690,1575)(690,1568)(695,1560)(695,1553)(701,1546)(701,1539)(706,1532)(706,1524)(712,1517)(712,1510)(717,1503)(723,1495)(723,1488)(728,1481)(728,1474)(734,1466)(740,1459)(745,1452)(745,1445)(751,1437)(756,1430)(756,1423)(762,1416)(767,1408)(773,1401)(773,1394)(778,1387)(784,1379)(789,1372)(795,1365)(795,1358)
\thinlines \path(795,1358)(800,1350)(806,1343)(811,1336)(817,1329)(822,1321)(822,1314)(828,1307)(833,1300)(839,1292)(844,1285)(850,1278)(855,1271)(861,1263)(866,1256)(872,1249)(872,1242)(877,1234)(883,1227)(888,1220)(894,1213)(899,1205)(905,1198)(910,1191)(916,1184)(922,1176)(927,1169)(933,1162)(938,1155)(944,1147)(955,1140)(960,1133)(966,1126)(971,1119)(977,1111)(982,1104)(988,1097)(993,1090)(999,1082)(1004,1075)(1010,1068)(1021,1061)(1026,1053)(1032,1046)(1037,1039)(1043,1032)(1048,1024)(1059,1017)(1065,1010)(1070,1003)(1076,995)
\thinlines \path(1076,995)(1081,988)(1092,981)(1098,974)(1103,966)(1109,959)(1120,952)(1126,945)(1131,937)(1142,930)(1148,923)(1153,916)(1164,908)(1170,901)(1175,894)(1186,887)(1192,879)(1197,872)(1208,865)(1214,858)(1225,850)(1230,843)(1241,836)(1247,829)(1258,821)(1263,814)(1274,807)(1280,800)(1291,792)(1297,785)(1308,778)(1319,771)(1324,763)(1335,756)(1346,749)(1352,742)(1363,734)(1374,727)(1385,720)(1390,713)(1401,705)(1412,698)(1423,691)(1434,684)(1445,677)(1456,669)(1467,662)(1479,655)(1490,648)(1501,640)(1512,633)
\thinlines \path(1512,633)(1528,626)(1539,619)(1550,611)(1567,604)(1578,597)(1594,590)(1605,582)(1622,575)(1638,568)(1649,561)(1666,553)(1683,546)(1699,539)(1716,532)(1738,524)(1754,517)(1776,510)(1793,503)(1815,495)(1837,488)(1859,481)(1887,474)(1909,466)(1936,459)(1969,452)(1997,445)(2030,437)(2063,430)(2102,423)(2140,416)(2179,408)(2229,401)(2278,394)(2328,387)(2388,379)(2455,372)(2526,365)(2576,358)(2576,350)
\put(1142,1176){\makebox(0,0){$\times$}}
{{\thinlines \path(1638,452)(1638,452)(1308,452)}}
{{\thinlines \dashline[-20]{19}(1638,249)(1638,249)(1638,1988)}}
{\thinlines \dashline[-20]{19}(1308,249)(1308,249)(1308,1988)}
\put(652,450){\makebox(0,0)[l]{{{$r_{EW} = 0.14 $}}}}
\end{picture}

\end{minipage}
\caption{The allowed region on $(r_{EW},  ~\cos \delta^{EW})$ and 
$(r_{EW}, ~r_T \cos(\delta^{EW}~+~\delta^{T}) \cos \phi_3)$ plane
at $1\sigma $ level varying the magnitude of
$r_T \cos(\delta^{EW}~+~\delta^{T}) \cos \phi_3 $ up to 1. The
dashed line shows the bound of  
$r_{T} \cos(\delta^{EW}+\delta^{T})\cos\phi_3$ 
for $r_T$ = 0.2 and $\phi_3 = 40^\circ $. }
    \label{fig:1}
\end{center}
\end{figure}
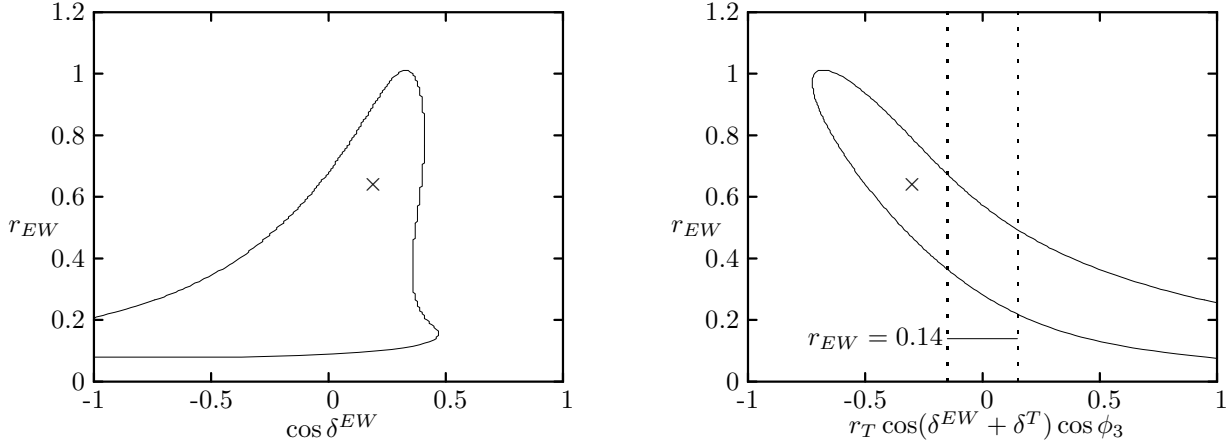

Under the assumption that 
all $r$ is smaller than $1$ and the $r^2$ terms including $r_C, r_A$ 
and $r_{EW}^C$ are neglected, the ratios among the decay rates of
$B\rightarrow K\pi $ are          
\bea
\frac{\bar{B}_K^{+-}}{2\bar{B}_K^{00}} &=& \left\{
  1 + 2 r_{EW} \cos\delta^{EW} + 2 r_{EW}^C \cos\delta^{EWC} 
    - 2 ( r_{T} \cos\delta^{T} + r_C \cos\delta^C )\cos\phi_3  
    + r_{T}^2 \right\}\nn \\
  & & ~~~ -r_{EW}^2 + 4 r_{EW}^2 \cos^2\delta^{EW}
          -4 r_{EW}r_T \cos\delta^{EW}\cos\delta^T\cos\phi_3\;, 
\label{B+-B00}\\[4mm]
\frac{2 \bar{B}_K^{+0}}{\bar{B}_K^{0+}} &=&
 \left\{ 1 + 2 r_{EW} \cos\delta^{EW} + 2 r_{EW}^C \cos\delta^{EWC} 
    - 2 ( r_{T} \cos\delta^{T} + r_C \cos\delta^C )\cos\phi_3  
  + r_{T}^2 \right\} \nn \\
  & & ~~~ + r_{EW}^2 
            - 2 r_{EW} r_T \cos(\delta^{EW}-\delta^T)\cos\phi_3\;,
\label{B+0B0+}\\[4mm]
\frac{\tau^+}{\tau^0}\frac{\bar{B}_K^{+-}}{\bar{B}_K^{0+}} &=&
  1 + 2 r_{EW}^C \cos\delta^{EWC} 
    - 2 ( r_{T} \cos\delta^{T} - r_A \cos\delta^A )\cos\phi_3  
    + r_{T}^2\;,  \label{B+-B0+}\\[4mm]  
\frac{\tau^0}{\tau^+}\frac{\bar{B}_K^{+0}}{\bar{B}_K^{00}} &=& 
 1 + 2 r_{EW}^C \cos\delta^{EWC} 
    - 2 ( r_{T} \cos\delta^{T} 
          + 2 r_C \cos\delta^C + r_A \cos\delta^A )\cos\phi_3 + r_{T}^2 
 \nn \\
  & & ~~~ + 4 r_{EW} \cos\delta^{EW}   + 8 r_{EW}^2 \cos^2\delta^{EW}
 \nn \\[2mm]
  & & ~~~ - 2 r_{EW} r_T \cos(\delta^{EW}-\delta^T)\cos\phi_3
        - 4 r_{EW} r_T \cos\delta^{EW}\cos\delta^T\cos\phi_3\;, 
\label{B+0B00} \\[4mm]
\frac{\tau^+}{\tau^0}\frac{2 \bar{B}_K^{00}}{\bar{B}_K^{0+}} &=&
   1 - 2 r_{EW} \cos\delta^{EW} 
       + 2 ( r_C \cos\delta^C + r_A \cos\delta^A )\cos\phi_3 + r_{EW}^2\;, 
\label{B00B0+} \\[4mm]
\frac{\tau^0}{\tau^+}\frac{2 \bar{B}_K^{+0}}{\bar{B}_K^{+-}} &=&
   1 + 2 r_{EW} \cos\delta^{EW} 
       - 2 ( r_C \cos\delta^C + r_A \cos\delta^A )\cos\phi_3 + r_{EW}^2
                               \nn \\
   & & ~~~ 
       + 2 r_{EW} r_T \cos(\delta^{EW}+\delta^T)\cos\phi_3\;. 
\label{B+0B+-}
\eea
If we neglect all $r^2$ terms, then there are a few relations among 
Eqs.~(\ref{B+-B00})--(\ref{B+0B+-}) as following 
\bea
R_c-R_n &\equiv & \frac{2 \bar{B}_K^{+0}}{\bar{B}_K^{0+}} - 
               \frac{\bar{B}_K^{+-}}{2\bar{B}_K^{00}}\, =\, 0\;,
                                   \\[3mm]
S ~~~ &\equiv & \frac{2 \bar{B}_K^{+0}}{\bar{B}_K^{0+}}
 - \frac{\tau^+}{\tau^0}\frac{\bar{B}_K^{+-}}{\bar{B}_K^{0+}} + 
\frac{\tau^+}{\tau^0}\frac{2 \bar{B}_K^{00}}{\bar{B}_K^{0+}} -1\, =\, 0\;, 
\label{BISO} \\[3mm] 
R_+ - 2  &\equiv & \frac{\tau^0}{\tau^+}
                  \frac{2 \bar{B}_K^{+0}}{\bar{B}_K^{+-}} + 
                  \frac{\tau^+}{\tau^{0}}
               \frac{2\bar{B}_K^{00}}{\bar{B}_K^{0+}} -2\, =\, 0\;.
\eea 
However, the experimental data listed in 
Eqs.~(\ref{Bpm00-Bp00p}), (\ref{datakpi2}) and (\ref{data+0})
do not satisfy these relations so well. According 
to the experimental data, 
$\frac{\bar{B}^{+-}_K}{2\bar{B}^{00}_K}$ seems to be smaller than 1 but 
$\frac{2 \bar{B}^{+0}_K}{\bar{B}^{0+}_K}$ be larger than 1. Thus it shows 
there is a discrepancy between them.  
The equations of 
$\frac{\bar{B}^{+-}_K}{2\bar{B}^{00}_K}$ and 
$\frac{2 \bar{B}^{+0}_K}{\bar{B}^{0+}_K}$ are same up to $r_T^2$ term 
and the difference comes from $r^2$ term including $r_{EW}$. 
The second relation corresponds to the isospin relation of 
Eq.~(\ref{isospinkpi}) at the first order of $r$.  
The discrepancy of relation (\ref{BISO}) from $0$ also comes from 
$r^2$ term including $r_{EW}$. 
The differences are 
\bea
R_c-R_n &= & \frac{2 \bar{B}_K^{+0}}{\bar{B}_K^{0+}} 
    - \frac{\bar{B}_K^{+-}}{2\bar{B}_K^{00}} \nn \\ 
        &=& 
   2 r_{EW}^2 + 2 r_{EW} r_{T} \cos(\delta^{EW}+\delta^{T})\cos\phi_3
              - 4 r_{EW}^2 \cos^2\delta^{EW}\, =\, 0.37 \pm 0.16\;,
\label{B+0B0+MB+-B00} \\[6mm]
S ~~~~  &= & \frac{2 \bar{B}_K^{+0}}{\bar{B}_K^{0+}}
 - \frac{\tau^+}{\tau^0}\frac{\bar{B}_K^{+-}}{\bar{B}_K^{0+}} + 
\frac{\tau^+}{\tau^0}\frac{2 \bar{B}_K^{00}}{\bar{B}_K^{0+}} -1 \nn \\
   &=& 
2 r_{EW}^2 - 2 r_{EW} r_{T} \cos(\delta^{EW}-\delta^{T})\cos\phi_3\, 
                                                      =\, 0.43 \pm 0.20\;, 
\label{B+0MB+-PB00M1} \\[6mm]
R_+ -2 ~ &=& 
\frac{\tau^0}{\tau^+}\frac{2\bar{B}_K^{+0}}{\bar{B}_K^{+-}} + 
\frac{\tau^+}{\tau^0}\frac{2\bar{B}_K^{00}}{\bar{B}_K^{0+}} - 2 \nn \\
&=& 2 r_{EW}^2 
          + 2 r_{EW} r_{T} \cos(\delta^{EW}+\delta^{T})\cos\phi_3\, 
                                                      =\, 0.43 \pm 0.21\;, 
\label{B+0MB00M2}
\eea 
so that one can find the EW penguin contributions may be large. 
All terms are including $r_{EW}$ and the deviation of the relation from
zero is finite. These deviations may be an evidence that the EW penguin is
larger than the estimation we expected within the SM. 
Here the errors are determined by adding quadratically all errors. 
We want to solve from these three relations but there are too
many parameters. To estimate the magnitude of the EW penguin
contribution satisfying the experimental data roughly, 
we use Eq.~(\ref{B00B0+}),
\bea
\frac{\tau^+}{\tau^0}\frac{2\bar{B}_K^{00}}{\bar{B}_K^{0+}} -1
&\simeq & - 2 r_{EW} \cos\delta^{EW} + r_{EW}^2\,  \simeq\, 0.17\pm 0.16\;,
\label{B00MB0+M1}   
\eea 
where the $r_C$ and $r_A$ terms were neglected to reduce the number of 
parameter, because the $r_{EW}$ in Eqs.~(\ref{B+0B0+MB+-B00}) and
(\ref{B+0MB+-PB00M1})
should be larger than the usual prediction, therefore, 
$r_C$ and $r_A$ must be quite smaller than $r_{EW}$. 
Using Eqs.~(\ref{B+0B0+MB+-B00}), (\ref{B+0MB00M2}) and
(\ref{B00MB0+M1}),
we can solve them in terms of $r_{EW}$ and if we can respect the central
values of experimental data,
the solution is
\bea
 & & \left(r_{EW},  ~\cos\delta^{EW}, 
   ~r_{T} \cos(\delta^{EW}+\delta^{T})\cos\phi_3 \right)\, 
=\, 
( 0.64,\,  0.19,\, -0.30 )\;. 
\label{solution}
\eea
This solution shows that the EW penguin contribution is too
large compared with the rough theoretical estimation and the strong
phase may not be close to zero.    
The allowed region of $r_{EW}$, 
$\cos\delta^{EW}$ and $r_{T} \cos(\delta^{EW}+\delta^{T})\cos\phi_3 $ at 
$1 \sigma $ level from the three relations, 
Eqs.~(\ref{B+0B0+MB+-B00}), (\ref{B+0MB00M2}) and (\ref{B00MB0+M1}) 
is shown in Fig.~\ref{fig:1}. 
The ``$\times $'' in these figures shows
the central values of the solution, Eq.~(\ref{solution}), and the
dotted lines in the right figure show the bound of 
$r_T \cos(\delta^{EW}+\delta^T)\cos\phi_3 $ for $r_T$ = 0.2 and $\phi_3 =
40^\circ$. The theoretical
prediction should be in the bound at $r_{EW} = 0.14$ but there is no
overlap region with the allowed one at $1 \sigma $ level.    

What we can expect at present are roughly $40^\circ < \phi_3 <
80^\circ $ from CKM fitting and $r_T = 0.2$ with $10\%$ error from the
estimation of the ratio $\frac{B_\pi^{+0}}{B_K^{0+}}$.
Hence the figure shows that $r_{EW}$ will be larger than 
0.2 while the theoretical prediction of $r_{EW}$ is 0.14, and a large 
strong phase difference between gluonic and EW penguins will be 
requested due to $\cos\delta^{EW} < 0.5$~\cite{Y2}. 
Accordingly, to explain the data we may need some contribution from new
physics in the EW-penguin-type contribution with a large phase. 
We have to also consider the possibility of large strong
phases. {}From Eq.~(\ref{B+0B0+MB+-B00}) one can extract more information
about strong phases to satisfy the experimental data. 
Eq.~(\ref{B+0B0+MB+-B00}) is  
\bea 
R_c-R_n\, =\, - 2 r_{EW}^2 \cos2\delta^{EW} 
   + 2 r_{EW} r_{T} \cos(\delta^{EW}+\delta^{T})\cos\phi_3\, =\,
   0.37\pm0.16\, >\, 0 \;.
\eea
The first term has negative sign but $R_c-R_n$ should be positive value. 
For this reason, negative $\cos2\delta^{EW}$, that is, $45^\circ <
|\delta^{EW}| < 135^\circ $ will be favored. 
Namely, it seems to show the strong phase difference should be large. 
Furthermore, considering $S$ in Eq.~(\ref{B+0MB+-PB00M1}), we can obtain
stronger constraint for 
the parameters because the second term has negative sign which is
differ from the case of Eq.~(\ref{B+0B0+MB+-B00}). If
$\cos(\delta^{EW}-\delta^T) \cos\phi_3 $ was positive value, $r_{EW}$
must be larger values to satisfy the condition for $S$.  
In Fig.~\ref{fig:2}, the allowed
regions for $r_{EW},\delta^{T}$ and $\delta^{EW}$ 
from three constraints, Eqs.~(\ref{B+0B0+MB+-B00}), (\ref{B+0MB+-PB00M1}) 
and (\ref{B+0MB00M2}), are plotted. Here we did not use
Eq.~(\ref{B00MB0+M1}) because it is including an assumption which $r_C$ can
be neglected.
One can find that to satisfy the experimental data at 1$\sigma $ level, 
$r_{EW}$ should be larger than about 0.3.   
Under exact flavor SU(3) symmetry, the strong phase difference between
the EW penguin and the color-favored tree, which is called as $\omega$,
\bea
\omega &\equiv& \delta^{EW}-\delta^T
\;,
\eea
should be close to zero because the diagrams are topologically same~\cite{NEU}
and effectively the difference is whether just only the exchanging weak
gauge boson is $W$ or $Z$. 
If it is correct, the constraint for
$\delta^T$ has to influence on $\delta^{EW}$ due to $\delta^{EW} \sim
\delta^T$.   
We consider the direct CP asymmetry to obtain the informations
about strong phase.       
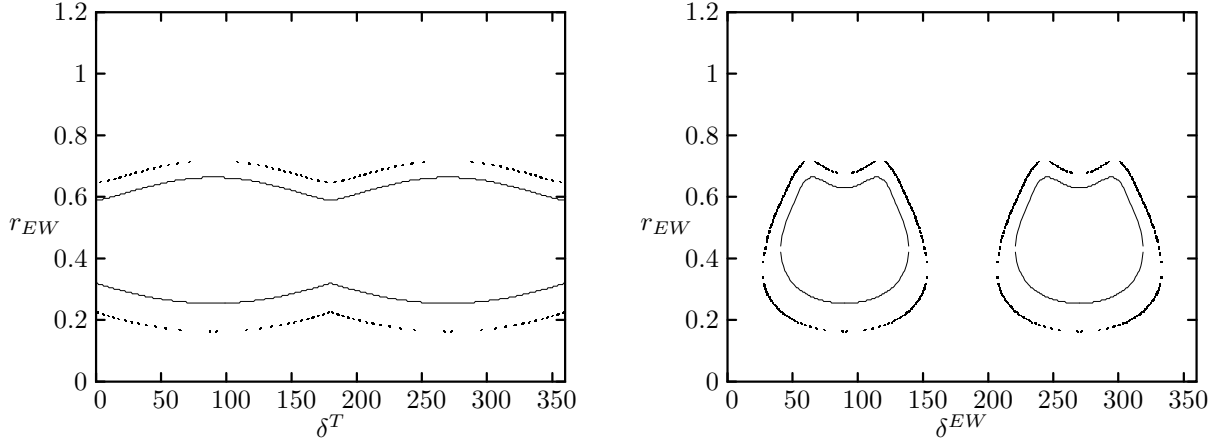
\begin{figure}[thb]
\begin{center}
\begin{minipage}[l]{3.0in}
\setlength{\unitlength}{0.080450pt}
\begin{picture}(2699,2069)(0,0)
\footnotesize
\thicklines \path(370,249)(411,249)
\thicklines \path(2576,249)(2535,249)
\put(329,249){\makebox(0,0)[r]{ 0}}
\thicklines \path(370,539)(411,539)
\thicklines \path(2576,539)(2535,539)
\put(329,539){\makebox(0,0)[r]{ 0.2}}
\thicklines \path(370,829)(411,829)
\thicklines \path(2576,829)(2535,829)
\put(329,829){\makebox(0,0)[r]{ 0.4}}
\thicklines \path(370,1119)(411,1119)
\thicklines \path(2576,1119)(2535,1119)
\put(329,1119){\makebox(0,0)[r]{ 0.6}}
\thicklines \path(370,1408)(411,1408)
\thicklines \path(2576,1408)(2535,1408)
\put(329,1408){\makebox(0,0)[r]{ 0.8}}
\thicklines \path(370,1698)(411,1698)
\thicklines \path(2576,1698)(2535,1698)
\put(329,1698){\makebox(0,0)[r]{ 1}}
\thicklines \path(370,1988)(411,1988)
\thicklines \path(2576,1988)(2535,1988)
\put(329,1988){\makebox(0,0)[r]{ 1.2}}
\thicklines \path(370,249)(370,290)
\thicklines \path(370,1988)(370,1947)
\put(370,166){\makebox(0,0){ 0}}
\thicklines \path(676,249)(676,290)
\thicklines \path(676,1988)(676,1947)
\put(676,166){\makebox(0,0){ 50}}
\thicklines \path(983,249)(983,290)
\thicklines \path(983,1988)(983,1947)
\put(983,166){\makebox(0,0){ 100}}
\thicklines \path(1289,249)(1289,290)
\thicklines \path(1289,1988)(1289,1947)
\put(1289,166){\makebox(0,0){ 150}}
\thicklines \path(1596,249)(1596,290)
\thicklines \path(1596,1988)(1596,1947)
\put(1596,166){\makebox(0,0){ 200}}
\thicklines \path(1902,249)(1902,290)
\thicklines \path(1902,1988)(1902,1947)
\put(1902,166){\makebox(0,0){ 250}}
\thicklines \path(2208,249)(2208,290)
\thicklines \path(2208,1988)(2208,1947)
\put(2208,166){\makebox(0,0){ 300}}
\thicklines \path(2515,249)(2515,290)
\thicklines \path(2515,1988)(2515,1947)
\put(2515,166){\makebox(0,0){ 350}}
\thicklines \path(370,249)(2576,249)(2576,1988)(370,1988)(370,249)
\put(-80,980){\makebox(0,0)[l]{{ $r_{EW}$}}}
\put(1473,42){\makebox(0,0){ $\delta^{T}$ }}
\thinlines \path(370,1104)(370,1104)(376,1104)(382,1104)(388,1104)(395,1104)(401,1104)(407,1111)(413,1111)(419,1111)(425,1111)(431,1111)(437,1119)(444,1119)(450,1119)(456,1119)(462,1126)(468,1126)(474,1126)(480,1126)(486,1133)(493,1133)(499,1133)(505,1133)(511,1140)(517,1140)(523,1140)(529,1140)(535,1147)(542,1147)(548,1147)(554,1147)(560,1155)(566,1155)(572,1155)(578,1155)(584,1155)(591,1162)(597,1162)(603,1162)(609,1162)(615,1169)(621,1169)(627,1169)(633,1169)(640,1169)(646,1176)(652,1176)(658,1176)(664,1176)(670,1176)
\thinlines \path(670,1176)(676,1184)(683,1184)(689,1184)(695,1184)(701,1184)(707,1191)(713,1191)(719,1191)(725,1191)(732,1191)(738,1191)(744,1198)(750,1198)(756,1198)(762,1198)(768,1198)(774,1198)(781,1198)(787,1198)(793,1205)(799,1205)(805,1205)(811,1205)(817,1205)(823,1205)(830,1205)(836,1205)(842,1205)(848,1205)(854,1205)(860,1205)(866,1205)(872,1213)(879,1213)(885,1213)(891,1213)(897,1213)(903,1213)(909,1213)(915,1213)(922,1213)(928,1213)(934,1213)(940,1213)(946,1213)(952,1213)(958,1213)(964,1213)(971,1213)(977,1205)
\thinlines \path(977,1205)(983,1205)(989,1205)(995,1205)(1001,1205)(1007,1205)(1013,1205)(1020,1205)(1026,1205)(1032,1205)(1038,1205)(1044,1205)(1050,1205)(1056,1198)(1062,1198)(1069,1198)(1075,1198)(1081,1198)(1087,1198)(1093,1198)(1099,1198)(1105,1191)(1111,1191)(1118,1191)(1124,1191)(1130,1191)(1136,1191)(1142,1184)(1148,1184)(1154,1184)(1160,1184)(1167,1184)(1173,1176)(1179,1176)(1185,1176)(1191,1176)(1197,1176)(1203,1169)(1210,1169)(1216,1169)(1222,1169)(1228,1169)(1234,1162)(1240,1162)(1246,1162)(1252,1162)(1259,1155)(1265,1155)(1271,1155)(1277,1155)(1283,1155)
\thinlines \path(1283,1155)(1289,1147)(1295,1147)(1301,1147)(1308,1147)(1314,1140)(1320,1140)(1326,1140)(1332,1140)(1338,1133)(1344,1133)(1350,1133)(1357,1133)(1363,1126)(1369,1126)(1375,1126)(1381,1126)(1387,1119)(1393,1119)(1399,1119)(1406,1119)(1412,1111)(1418,1111)(1424,1111)(1430,1111)(1436,1111)(1442,1104)(1448,1104)(1455,1104)(1461,1104)(1467,1104)(1473,1104)(1479,1104)(1485,1104)(1491,1104)(1498,1104)(1504,1104)(1510,1111)(1516,1111)(1522,1111)(1528,1111)(1534,1111)(1540,1119)(1547,1119)(1553,1119)(1559,1119)(1565,1126)(1571,1126)(1577,1126)(1583,1126)(1589,1133)
\thinlines \path(1589,1133)(1596,1133)(1602,1133)(1608,1133)(1614,1140)(1620,1140)(1626,1140)(1632,1140)(1638,1147)(1645,1147)(1651,1147)(1657,1147)(1663,1155)(1669,1155)(1675,1155)(1681,1155)(1687,1155)(1694,1162)(1700,1162)(1706,1162)(1712,1162)(1718,1169)(1724,1169)(1730,1169)(1736,1169)(1743,1169)(1749,1176)(1755,1176)(1761,1176)(1767,1176)(1773,1176)(1779,1184)(1786,1184)(1792,1184)(1798,1184)(1804,1184)(1810,1191)(1816,1191)(1822,1191)(1828,1191)(1835,1191)(1841,1191)(1847,1198)(1853,1198)(1859,1198)(1865,1198)(1871,1198)(1877,1198)(1884,1198)(1890,1198)(1896,1205)
\thinlines \path(1896,1205)(1902,1205)(1908,1205)(1914,1205)(1920,1205)(1926,1205)(1933,1205)(1939,1205)(1945,1205)(1951,1205)(1957,1205)(1963,1205)(1969,1205)(1975,1213)(1982,1213)(1988,1213)(1994,1213)(2000,1213)(2006,1213)(2012,1213)(2018,1213)(2025,1213)(2031,1213)(2037,1213)(2043,1213)(2049,1213)(2055,1213)(2061,1213)(2067,1213)(2074,1213)(2080,1205)(2086,1205)(2092,1205)(2098,1205)(2104,1205)(2110,1205)(2116,1205)(2123,1205)(2129,1205)(2135,1205)(2141,1205)(2147,1205)(2153,1205)(2159,1198)(2165,1198)(2172,1198)(2178,1198)(2184,1198)(2190,1198)(2196,1198)(2202,1198)
\thinlines \path(2202,1198)(2208,1191)(2214,1191)(2221,1191)(2227,1191)(2233,1191)(2239,1191)(2245,1184)(2251,1184)(2257,1184)(2263,1184)(2270,1184)(2276,1176)(2282,1176)(2288,1176)(2294,1176)(2300,1176)(2306,1169)(2313,1169)(2319,1169)(2325,1169)(2331,1169)(2337,1162)(2343,1162)(2349,1162)(2355,1162)(2362,1155)(2368,1155)(2374,1155)(2380,1155)(2386,1155)(2392,1147)(2398,1147)(2404,1147)(2411,1147)(2417,1140)(2423,1140)(2429,1140)(2435,1140)(2441,1133)(2447,1133)(2453,1133)(2460,1133)(2466,1126)(2472,1126)(2478,1126)(2484,1126)(2490,1119)(2496,1119)(2502,1119)(2509,1119)
\thinlines \path(2509,1119)(2515,1111)(2521,1111)(2527,1111)(2533,1111)(2539,1111)(2545,1104)(2551,1104)(2558,1104)(2564,1104)(2570,1104)(2576,1104)
\thinlines \path(370,713)(370,713)(376,713)(382,705)(388,705)(395,705)(401,705)(407,698)(413,698)(419,698)(425,698)(431,691)(437,691)(444,691)(450,691)(456,684)(462,684)(468,684)(474,684)(480,684)(486,677)(493,677)(499,677)(505,677)(511,669)(517,669)(523,669)(529,669)(535,669)(542,662)(548,662)(554,662)(560,662)(566,662)(572,655)(578,655)(584,655)(591,655)(597,655)(603,648)(609,648)(615,648)(621,648)(627,648)(633,648)(640,640)(646,640)(652,640)(658,640)(664,640)(670,640)
\thinlines \path(670,640)(676,640)(683,633)(689,633)(695,633)(701,633)(707,633)(713,633)(719,633)(725,633)(732,626)(738,626)(744,626)(750,626)(756,626)(762,626)(768,626)(774,626)(781,626)(787,626)(793,626)(799,619)(805,619)(811,619)(817,619)(823,619)(830,619)(836,619)(842,619)(848,619)(854,619)(860,619)(866,619)(872,619)(879,619)(885,619)(891,619)(897,619)(903,619)(909,619)(915,619)(922,619)(928,619)(934,619)(940,619)(946,619)(952,619)(958,619)(964,619)(971,619)(977,619)
\thinlines \path(977,619)(983,619)(989,619)(995,619)(1001,619)(1007,619)(1013,619)(1020,619)(1026,619)(1032,619)(1038,619)(1044,619)(1050,626)(1056,626)(1062,626)(1069,626)(1075,626)(1081,626)(1087,626)(1093,626)(1099,626)(1105,626)(1111,626)(1118,633)(1124,633)(1130,633)(1136,633)(1142,633)(1148,633)(1154,633)(1160,633)(1167,640)(1173,640)(1179,640)(1185,640)(1191,640)(1197,640)(1203,640)(1210,648)(1216,648)(1222,648)(1228,648)(1234,648)(1240,648)(1246,655)(1252,655)(1259,655)(1265,655)(1271,655)(1277,662)(1283,662)
\thinlines \path(1283,662)(1289,662)(1295,662)(1301,662)(1308,669)(1314,669)(1320,669)(1326,669)(1332,669)(1338,677)(1344,677)(1350,677)(1357,677)(1363,684)(1369,684)(1375,684)(1381,684)(1387,684)(1393,691)(1399,691)(1406,691)(1412,691)(1418,698)(1424,698)(1430,698)(1436,698)(1442,705)(1448,705)(1455,705)(1461,705)(1467,713)(1473,713)(1479,713)(1485,705)(1491,705)(1498,705)(1504,705)(1510,698)(1516,698)(1522,698)(1528,698)(1534,691)(1540,691)(1547,691)(1553,691)(1559,684)(1565,684)(1571,684)(1577,684)(1583,684)(1589,677)
\thinlines \path(1589,677)(1596,677)(1602,677)(1608,677)(1614,669)(1620,669)(1626,669)(1632,669)(1638,669)(1645,662)(1651,662)(1657,662)(1663,662)(1669,662)(1675,655)(1681,655)(1687,655)(1694,655)(1700,655)(1706,648)(1712,648)(1718,648)(1724,648)(1730,648)(1736,648)(1743,640)(1749,640)(1755,640)(1761,640)(1767,640)(1773,640)(1779,640)(1786,633)(1792,633)(1798,633)(1804,633)(1810,633)(1816,633)(1822,633)(1828,633)(1835,626)(1841,626)(1847,626)(1853,626)(1859,626)(1865,626)(1871,626)(1877,626)(1884,626)(1890,626)(1896,626)
\thinlines \path(1896,626)(1902,619)(1908,619)(1914,619)(1920,619)(1926,619)(1933,619)(1939,619)(1945,619)(1951,619)(1957,619)(1963,619)(1969,619)(1975,619)(1982,619)(1988,619)(1994,619)(2000,619)(2006,619)(2012,619)(2018,619)(2025,619)(2031,619)(2037,619)(2043,619)(2049,619)(2055,619)(2061,619)(2067,619)(2074,619)(2080,619)(2086,619)(2092,619)(2098,619)(2104,619)(2110,619)(2116,619)(2123,619)(2129,619)(2135,619)(2141,619)(2147,619)(2153,626)(2159,626)(2165,626)(2172,626)(2178,626)(2184,626)(2190,626)(2196,626)(2202,626)
\thinlines \path(2202,626)(2208,626)(2214,626)(2221,633)(2227,633)(2233,633)(2239,633)(2245,633)(2251,633)(2257,633)(2263,633)(2270,640)(2276,640)(2282,640)(2288,640)(2294,640)(2300,640)(2306,640)(2313,648)(2319,648)(2325,648)(2331,648)(2337,648)(2343,648)(2349,655)(2355,655)(2362,655)(2368,655)(2374,655)(2380,662)(2386,662)(2392,662)(2398,662)(2404,662)(2411,669)(2417,669)(2423,669)(2429,669)(2435,669)(2441,677)(2447,677)(2453,677)(2460,677)(2466,684)(2472,684)(2478,684)(2484,684)(2490,684)(2496,691)(2502,691)(2509,691)
\thinlines \path(2509,691)(2515,691)(2521,698)(2527,698)(2533,698)(2539,698)(2545,705)(2551,705)(2558,705)(2564,705)(2570,713)(2576,713)
\thinlines \dashline[-20]{12}(370,1184)(370,1184)(376,1184)(382,1184)(388,1184)(395,1191)(401,1191)(407,1191)(413,1191)(419,1191)(425,1191)(431,1198)(437,1198)(444,1198)(450,1198)(456,1198)(462,1205)(468,1205)(474,1205)(480,1205)(486,1213)(493,1213)(499,1213)(505,1213)(511,1220)(517,1220)(523,1220)(529,1220)(535,1227)(542,1227)(548,1227)(554,1227)(560,1234)(566,1234)(572,1234)(578,1234)(584,1234)(591,1242)(597,1242)(603,1242)(609,1242)(615,1249)(621,1249)(627,1249)(633,1249)(640,1249)(646,1256)(652,1256)(658,1256)(664,1256)(670,1256)
\thinlines \dashline[-20]{12}(670,1256)(676,1263)(683,1263)(689,1263)(695,1263)(701,1263)(707,1263)(713,1271)(719,1271)(725,1271)(732,1271)(738,1271)(744,1271)(750,1271)(756,1278)(762,1278)(768,1278)(774,1278)(781,1278)(787,1278)(793,1278)(799,1278)(805,1278)(811,1285)(817,1285)(823,1285)(830,1285)(836,1285)(842,1285)(848,1285)(854,1285)(860,1285)(866,1285)(872,1285)(879,1285)(885,1285)(891,1285)(897,1285)(903,1285)(909,1285)(915,1285)(922,1285)(928,1285)(934,1285)(940,1285)(946,1285)(952,1285)(958,1285)(964,1285)(971,1285)(977,1285)
\thinlines \dashline[-20]{12}(977,1285)(983,1285)(989,1285)(995,1285)(1001,1285)(1007,1285)(1013,1285)(1020,1285)(1026,1285)(1032,1285)(1038,1278)(1044,1278)(1050,1278)(1056,1278)(1062,1278)(1069,1278)(1075,1278)(1081,1278)(1087,1278)(1093,1271)(1099,1271)(1105,1271)(1111,1271)(1118,1271)(1124,1271)(1130,1271)(1136,1263)(1142,1263)(1148,1263)(1154,1263)(1160,1263)(1167,1263)(1173,1256)(1179,1256)(1185,1256)(1191,1256)(1197,1256)(1203,1249)(1210,1249)(1216,1249)(1222,1249)(1228,1249)(1234,1242)(1240,1242)(1246,1242)(1252,1242)(1259,1234)(1265,1234)(1271,1234)(1277,1234)(1283,1234)
\thinlines \dashline[-20]{12}(1283,1234)(1289,1227)(1295,1227)(1301,1227)(1308,1227)(1314,1220)(1320,1220)(1326,1220)(1332,1220)(1338,1213)(1344,1213)(1350,1213)(1357,1213)(1363,1205)(1369,1205)(1375,1205)(1381,1205)(1387,1198)(1393,1198)(1399,1198)(1406,1198)(1412,1198)(1418,1191)(1424,1191)(1430,1191)(1436,1191)(1442,1191)(1448,1191)(1455,1184)(1461,1184)(1467,1184)(1473,1184)(1479,1184)(1485,1184)(1491,1184)(1498,1191)(1504,1191)(1510,1191)(1516,1191)(1522,1191)(1528,1191)(1534,1198)(1540,1198)(1547,1198)(1553,1198)(1559,1198)(1565,1205)(1571,1205)(1577,1205)(1583,1205)(1589,1213)
\thinlines \dashline[-20]{12}(1589,1213)(1596,1213)(1602,1213)(1608,1213)(1614,1220)(1620,1220)(1626,1220)(1632,1220)(1638,1227)(1645,1227)(1651,1227)(1657,1227)(1663,1234)(1669,1234)(1675,1234)(1681,1234)(1687,1234)(1694,1242)(1700,1242)(1706,1242)(1712,1242)(1718,1249)(1724,1249)(1730,1249)(1736,1249)(1743,1249)(1749,1256)(1755,1256)(1761,1256)(1767,1256)(1773,1256)(1779,1263)(1786,1263)(1792,1263)(1798,1263)(1804,1263)(1810,1263)(1816,1271)(1822,1271)(1828,1271)(1835,1271)(1841,1271)(1847,1271)(1853,1271)(1859,1278)(1865,1278)(1871,1278)(1877,1278)(1884,1278)(1890,1278)(1896,1278)
\thinlines \dashline[-20]{12}(1896,1278)(1902,1278)(1908,1278)(1914,1285)(1920,1285)(1926,1285)(1933,1285)(1939,1285)(1945,1285)(1951,1285)(1957,1285)(1963,1285)(1969,1285)(1975,1285)(1982,1285)(1988,1285)(1994,1285)(2000,1285)(2006,1285)(2012,1285)(2018,1285)(2025,1285)(2031,1285)(2037,1285)(2043,1285)(2049,1285)(2055,1285)(2061,1285)(2067,1285)(2074,1285)(2080,1285)(2086,1285)(2092,1285)(2098,1285)(2104,1285)(2110,1285)(2116,1285)(2123,1285)(2129,1285)(2135,1285)(2141,1278)(2147,1278)(2153,1278)(2159,1278)(2165,1278)(2172,1278)(2178,1278)(2184,1278)(2190,1278)(2196,1271)(2202,1271)
\thinlines \dashline[-20]{12}(2202,1271)(2208,1271)(2214,1271)(2221,1271)(2227,1271)(2233,1271)(2239,1263)(2245,1263)(2251,1263)(2257,1263)(2263,1263)(2270,1263)(2276,1256)(2282,1256)(2288,1256)(2294,1256)(2300,1256)(2306,1249)(2313,1249)(2319,1249)(2325,1249)(2331,1249)(2337,1242)(2343,1242)(2349,1242)(2355,1242)(2362,1234)(2368,1234)(2374,1234)(2380,1234)(2386,1234)(2392,1227)(2398,1227)(2404,1227)(2411,1227)(2417,1220)(2423,1220)(2429,1220)(2435,1220)(2441,1213)(2447,1213)(2453,1213)(2460,1213)(2466,1205)(2472,1205)(2478,1205)(2484,1205)(2490,1198)(2496,1198)(2502,1198)(2509,1198)
\thinlines \dashline[-20]{12}(2509,1198)(2515,1198)(2521,1191)(2527,1191)(2533,1191)(2539,1191)(2545,1191)(2551,1191)(2558,1184)(2564,1184)(2570,1184)(2576,1184)
\thinlines \dashline[-20]{12}(370,575)(370,575)(376,575)(382,568)(388,568)(395,568)(401,568)(407,561)(413,561)(419,561)(425,561)(431,553)(437,553)(444,553)(450,553)(456,546)(462,546)(468,546)(474,546)(480,539)(486,539)(493,539)(499,539)(505,539)(511,532)(517,532)(523,532)(529,532)(535,532)(542,524)(548,524)(554,524)(560,524)(566,524)(572,517)(578,517)(584,517)(591,517)(597,517)(603,517)(609,510)(615,510)(621,510)(627,510)(633,510)(640,510)(646,510)(652,503)(658,503)(664,503)(670,503)
\thinlines \dashline[-20]{12}(670,503)(676,503)(683,503)(689,503)(695,503)(701,495)(707,495)(713,495)(719,495)(725,495)(732,495)(738,495)(744,495)(750,495)(756,495)(762,495)(768,488)(774,488)(781,488)(787,488)(793,488)(799,488)(805,488)(811,488)(817,488)(823,488)(830,488)(836,488)(842,488)(848,488)(854,488)(860,488)(866,488)(872,488)(879,488)(885,488)(891,488)(897,488)(903,488)(909,481)(915,481)(922,481)(928,481)(934,481)(940,488)(946,488)(952,488)(958,488)(964,488)(971,488)(977,488)
\thinlines \dashline[-20]{12}(977,488)(983,488)(989,488)(995,488)(1001,488)(1007,488)(1013,488)(1020,488)(1026,488)(1032,488)(1038,488)(1044,488)(1050,488)(1056,488)(1062,488)(1069,488)(1075,488)(1081,495)(1087,495)(1093,495)(1099,495)(1105,495)(1111,495)(1118,495)(1124,495)(1130,495)(1136,495)(1142,495)(1148,503)(1154,503)(1160,503)(1167,503)(1173,503)(1179,503)(1185,503)(1191,503)(1197,510)(1203,510)(1210,510)(1216,510)(1222,510)(1228,510)(1234,510)(1240,517)(1246,517)(1252,517)(1259,517)(1265,517)(1271,517)(1277,524)(1283,524)
\thinlines \dashline[-20]{12}(1283,524)(1289,524)(1295,524)(1301,524)(1308,532)(1314,532)(1320,532)(1326,532)(1332,532)(1338,539)(1344,539)(1350,539)(1357,539)(1363,539)(1369,546)(1375,546)(1381,546)(1387,546)(1393,553)(1399,553)(1406,553)(1412,553)(1418,561)(1424,561)(1430,561)(1436,561)(1442,568)(1448,568)(1455,568)(1461,568)(1467,575)(1473,575)(1479,575)(1485,568)(1491,568)(1498,568)(1504,568)(1510,561)(1516,561)(1522,561)(1528,561)(1534,553)(1540,553)(1547,553)(1553,553)(1559,546)(1565,546)(1571,546)(1577,546)(1583,539)(1589,539)
\thinlines \dashline[-20]{12}(1589,539)(1596,539)(1602,539)(1608,539)(1614,532)(1620,532)(1626,532)(1632,532)(1638,532)(1645,524)(1651,524)(1657,524)(1663,524)(1669,524)(1675,517)(1681,517)(1687,517)(1694,517)(1700,517)(1706,517)(1712,510)(1718,510)(1724,510)(1730,510)(1736,510)(1743,510)(1749,510)(1755,503)(1761,503)(1767,503)(1773,503)(1779,503)(1786,503)(1792,503)(1798,503)(1804,495)(1810,495)(1816,495)(1822,495)(1828,495)(1835,495)(1841,495)(1847,495)(1853,495)(1859,495)(1865,495)(1871,488)(1877,488)(1884,488)(1890,488)(1896,488)
\thinlines \dashline[-20]{12}(1896,488)(1902,488)(1908,488)(1914,488)(1920,488)(1926,488)(1933,488)(1939,488)(1945,488)(1951,488)(1957,488)(1963,488)(1969,488)(1975,488)(1982,488)(1988,488)(1994,488)(2000,488)(2006,488)(2012,481)(2018,481)(2025,481)(2031,481)(2037,481)(2043,488)(2049,488)(2055,488)(2061,488)(2067,488)(2074,488)(2080,488)(2086,488)(2092,488)(2098,488)(2104,488)(2110,488)(2116,488)(2123,488)(2129,488)(2135,488)(2141,488)(2147,488)(2153,488)(2159,488)(2165,488)(2172,488)(2178,488)(2184,495)(2190,495)(2196,495)(2202,495)
\thinlines \dashline[-20]{12}(2202,495)(2208,495)(2214,495)(2221,495)(2227,495)(2233,495)(2239,495)(2245,495)(2251,503)(2257,503)(2263,503)(2270,503)(2276,503)(2282,503)(2288,503)(2294,503)(2300,510)(2306,510)(2313,510)(2319,510)(2325,510)(2331,510)(2337,510)(2343,517)(2349,517)(2355,517)(2362,517)(2368,517)(2374,517)(2380,524)(2386,524)(2392,524)(2398,524)(2404,524)(2411,532)(2417,532)(2423,532)(2429,532)(2435,532)(2441,539)(2447,539)(2453,539)(2460,539)(2466,539)(2472,546)(2478,546)(2484,546)(2490,546)(2496,553)(2502,553)(2509,553)
\thinlines \dashline[-20]{12}(2509,553)(2515,553)(2521,561)(2527,561)(2533,561)(2539,561)(2545,568)(2551,568)(2558,568)(2564,568)(2570,575)(2576,575)
\end{picture}
\end{minipage}
\hspace*{5mm}
\begin{minipage}[r]{3.0in}
\setlength{\unitlength}{0.080450pt}
\begin{picture}(2699,2069)(0,0)
\footnotesize
\thicklines \path(370,249)(411,249)
\thicklines \path(2576,249)(2535,249)
\put(329,249){\makebox(0,0)[r]{ 0}}
\thicklines \path(370,539)(411,539)
\thicklines \path(2576,539)(2535,539)
\put(329,539){\makebox(0,0)[r]{ 0.2}}
\thicklines \path(370,829)(411,829)
\thicklines \path(2576,829)(2535,829)
\put(329,829){\makebox(0,0)[r]{ 0.4}}
\thicklines \path(370,1119)(411,1119)
\thicklines \path(2576,1119)(2535,1119)
\put(329,1119){\makebox(0,0)[r]{ 0.6}}
\thicklines \path(370,1408)(411,1408)
\thicklines \path(2576,1408)(2535,1408)
\put(329,1408){\makebox(0,0)[r]{ 0.8}}
\thicklines \path(370,1698)(411,1698)
\thicklines \path(2576,1698)(2535,1698)
\put(329,1698){\makebox(0,0)[r]{ 1}}
\thicklines \path(370,1988)(411,1988)
\thicklines \path(2576,1988)(2535,1988)
\put(329,1988){\makebox(0,0)[r]{ 1.2}}
\thicklines \path(370,249)(370,290)
\thicklines \path(370,1988)(370,1947)
\put(370,166){\makebox(0,0){ 0}}
\thicklines \path(676,249)(676,290)
\thicklines \path(676,1988)(676,1947)
\put(676,166){\makebox(0,0){ 50}}
\thicklines \path(983,249)(983,290)
\thicklines \path(983,1988)(983,1947)
\put(983,166){\makebox(0,0){ 100}}
\thicklines \path(1289,249)(1289,290)
\thicklines \path(1289,1988)(1289,1947)
\put(1289,166){\makebox(0,0){ 150}}
\thicklines \path(1596,249)(1596,290)
\thicklines \path(1596,1988)(1596,1947)
\put(1596,166){\makebox(0,0){ 200}}
\thicklines \path(1902,249)(1902,290)
\thicklines \path(1902,1988)(1902,1947)
\put(1902,166){\makebox(0,0){ 250}}
\thicklines \path(2208,249)(2208,290)
\thicklines \path(2208,1988)(2208,1947)
\put(2208,166){\makebox(0,0){ 300}}
\thicklines \path(2515,249)(2515,290)
\thicklines \path(2515,1988)(2515,1947)
\put(2515,166){\makebox(0,0){ 350}}
\thicklines \path(370,249)(2576,249)(2576,1988)(370,1988)(370,249)
\put(-80,980){\makebox(0,0)[l]{{ $r_{EW}$}}}
\put(1473,42){\makebox(0,0){ $\delta^{EW}$ }}
\thinlines \path(621,887)(621,887)(627,937)(633,959)(640,981)(646,1003)(652,1017)(658,1032)(664,1046)(670,1061)(676,1075)(683,1090)(689,1104)(695,1119)(701,1133)(707,1147)(713,1162)(719,1169)(725,1184)(732,1191)(738,1198)(744,1198)(750,1205)(756,1205)(762,1213)(768,1213)(774,1213)(781,1213)(787,1205)(793,1205)(799,1205)(805,1198)(811,1198)(817,1198)(823,1191)(830,1191)(836,1184)(842,1184)(848,1176)(854,1176)(860,1176)(866,1169)(872,1169)(879,1169)(885,1162)(891,1162)(897,1162)(903,1162)(909,1162)(915,1162)(922,1162)
\thinlines \path(922,1162)(928,1162)(934,1162)(940,1162)(946,1162)(952,1162)(958,1162)(964,1169)(971,1169)(977,1169)(983,1176)(989,1176)(995,1176)(1001,1184)(1007,1184)(1013,1191)(1020,1191)(1026,1198)(1032,1198)(1038,1198)(1044,1205)(1050,1205)(1056,1205)(1062,1213)(1069,1213)(1075,1213)(1081,1213)(1087,1205)(1093,1205)(1099,1198)(1105,1198)(1111,1191)(1118,1184)(1124,1169)(1130,1162)(1136,1147)(1142,1133)(1148,1119)(1154,1104)(1160,1090)(1167,1075)(1173,1061)(1179,1046)(1185,1032)(1191,1017)(1197,1003)(1203,981)(1210,959)(1216,937)(1222,887)
\thinlines \path(1724,887)(1724,887)(1730,937)(1736,959)(1743,981)(1749,1003)(1755,1017)(1761,1032)(1767,1046)(1773,1061)(1779,1075)(1786,1090)(1792,1104)(1798,1119)(1804,1133)(1810,1147)(1816,1162)(1822,1169)(1828,1184)(1835,1191)(1841,1198)(1847,1198)(1853,1205)(1859,1205)(1865,1213)(1871,1213)(1877,1213)(1884,1213)(1890,1205)(1896,1205)(1902,1205)(1908,1198)(1914,1198)(1920,1198)(1926,1191)(1933,1191)(1939,1184)(1945,1184)(1951,1176)(1957,1176)(1963,1176)(1969,1169)(1975,1169)(1982,1169)(1988,1162)(1994,1162)(2000,1162)(2006,1162)(2012,1162)(2018,1162)(2025,1162)
\thinlines \path(2025,1162)(2031,1162)(2037,1162)(2043,1162)(2049,1162)(2055,1162)(2061,1162)(2067,1169)(2074,1169)(2080,1169)(2086,1176)(2092,1176)(2098,1176)(2104,1184)(2110,1184)(2116,1191)(2123,1191)(2129,1198)(2135,1198)(2141,1198)(2147,1205)(2153,1205)(2159,1205)(2165,1213)(2172,1213)(2178,1213)(2184,1213)(2190,1205)(2196,1205)(2202,1198)(2208,1198)(2214,1191)(2221,1184)(2227,1169)(2233,1162)(2239,1147)(2245,1133)(2251,1119)(2257,1104)(2263,1090)(2270,1075)(2276,1061)(2282,1046)(2288,1032)(2294,1017)(2300,1003)(2306,981)(2313,959)(2319,937)(2325,887)
\thinlines \path(621,858)(621,858)(627,814)(633,792)(640,771)(646,763)(652,749)(658,742)(664,727)(670,720)(676,713)(683,705)(689,698)(695,691)(701,691)(707,684)(713,677)(719,677)(725,669)(732,669)(738,662)(744,662)(750,655)(756,655)(762,648)(768,648)(774,648)(781,640)(787,640)(793,640)(799,633)(805,633)(811,633)(817,633)(823,626)(830,626)(836,626)(842,626)(848,626)(854,626)(860,619)(866,619)(872,619)(879,619)(885,619)(891,619)(897,619)(903,619)(909,619)(915,619)(922,619)
\thinlines \path(922,619)(928,619)(934,619)(940,619)(946,619)(952,619)(958,619)(964,619)(971,619)(977,619)(983,619)(989,626)(995,626)(1001,626)(1007,626)(1013,626)(1020,626)(1026,633)(1032,633)(1038,633)(1044,633)(1050,640)(1056,640)(1062,640)(1069,648)(1075,648)(1081,648)(1087,655)(1093,655)(1099,662)(1105,662)(1111,669)(1118,669)(1124,677)(1130,677)(1136,684)(1142,691)(1148,691)(1154,698)(1160,705)(1167,713)(1173,720)(1179,727)(1185,742)(1191,749)(1197,763)(1203,771)(1210,792)(1216,814)(1222,858)
\thinlines \path(1724,858)(1724,858)(1730,814)(1736,792)(1743,771)(1749,763)(1755,749)(1761,742)(1767,727)(1773,720)(1779,713)(1786,705)(1792,698)(1798,691)(1804,691)(1810,684)(1816,677)(1822,677)(1828,669)(1835,669)(1841,662)(1847,662)(1853,655)(1859,655)(1865,648)(1871,648)(1877,648)(1884,640)(1890,640)(1896,640)(1902,633)(1908,633)(1914,633)(1920,633)(1926,626)(1933,626)(1939,626)(1945,626)(1951,626)(1957,626)(1963,619)(1969,619)(1975,619)(1982,619)(1988,619)(1994,619)(2000,619)(2006,619)(2012,619)(2018,619)(2025,619)
\thinlines \path(2025,619)(2031,619)(2037,619)(2043,619)(2049,619)(2055,619)(2061,619)(2067,619)(2074,619)(2080,619)(2086,619)(2092,626)(2098,626)(2104,626)(2110,626)(2116,626)(2123,626)(2129,633)(2135,633)(2141,633)(2147,633)(2153,640)(2159,640)(2165,640)(2172,648)(2178,648)(2184,648)(2190,655)(2196,655)(2202,662)(2208,662)(2214,669)(2221,669)(2227,677)(2233,677)(2239,684)(2245,691)(2251,691)(2257,698)(2263,705)(2270,713)(2276,720)(2282,727)(2288,742)(2294,749)(2300,763)(2306,771)(2313,792)(2319,814)(2325,858)
\thinlines
\dashline[-20]{12}(535,807)(535,807)(542,850)(548,879)(554,908)(560,930)(566,952)(572,974)(578,988)(584,1003)(591,1024)(597,1039)(603,1053)(609,1068)(615,1082)(621,1090)(627,1104)(633,1119)(640,1133)(646,1147)(652,1162)(658,1169)(664,1184)(670,1198)(676,1213)(683,1220)(689,1234)(695,1242)(701,1249)(707,1256)(713,1263)(719,1271)(725,1278)(732,1285)(738,1285)(744,1285)(750,1285)(756,1285)(762,1285)(768,1285)(774,1285)(781,1285)(787,1278)(793,1278)(799,1271)(805,1271)(811,1263)(817,1263)(823,1256)(830,1256)(836,1249)
\thinlines \dashline[-20]{12}(836,1249)(842,1249)(848,1242)(854,1242)(860,1242)(866,1234)(872,1234)(879,1234)(885,1227)(891,1227)(897,1227)(903,1227)(909,1227)(915,1227)(922,1227)(928,1227)(934,1227)(940,1227)(946,1227)(952,1227)(958,1227)(964,1234)(971,1234)(977,1234)(983,1242)(989,1242)(995,1242)(1001,1249)(1007,1249)(1013,1256)(1020,1256)(1026,1263)(1032,1263)(1038,1271)(1044,1271)(1050,1278)(1056,1278)(1062,1285)(1069,1285)(1075,1285)(1081,1285)(1087,1285)(1093,1285)(1099,1285)(1105,1285)(1111,1285)(1118,1278)(1124,1271)(1130,1263)(1136,1256)(1142,1249)
\thinlines \dashline[-20]{12}(1142,1249)(1148,1242)(1154,1234)(1160,1220)(1167,1213)(1173,1198)(1179,1184)(1185,1169)(1191,1162)(1197,1147)(1203,1133)(1210,1119)(1216,1104)(1222,1090)(1228,1082)(1234,1068)(1240,1053)(1246,1039)(1252,1024)(1259,1003)(1265,988)(1271,974)(1277,952)(1283,930)(1289,908)(1295,879)(1301,850)(1308,807)
\thinlines \dashline[-20]{12}(1638,807)(1638,807)(1645,850)(1651,879)(1657,908)(1663,930)(1669,952)(1675,974)(1681,988)(1687,1003)(1694,1024)(1700,1039)(1706,1053)(1712,1068)(1718,1082)(1724,1090)(1730,1104)(1736,1119)(1743,1133)(1749,1147)(1755,1162)(1761,1169)(1767,1184)(1773,1198)(1779,1213)(1786,1220)(1792,1234)(1798,1242)(1804,1249)(1810,1256)(1816,1263)(1822,1271)(1828,1278)(1835,1285)(1841,1285)(1847,1285)(1853,1285)(1859,1285)(1865,1285)(1871,1285)(1877,1285)(1884,1285)(1890,1278)(1896,1278)(1902,1271)(1908,1271)(1914,1263)(1920,1263)(1926,1256)(1933,1256)(1939,1249)
\thinlines \dashline[-20]{12}(1939,1249)(1945,1249)(1951,1242)(1957,1242)(1963,1242)(1969,1234)(1975,1234)(1982,1234)(1988,1227)(1994,1227)(2000,1227)(2006,1227)(2012,1227)(2018,1227)(2025,1227)(2031,1227)(2037,1227)(2043,1227)(2049,1227)(2055,1227)(2061,1227)(2067,1234)(2074,1234)(2080,1234)(2086,1242)(2092,1242)(2098,1242)(2104,1249)(2110,1249)(2116,1256)(2123,1256)(2129,1263)(2135,1263)(2141,1271)(2147,1271)(2153,1278)(2159,1278)(2165,1285)(2172,1285)(2178,1285)(2184,1285)(2190,1285)(2196,1285)(2202,1285)(2208,1285)(2214,1285)(2221,1278)(2227,1271)(2233,1263)(2239,1256)(2245,1249)
\thinlines\dashline[-20]{12}(2245,1249)(2251,1242)(2257,1234)(2263,1220)(2270,1213)(2276,1198)(2282,1184)(2288,1169)(2294,1162)(2300,1147)(2306,1133)(2313,1119)(2319,1104)(2325,1090)(2331,1082)(2337,1068)(2343,1053)(2349,1039)(2355,1024)(2362,1003)(2368,988)(2374,974)(2380,952)(2386,930)(2392,908)(2398,879)(2404,850)(2411,807)
\thinlines \dashline[-20]{12}(535,742)(535,742)(542,705)(548,684)(554,669)(560,655)(566,648)(572,633)(578,626)(584,619)(591,611)(597,604)(603,597)(609,590)(615,582)(621,575)(627,575)(633,568)(640,561)(646,561)(652,553)(658,553)(664,546)(670,546)(676,539)(683,539)(689,532)(695,532)(701,524)(707,524)(713,524)(719,517)(725,517)(732,517)(738,510)(744,510)(750,510)(756,510)(762,503)(768,503)(774,503)(781,503)(787,503)(793,495)(799,495)(805,495)(811,495)(817,495)(823,495)(830,488)(836,488)
\thinlines \dashline[-20]{12}(836,488)(842,488)(848,488)(854,488)(860,488)(866,488)(872,488)(879,488)(885,488)(891,488)(897,488)(903,488)(909,488)(915,481)(922,481)(928,481)(934,488)(940,488)(946,488)(952,488)(958,488)(964,488)(971,488)(977,488)(983,488)(989,488)(995,488)(1001,488)(1007,488)(1013,488)(1020,495)(1026,495)(1032,495)(1038,495)(1044,495)(1050,495)(1056,503)(1062,503)(1069,503)(1075,503)(1081,503)(1087,510)(1093,510)(1099,510)(1105,510)(1111,517)(1118,517)(1124,517)(1130,524)(1136,524)(1142,524)
\thinlines \dashline[-20]{12}(1142,524)(1148,532)(1154,532)(1160,539)(1167,539)(1173,546)(1179,546)(1185,553)(1191,553)(1197,561)(1203,561)(1210,568)(1216,575)(1222,575)(1228,582)(1234,590)(1240,597)(1246,604)(1252,611)(1259,619)(1265,626)(1271,633)(1277,648)(1283,655)(1289,669)(1295,684)(1301,705)(1308,742)
\thinlines \dashline[-20]{12}(1638,742)(1638,742)(1645,705)(1651,684)(1657,669)(1663,655)(1669,648)(1675,633)(1681,626)(1687,619)(1694,611)(1700,604)(1706,597)(1712,590)(1718,582)(1724,575)(1730,575)(1736,568)(1743,561)(1749,561)(1755,553)(1761,553)(1767,546)(1773,546)(1779,539)(1786,539)(1792,532)(1798,532)(1804,524)(1810,524)(1816,524)(1822,517)(1828,517)(1835,517)(1841,510)(1847,510)(1853,510)(1859,510)(1865,503)(1871,503)(1877,503)(1884,503)(1890,503)(1896,495)(1902,495)(1908,495)(1914,495)(1920,495)(1926,495)(1933,488)(1939,488)
\thinlines \dashline[-20]{12}(1939,488)(1945,488)(1951,488)(1957,488)(1963,488)(1969,488)(1975,488)(1982,488)(1988,488)(1994,488)(2000,488)(2006,488)(2012,488)(2018,481)(2025,481)(2031,481)(2037,488)(2043,488)(2049,488)(2055,488)(2061,488)(2067,488)(2074,488)(2080,488)(2086,488)(2092,488)(2098,488)(2104,488)(2110,488)(2116,488)(2123,495)(2129,495)(2135,495)(2141,495)(2147,495)(2153,495)(2159,503)(2165,503)(2172,503)(2178,503)(2184,503)(2190,510)(2196,510)(2202,510)(2208,510)(2214,517)(2221,517)(2227,517)(2233,524)(2239,524)(2245,524)
\thinlines \dashline[-20]{12}(2245,524)(2251,532)(2257,532)(2263,539)(2270,539)(2276,546)(2282,546)(2288,553)(2294,553)(2300,561)(2306,561)(2313,568)(2319,575)(2325,575)(2331,582)(2337,590)(2343,597)(2349,604)(2355,611)(2362,619)(2368,626)(2374,633)(2380,648)(2386,655)(2392,669)(2398,684)(2404,705)(2411,742)
\end{picture}
\end{minipage} 
\caption{{}From Eqs.~(\ref{B+0B0+MB+-B00}), (\ref{B+0MB+-PB00M1}) 
and (\ref{B+0MB00M2}) the allowed regions for $r_{EW}, \delta^T$ and
    $\delta^{EW}$ 
    at $r_T = 0.2 $ and  
$40^\circ <
    \phi_3 <
     80^\circ $. The solid (dashed) line shows the $1\sigma $
    ($2\sigma $) bound.}
    \label{fig:2}
\end{center}              
\end{figure}

\begin{figure}[thb]
\begin{center}
\setlength{\unitlength}{0.115450pt}
\begin{picture}(2400,1800)(0,0)
\footnotesize
\thicklines \path(411,249)(452,249)
\thicklines \path(2276,249)(2235,249)
\put(370,249){\makebox(0,0)[r]{-0.4}}
\thicklines \path(411,433)(452,433)
\thicklines \path(2276,433)(2235,433)
\put(370,433){\makebox(0,0)[r]{-0.35}}
\thicklines \path(411,616)(452,616)
\thicklines \path(2276,616)(2235,616)
\put(370,616){\makebox(0,0)[r]{-0.3}}
\thicklines \path(411,800)(452,800)
\thicklines \path(2276,800)(2235,800)
\put(370,800){\makebox(0,0)[r]{-0.25}}
\thicklines \path(411,983)(452,983)
\thicklines \path(2276,983)(2235,983)
\put(370,983){\makebox(0,0)[r]{-0.2}}
\thicklines \path(411,1167)(452,1167)
\thicklines \path(2276,1167)(2235,1167)
\put(370,1167){\makebox(0,0)[r]{-0.15}}
\thicklines \path(411,1351)(452,1351)
\thicklines \path(2276,1351)(2235,1351)
\put(370,1351){\makebox(0,0)[r]{-0.1}}
\thicklines \path(411,1534)(452,1534)
\thicklines \path(2276,1534)(2235,1534)
\put(370,1534){\makebox(0,0)[r]{-0.05}}
\thicklines \path(411,1718)(452,1718)
\thicklines \path(2276,1718)(2235,1718)
\put(370,1718){\makebox(0,0)[r]{ 0}}
\thicklines \path(411,249)(411,290)
\thicklines \path(411,1718)(411,1677)
\put(411,166){\makebox(0,0){ 0}}
\thicklines \path(618,249)(618,290)
\thicklines \path(618,1718)(618,1677)
\put(618,166){\makebox(0,0){ 20}}
\thicklines \path(825,249)(825,290)
\thicklines \path(825,1718)(825,1677)
\put(825,166){\makebox(0,0){ 40}}
\thicklines \path(1033,249)(1033,290)
\thicklines \path(1033,1718)(1033,1677)
\put(1033,166){\makebox(0,0){ 60}}
\thicklines \path(1240,249)(1240,290)
\thicklines \path(1240,1718)(1240,1677)
\put(1240,166){\makebox(0,0){ 80}}
\thicklines \path(1447,249)(1447,290)
\thicklines \path(1447,1718)(1447,1677)
\put(1447,166){\makebox(0,0){ 100}}
\thicklines \path(1654,249)(1654,290)
\thicklines \path(1654,1718)(1654,1677)
\put(1654,166){\makebox(0,0){ 120}}
\thicklines \path(1862,249)(1862,290)
\thicklines \path(1862,1718)(1862,1677)
\put(1862,166){\makebox(0,0){ 140}}
\thicklines \path(2069,249)(2069,290)
\thicklines \path(2069,1718)(2069,1677)
\put(2069,166){\makebox(0,0){ 160}}
\thicklines \path(2276,249)(2276,290)
\thicklines \path(2276,1718)(2276,1677)
\put(2276,166){\makebox(0,0){ 180}}
\thicklines \path(411,249)(2276,249)(2276,1718)(411,1718)(411,249)
\put(-20,1000){\makebox(0,0)[l]{{$A_{CP}^{\pm }$}}}
\put(1240,983){\makebox(0,0)[l]{{$\phi_3=40^\circ$}}}
\put(1470,683){\makebox(0,0)[l]{{$60^\circ $}}}
\put(1570,343){\makebox(0,0)[l]{{$80^\circ $}}}
\put(1343,0){\makebox(0,0){$\delta^T $ }}
\thinlines \path(411,1718)(411,1718)(487,1562)(569,1396)(647,1250)(721,1122)(801,1002)(877,906)(916,865)(958,827)(995,798)(1036,773)(1075,755)(1096,747)(1116,742)(1126,740)(1137,738)(1146,736)(1156,735)(1161,735)(1166,734)(1170,734)(1172,734)(1175,734)(1177,734)(1179,734)(1181,734)(1184,734)(1186,734)(1188,734)(1190,734)(1192,734)(1195,734)(1197,734)(1202,734)(1206,734)(1211,735)(1220,736)(1231,737)(1252,741)(1272,746)(1312,760)(1348,776)(1427,823)(1502,881)(1582,953)(1659,1031)(1734,1110)(1813,1198)(1889,1284)(1969,1376)
\thinlines \path(1969,1376)(2047,1464)(2122,1548)(2202,1636)(2276,1718)
\thinlines \path(411,1718)(411,1718)(487,1524)(569,1318)(647,1135)(721,973)(801,819)(877,692)(958,582)(999,538)(1036,503)(1075,474)(1111,452)(1148,436)(1168,430)(1178,428)(1189,425)(1198,424)(1208,423)(1214,422)(1217,422)(1220,422)(1222,422)(1225,422)(1226,422)(1227,422)(1229,422)(1230,422)(1231,422)(1233,422)(1235,422)(1236,422)(1238,422)(1239,422)(1240,422)(1243,422)(1246,422)(1250,422)(1255,423)(1259,423)(1269,424)(1279,426)(1290,428)(1309,433)(1346,447)(1385,466)(1426,493)(1500,554)(1579,635)(1655,727)(1736,837)
\thinlines \path(1736,837)(1814,952)(1889,1068)(1968,1197)(2045,1325)(2127,1463)(2205,1597)(2276,1718)
\thinlines \path(411,1718)(411,1718)(487,1521)(569,1312)(647,1123)(721,953)(801,787)(877,645)(958,517)(1036,418)(1111,345)(1150,317)(1170,305)(1191,294)(1211,285)(1230,279)(1248,274)(1257,272)(1266,271)(1275,269)(1280,269)(1284,269)(1287,268)(1289,268)(1292,268)(1294,268)(1297,268)(1298,268)(1299,268)(1302,268)(1303,268)(1304,268)(1306,268)(1309,268)(1311,268)(1313,268)(1317,268)(1322,269)(1331,269)(1340,271)(1351,273)(1360,275)(1379,280)(1398,287)(1420,296)(1459,317)(1500,347)(1574,414)(1653,507)(1729,614)(1810,746)
\thinlines \path(1810,746)(1888,887)(1963,1033)(2042,1199)(2119,1365)(2201,1548)(2276,1718)
\thinlines \dashline[-20]{11}(411,1266)(411,1266)(487,1266)(569,1266)(647,1266)(721,1266)(801,1266)(877,1266)(958,1266)(1036,1266)(1111,1266)(1191,1266)(1269,1266)(1343,1266)(1422,1266)(1497,1266)(1578,1266)(1656,1266)(1731,1266)(1810,1266)(1887,1266)(1968,1266)(2047,1266)(2122,1266)(2203,1266)(2276,1266)
\thinlines \dashline[-20]{11}(411,1472)(411,1472)(487,1472)(569,1472)(647,1472)(721,1472)(801,1472)(877,1472)(958,1472)(1036,1472)(1111,1472)(1191,1472)(1269,1472)(1343,1472)(1422,1472)(1497,1472)(1578,1472)(1656,1472)(1731,1472)(1810,1472)(1887,1472)(1968,1472)(2047,1472)(2122,1472)(2203,1472)(2276,1472)
\end{picture}
\caption{ $A_{CP}^{+-}$ as a function of $\delta^T $.   }
    \label{fig:3}
\end{center}
\end{figure}
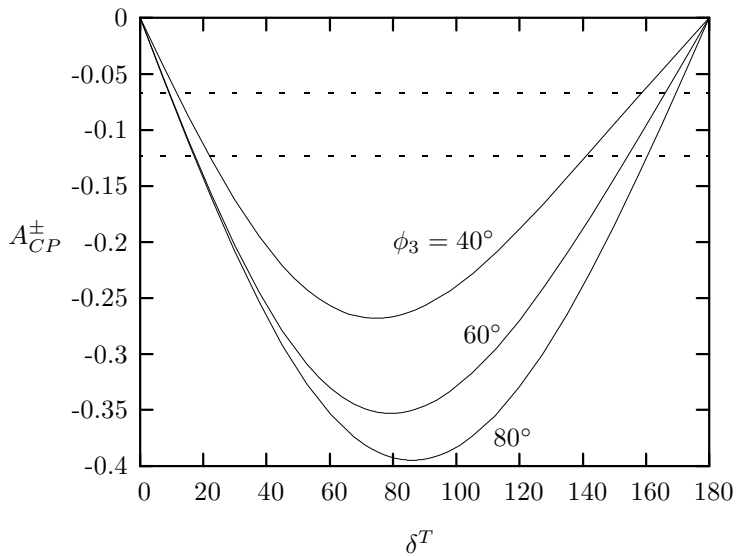

The direct CP asymmetries under the same assumption which we neglect the
terms of $O(0.001)$ are
\bea
A_{CP}^{0+} &\equiv & \frac{|A_K^{0-}|^2 - |A_K^{0+}|^2}{|A_K^{0-}|^2 
                + |A_K^{0+}|^2}\,
             =\, - 2 r_A \sin\delta^A \sin\phi_3\, \sim\, 0.0\;, \\[1mm]
A_{CP}^{00} &\equiv & \frac{|\bar{A}_K^{00}|^2 - |A_K^{00}|^2}
                     {|\bar{A}_K^{00}|^2 + |A_K^{00}|^2}\,
             =\, 2 r_C \sin\delta^C \sin\phi_3\, \sim\, O(0.01)\;, \\[1mm]
A_{CP}^{+-} &\equiv & \frac{|A_K^{-+}|^2 
                      - |A_K^{+-}|^2}{|A_K^{-+}|^2 + |A_K^{+-}|^2}\,
             =\, - 2 r_T \sin\delta^T \sin\phi_3
             - r_T^2 \sin2\delta^T \sin2\phi_3 \;, \\[1mm]
A_{CP}^{+0} &\equiv & \frac{|A_K^{-0}|^2 - |A_K^{+0}|^2}
                           {|A_K^{-0}|^2 + |A_K^{+0}|^2}\,
             =\, - 2 (r_T \sin\delta^T + r_C \sin\delta^C
                  + r_A \sin\delta^A )\sin\phi_3 \nn \\
            & & ~~~~~~~~~~~~~~
         + 2 r_{EW} r_{T} \sin(\delta^T + \delta^{EW} )\sin\phi_3
               - r_T^2 \sin2\delta^T \sin2\phi_3\;. 
\eea
Up to the first order of $r$, there is a
relation among the CP asymmetries as follows:
\bea
A_{CP}^{+0} - A_{CP}^{+-} + A_{CP}^{00} - A_{CP}^{0+}\, =\, 0\;.
\eea
The discrepancy of this relation is caused from the cross term of $r_T$
and $r_{EW}$. 
\bea  
A_{CP}^{+0} - A_{CP}^{+-} + A_{CP}^{00} - A_{CP}^{0+}\, =\, 
         2 r_T  r_{EW} \sin(\delta^T + \delta^{EW}) \sin\phi_3\, 
                                               =\, 0.18\pm 0.24\;. 
\eea
This may also give us some useful informations about $r_{EW}$ and the
strong phases but the data of $A_{CP}^{00}$ still has quite large
error, as shown in Table.~\ref{tab:2}, so that one can not extract from it
at present time. 
\begin{table}
\begin{center} {\small
\begin{tabular}{|c|c|c|c|c|}\hline
  & CLEO
  & Belle & BaBar & Average \\
\hline
$A_{CP}^{0+} $ & 0.18 $\pm $ 0.24 $\pm $ 0.02 
             & 0.05 $\pm $ 0.05 $\pm$ 0.01
                          & -0.05 $\pm$ 0.08
                          $\pm$ 0.01
                          & 0.03 $\pm$ 0.04 \\[2mm]
$A_{CP}^{00}  $ & -
             & 0.16 $\pm$ 0.29 $\pm$ 0.05
                          & 0.03 $\pm$ 0.36
                          $\pm$ 0.11
                          & 0.11 $\pm$ 0.23  \\[2mm]
$A_{CP}^{+-} $ & -0.04 $\pm$ 0.16 $\pm $ 0.02
             & -0.088 $\pm$ 0.035 $\pm$ 0.018
                          & -0.107
                          $\pm$ 0.041 $\pm$ 0.013
                          & -0.095 $\pm$ 0.028  \\[2mm]
$A_{CP}^{+0}  $ & -0.29 $\pm$ 0.24 $\pm$ 0.02
             & 0.06 $\pm$ 0.06 $\pm$ 0.02
                          & -0.09 $\pm$ 0.09
                          $\pm $ 0.01
                          & 0.00 $\pm$ 0.05 \\
\hline
\end{tabular} }
\caption{The experimental data of the direct CP asymmetry~\cite{TOMURA,
 NBABAR, AKPBELLE2} and the 
             averaged values~\cite{HFAG}.}
 \label{tab:2}
\end{center}
\end{table}
We need more accurate data to use this relation. 
For this reason, we use only $A_{CP}^{+-}$ because it is 
accurate measurement and will give some constraint to 
$\delta^T $. 
In Fig.~\ref{fig:3}, we plot $A_{CP}^{+-}$ as a function of
$\delta^T $. {}From this figure, we can find the constraint for
$\delta^T $ at  $-0.123 < A_{CP}^{+-} < -0.067 $. It tells us that the
small $\delta^T$ is favored and $\delta^T $ should be around $15^\circ $
or $160^\circ$.

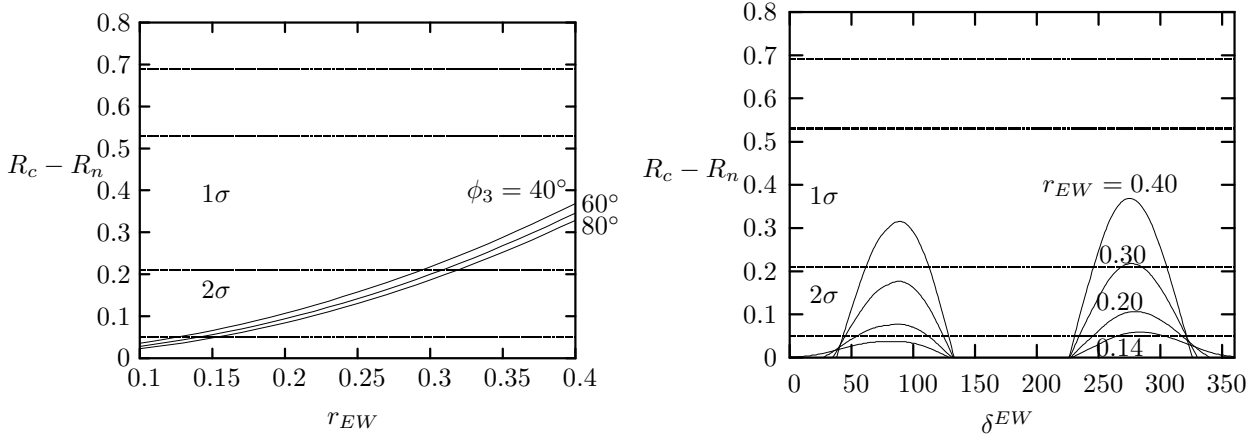
\begin{figure}[thb]
\bec
\hspace*{3mm}
\begin{minipage}[l]{3.0in} 
\begin{center}
\setlength{\unitlength}{0.086450pt}
\begin{picture}(2400,1800)(0,0)
\footnotesize
\thicklines \path(370,249)(411,249)
\thicklines \path(2276,249)(2235,249)
\put(329,249){\makebox(0,0)[r]{ 0}}
\thicklines \path(370,433)(411,433)
\thicklines \path(2276,433)(2235,433)
\put(329,433){\makebox(0,0)[r]{ 0.1}}
\thicklines \path(370,616)(411,616)
\thicklines \path(2276,616)(2235,616)
\put(329,616){\makebox(0,0)[r]{ 0.2}}
\thicklines \path(370,800)(411,800)
\thicklines \path(2276,800)(2235,800)
\put(329,800){\makebox(0,0)[r]{ 0.3}}
\thicklines \path(370,984)(411,984)
\thicklines \path(2276,984)(2235,984)
\put(329,984){\makebox(0,0)[r]{ 0.4}}
\thicklines \path(370,1167)(411,1167)
\thicklines \path(2276,1167)(2235,1167)
\put(329,1167){\makebox(0,0)[r]{ 0.5}}
\thicklines \path(370,1351)(411,1351)
\thicklines \path(2276,1351)(2235,1351)
\put(329,1351){\makebox(0,0)[r]{ 0.6}}
\thicklines \path(370,1534)(411,1534)
\thicklines \path(2276,1534)(2235,1534)
\put(329,1534){\makebox(0,0)[r]{ 0.7}}
\thicklines \path(370,1718)(411,1718)
\thicklines \path(2276,1718)(2235,1718)
\put(329,1718){\makebox(0,0)[r]{ 0.8}}
\thicklines \path(370,249)(370,290)
\thicklines \path(370,1718)(370,1677)
\put(370,166){\makebox(0,0){ 0.1}}
\thicklines \path(688,249)(688,290)
\thicklines \path(688,1718)(688,1677)
\put(688,166){\makebox(0,0){ 0.15}}
\thicklines \path(1005,249)(1005,290)
\thicklines \path(1005,1718)(1005,1677)
\put(1005,166){\makebox(0,0){ 0.2}}
\thicklines \path(1323,249)(1323,290)
\thicklines \path(1323,1718)(1323,1677)
\put(1323,166){\makebox(0,0){ 0.25}}
\thicklines \path(1641,249)(1641,290)
\thicklines \path(1641,1718)(1641,1677)
\put(1641,166){\makebox(0,0){ 0.3}}
\thicklines \path(1958,249)(1958,290)
\thicklines \path(1958,1718)(1958,1677)
\put(1958,166){\makebox(0,0){ 0.35}}
\thicklines \path(2276,249)(2276,290)
\thicklines \path(2276,1718)(2276,1677)
\put(2276,166){\makebox(0,0){ 0.4}}
\thicklines \path(370,249)(2276,249)(2276,1718)(370,1718)(370,249)
\put(-210,1090){\makebox(0,0)[l]{{$R_c-R_n $}}}
\put(640,550){\makebox(0,0)[l]{$2 \sigma $  }}
\put(640,970){\makebox(0,0)[l]{$1 \sigma $  }}
\put(1323,-25){\makebox(0,0){{$ r_{EW} $} }}
\put(1800,980){\makebox(0,0)[l]{$\phi_3 = 40^\circ $  }}
\put(2110,928){\makebox(0,0)[l]{$~~~~~60^\circ $  }}
\put(2110,830){\makebox(0,0)[l]{$~~~~~80^\circ $  }}
\thinlines \path(370,314)(370,314)(434,324)(497,334)(561,346)(624,358)(688,370)(751,384)(815,398)(878,413)(942,429)(1005,445)(1069,462)(1132,480)(1196,498)(1259,518)(1323,538)(1387,559)(1450,580)(1514,602)(1577,625)(1641,649)(1704,673)(1768,699)(1831,724)(1895,751)(1958,778)(2022,807)(2085,835)(2149,865)(2212,895)(2276,926)
\thinlines \path(370,300)(370,300)(434,309)(497,319)(561,329)(624,340)(688,352)(751,364)(815,377)(878,391)(942,406)(1005,421)(1069,438)(1132,454)(1196,472)(1259,490)(1323,509)(1387,529)(1450,550)(1514,571)(1577,593)(1641,616)(1704,639)(1768,664)(1831,688)(1895,714)(1958,741)(2022,768)(2085,796)(2149,824)(2212,854)(2276,884)
\thinlines \path(370,290)(370,290)(434,298)(497,307)(561,316)(624,327)(688,338)(751,349)(815,362)(878,375)(942,389)(1005,404)(1069,419)(1132,435)(1196,452)(1259,470)(1323,488)(1387,507)(1450,527)(1514,548)(1577,569)(1641,591)(1704,614)(1768,637)(1831,662)(1895,687)(1958,712)(2022,739)(2085,766)(2149,794)(2212,823)(2276,852)
\thinlines \dashline[-20]{16}(370,1222)(370,1222)(389,1222)(409,1222)(428,1222)(447,1222)(466,1222)(486,1222)(505,1222)(524,1222)(543,1222)(563,1222)(582,1222)(601,1222)(620,1222)(640,1222)(659,1222)(678,1222)(697,1222)(717,1222)(736,1222)(755,1222)(774,1222)(794,1222)(813,1222)(832,1222)(851,1222)(871,1222)(890,1222)(909,1222)(928,1222)(948,1222)(967,1222)(986,1222)(1005,1222)(1025,1222)(1044,1222)(1063,1222)(1082,1222)(1102,1222)(1121,1222)(1140,1222)(1159,1222)(1179,1222)(1198,1222)(1217,1222)(1236,1222)(1256,1222)(1275,1222)(1294,1222)(1313,1222)
\thinlines \dashline[-20]{16}(1313,1222)(1333,1222)(1352,1222)(1371,1222)(1390,1222)(1410,1222)(1429,1222)(1448,1222)(1467,1222)(1487,1222)(1506,1222)(1525,1222)(1544,1222)(1564,1222)(1583,1222)(1602,1222)(1621,1222)(1641,1222)(1660,1222)(1679,1222)(1698,1222)(1718,1222)(1737,1222)(1756,1222)(1775,1222)(1795,1222)(1814,1222)(1833,1222)(1852,1222)(1872,1222)(1891,1222)(1910,1222)(1929,1222)(1949,1222)(1968,1222)(1987,1222)(2006,1222)(2026,1222)(2045,1222)(2064,1222)(2083,1222)(2103,1222)(2122,1222)(2141,1222)(2160,1222)(2180,1222)(2199,1222)(2218,1222)(2237,1222)(2257,1222)(2276,1222)
\thinlines \dashline[-20]{16}(370,635)(370,635)(389,635)(409,635)(428,635)(447,635)(466,635)(486,635)(505,635)(524,635)(543,635)(563,635)(582,635)(601,635)(620,635)(640,635)(659,635)(678,635)(697,635)(717,635)(736,635)(755,635)(774,635)(794,635)(813,635)(832,635)(851,635)(871,635)(890,635)(909,635)(928,635)(948,635)(967,635)(986,635)(1005,635)(1025,635)(1044,635)(1063,635)(1082,635)(1102,635)(1121,635)(1140,635)(1159,635)(1179,635)(1198,635)(1217,635)(1236,635)(1256,635)(1275,635)(1294,635)(1313,635)
\thinlines \dashline[-20]{16}(1313,635)(1333,635)(1352,635)(1371,635)(1390,635)(1410,635)(1429,635)(1448,635)(1467,635)(1487,635)(1506,635)(1525,635)(1544,635)(1564,635)(1583,635)(1602,635)(1621,635)(1641,635)(1660,635)(1679,635)(1698,635)(1718,635)(1737,635)(1756,635)(1775,635)(1795,635)(1814,635)(1833,635)(1852,635)(1872,635)(1891,635)(1910,635)(1929,635)(1949,635)(1968,635)(1987,635)(2006,635)(2026,635)(2045,635)(2064,635)(2083,635)(2103,635)(2122,635)(2141,635)(2160,635)(2180,635)(2199,635)(2218,635)(2237,635)(2257,635)(2276,635)
\thinlines \dashline[-20]{11}(370,1516)(370,1516)(389,1516)(409,1516)(428,1516)(447,1516)(466,1516)(486,1516)(505,1516)(524,1516)(543,1516)(563,1516)(582,1516)(601,1516)(620,1516)(640,1516)(659,1516)(678,1516)(697,1516)(717,1516)(736,1516)(755,1516)(774,1516)(794,1516)(813,1516)(832,1516)(851,1516)(871,1516)(890,1516)(909,1516)(928,1516)(948,1516)(967,1516)(986,1516)(1005,1516)(1025,1516)(1044,1516)(1063,1516)(1082,1516)(1102,1516)(1121,1516)(1140,1516)(1159,1516)(1179,1516)(1198,1516)(1217,1516)(1236,1516)(1256,1516)(1275,1516)(1294,1516)(1313,1516)
\thinlines \dashline[-20]{11}(1313,1516)(1333,1516)(1352,1516)(1371,1516)(1390,1516)(1410,1516)(1429,1516)(1448,1516)(1467,1516)(1487,1516)(1506,1516)(1525,1516)(1544,1516)(1564,1516)(1583,1516)(1602,1516)(1621,1516)(1641,1516)(1660,1516)(1679,1516)(1698,1516)(1718,1516)(1737,1516)(1756,1516)(1775,1516)(1795,1516)(1814,1516)(1833,1516)(1852,1516)(1872,1516)(1891,1516)(1910,1516)(1929,1516)(1949,1516)(1968,1516)(1987,1516)(2006,1516)(2026,1516)(2045,1516)(2064,1516)(2083,1516)(2103,1516)(2122,1516)(2141,1516)(2160,1516)(2180,1516)(2199,1516)(2218,1516)(2237,1516)(2257,1516)(2276,1516)
\thinlines \dashline[-20]{11}(370,341)(370,341)(389,341)(409,341)(428,341)(447,341)(466,341)(486,341)(505,341)(524,341)(543,341)(563,341)(582,341)(601,341)(620,341)(640,341)(659,341)(678,341)(697,341)(717,341)(736,341)(755,341)(774,341)(794,341)(813,341)(832,341)(851,341)(871,341)(890,341)(909,341)(928,341)(948,341)(967,341)(986,341)(1005,341)(1025,341)(1044,341)(1063,341)(1082,341)(1102,341)(1121,341)(1140,341)(1159,341)(1179,341)(1198,341)(1217,341)(1236,341)(1256,341)(1275,341)(1294,341)(1313,341)
\thinlines \dashline[-20]{11}(1313,341)(1333,341)(1352,341)(1371,341)(1390,341)(1410,341)(1429,341)(1448,341)(1467,341)(1487,341)(1506,341)(1525,341)(1544,341)(1564,341)(1583,341)(1602,341)(1621,341)(1641,341)(1660,341)(1679,341)(1698,341)(1718,341)(1737,341)(1756,341)(1775,341)(1795,341)(1814,341)(1833,341)(1852,341)(1872,341)(1891,341)(1910,341)(1929,341)(1949,341)(1968,341)(1987,341)(2006,341)(2026,341)(2045,341)(2064,341)(2083,341)(2103,341)(2122,341)(2141,341)(2160,341)(2180,341)(2199,341)(2218,341)(2237,341)(2257,341)(2276,341)
\end{picture}
\end{center} 
\end{minipage} 
\hspace*{5mm}
\begin{minipage}[r]{3.0in}
\begin{center}
\setlength{\unitlength}{0.118pt}
\begin{picture}(1919,1440)(0,0)
\footnotesize
\thicklines \path(370,249)(411,249)
\thicklines \path(1796,249)(1755,249)
\put(329,249){\makebox(0,0)[r]{ 0}}
\thicklines \path(370,388)(411,388)
\thicklines \path(1796,388)(1755,388)
\put(329,388){\makebox(0,0)[r]{ 0.1}}
\thicklines \path(370,526)(411,526)
\thicklines \path(1796,526)(1755,526)
\put(329,526){\makebox(0,0)[r]{ 0.2}}
\thicklines \path(370,665)(411,665)
\thicklines \path(1796,665)(1755,665)
\put(329,665){\makebox(0,0)[r]{ 0.3}}
\thicklines \path(370,804)(411,804)
\thicklines \path(1796,804)(1755,804)
\put(329,804){\makebox(0,0)[r]{ 0.4}}
\thicklines \path(370,942)(411,942)
\thicklines \path(1796,942)(1755,942)
\put(329,942){\makebox(0,0)[r]{ 0.5}}
\thicklines \path(370,1081)(411,1081)
\thicklines \path(1796,1081)(1755,1081)
\put(329,1081){\makebox(0,0)[r]{ 0.6}}
\thicklines \path(370,1219)(411,1219)
\thicklines \path(1796,1219)(1755,1219)
\put(329,1219){\makebox(0,0)[r]{ 0.7}}
\thicklines \path(370,1358)(411,1358)
\thicklines \path(1796,1358)(1755,1358)
\put(329,1358){\makebox(0,0)[r]{ 0.8}}
\thicklines \path(370,249)(370,290)
\thicklines \path(370,1358)(370,1317)
\put(370,166){\makebox(0,0){ 0}}
\thicklines \path(568,249)(568,290)
\thicklines \path(568,1358)(568,1317)
\put(568,166){\makebox(0,0){ 50}}
\thicklines \path(766,249)(766,290)
\thicklines \path(766,1358)(766,1317)
\put(766,166){\makebox(0,0){ 100}}
\thicklines \path(964,249)(964,290)
\thicklines \path(964,1358)(964,1317)
\put(964,166){\makebox(0,0){ 150}}
\thicklines \path(1162,249)(1162,290)
\thicklines \path(1162,1358)(1162,1317)
\put(1162,166){\makebox(0,0){ 200}}
\thicklines \path(1360,249)(1360,290)
\thicklines \path(1360,1358)(1360,1317)
\put(1360,166){\makebox(0,0){ 250}}
\thicklines \path(1558,249)(1558,290)
\thicklines \path(1558,1358)(1558,1317)
\put(1558,166){\makebox(0,0){ 300}}
\thicklines \path(1756,249)(1756,290)
\thicklines \path(1756,1358)(1756,1317)
\put(1756,166){\makebox(0,0){ 350}}
\thicklines \path(370,249)(1796,249)(1796,1358)(370,1358)(370,249)
\put(-130,850){\makebox(0,0)[l]{{ $R_c-R_n$}  }}
\put(433,770){\makebox(0,0)[l]{$1 \sigma $  }}
\put(433,450){\makebox(0,0)[l]{$2 \sigma $  }}
\put(1180,800){\makebox(0,0)[l]{$r_{EW} = 0.40 $  }}
\put(1220,580){\makebox(0,0)[l]{$~~~~~0.30 $  }}
\put(1210,430){\makebox(0,0)[l]{$~~~~~0.20 $  }}
\put(1210,280){\makebox(0,0)[l]{$~~~~~0.14 $  }}
\put(1083,42){\makebox(0,0){{$ \delta^{EW} $} }}
\thinlines \path(370,253)(370,253)(374,253)(378,252)(382,252)(386,252)(390,252)(394,252)(398,253)(402,253)(406,253)(410,253)(414,253)(418,254)(421,254)(425,254)(429,255)(433,255)(437,256)(441,257)(445,257)(449,258)(453,258)(457,259)(461,260)(465,261)(469,262)(473,262)(477,263)(481,264)(485,265)(489,266)(493,267)(497,268)(501,269)(505,270)(509,271)(513,272)(517,273)(521,274)(524,275)(528,276)(532,277)(536,278)(540,279)(544,280)(548,281)(552,282)(556,283)(560,284)(564,285)
\thinlines \path(564,285)(568,286)(572,287)(576,288)(580,289)(584,290)(588,291)(592,292)(596,292)(600,293)(604,294)(608,295)(612,295)(616,296)(620,297)(624,297)(627,298)(631,298)(635,299)(639,299)(643,299)(647,300)(651,300)(655,300)(659,300)(663,300)(667,300)(671,300)(675,300)(679,300)(683,300)(687,300)(691,300)(695,301)(699,301)(703,301)(707,301)(711,301)(715,301)(719,301)(723,301)(727,301)(730,301)(734,300)(738,300)(742,300)(746,299)(750,298)(754,298)(758,297)(762,296)
\thinlines \path(762,296)(766,295)(770,295)(774,294)(778,293)(782,291)(786,290)(790,289)(794,288)(798,287)(802,285)(806,284)(810,282)(814,281)(818,279)(822,278)(826,276)(829,275)(833,273)(837,271)(841,269)(845,268)(849,266)(853,264)(857,262)(861,260)(865,258)(869,256)(873,254)(877,252)(881,250)(885,249)(889,249)(893,249)(897,249)(901,249)(905,249)(909,249)(913,249)(917,249)(921,249)(925,249)(929,249)(932,249)(936,249)(940,249)(944,249)(948,249)(952,249)(956,249)(960,249)
\thinlines \path(960,249)(964,249)(968,249)(972,249)(976,249)(980,249)(984,249)(988,249)(992,249)(996,249)(1000,249)(1004,249)(1008,249)(1012,249)(1016,249)(1020,249)(1024,249)(1028,249)(1032,249)(1035,249)(1039,249)(1043,249)(1047,249)(1051,249)(1055,249)(1059,249)(1063,249)(1067,249)(1071,249)(1075,249)(1079,249)(1083,249)(1087,249)(1091,249)(1095,249)(1099,249)(1103,249)(1107,249)(1111,249)(1115,249)(1119,249)(1123,249)(1127,249)(1131,249)(1134,249)(1138,249)(1142,249)(1146,249)(1150,249)(1154,249)(1158,249)
\thinlines \path(1158,249)(1162,249)(1166,249)(1170,249)(1174,249)(1178,249)(1182,249)(1186,249)(1190,249)(1194,249)(1198,249)(1202,249)(1206,249)(1210,249)(1214,249)(1218,249)(1222,249)(1226,249)(1230,249)(1234,249)(1237,249)(1241,249)(1245,249)(1249,249)(1253,249)(1257,249)(1261,249)(1265,249)(1269,249)(1273,249)(1277,251)(1281,253)(1285,255)(1289,257)(1293,259)(1297,261)(1301,263)(1305,265)(1309,267)(1313,269)(1317,271)(1321,273)(1325,275)(1329,277)(1333,279)(1337,280)(1340,282)(1344,284)(1348,285)(1352,287)(1356,289)
\thinlines \path(1356,289)(1360,292)(1364,294)(1368,296)(1372,298)(1376,300)(1380,302)(1384,304)(1388,306)(1392,308)(1396,310)(1400,311)(1404,313)(1408,315)(1412,316)(1416,318)(1420,319)(1424,320)(1428,321)(1432,323)(1436,324)(1440,325)(1443,326)(1447,326)(1451,327)(1455,328)(1459,329)(1463,329)(1467,330)(1471,330)(1475,330)(1479,331)(1483,331)(1487,331)(1491,331)(1495,331)(1499,331)(1503,331)(1507,331)(1511,330)(1515,330)(1519,330)(1523,329)(1527,329)(1531,328)(1535,327)(1539,327)(1542,326)(1546,325)(1550,324)(1554,323)
\thinlines \path(1554,323)(1558,322)(1562,321)(1566,320)(1570,319)(1574,318)(1578,317)(1582,316)(1586,314)(1590,313)(1594,312)(1598,310)(1602,309)(1606,308)(1610,306)(1614,305)(1618,303)(1622,302)(1626,300)(1630,299)(1634,297)(1638,296)(1642,294)(1645,293)(1649,291)(1653,290)(1657,288)(1661,287)(1665,285)(1669,284)(1673,282)(1677,281)(1681,279)(1685,278)(1689,276)(1693,275)(1697,273)(1701,272)(1705,271)(1709,269)(1713,268)(1717,267)(1721,266)(1725,265)(1729,263)(1733,262)(1737,261)(1741,260)(1745,260)(1748,259)(1752,258)
\thinlines \path(1752,258)(1756,257)(1760,257)(1764,256)(1768,256)(1772,255)(1776,255)(1780,254)(1784,254)(1788,253)(1792,253)(1796,253)
\thinlines \path(370,249)(370,249)(374,249)(378,249)(382,249)(386,249)(390,249)(394,249)(398,249)(402,249)(406,249)(410,249)(414,249)(418,249)(421,249)(425,249)(429,249)(433,249)(437,249)(441,249)(445,249)(449,249)(453,249)(457,249)(461,249)(465,249)(469,249)(473,249)(477,250)(481,252)(485,254)(489,257)(493,259)(497,261)(501,264)(505,266)(509,269)(513,272)(517,274)(521,277)(524,279)(528,282)(532,285)(536,287)(540,290)(544,293)(548,295)(552,298)(556,300)(560,303)(564,306)
\thinlines \path(564,306)(568,308)(572,311)(576,313)(580,315)(584,318)(588,320)(592,322)(596,325)(600,327)(604,329)(608,331)(612,333)(616,335)(620,337)(624,338)(627,340)(631,341)(635,343)(639,344)(643,345)(647,347)(651,348)(655,349)(659,349)(663,350)(667,351)(671,352)(675,352)(679,353)(683,353)(687,353)(691,354)(695,355)(699,356)(703,356)(707,357)(711,357)(715,357)(719,357)(723,357)(727,357)(730,356)(734,356)(738,355)(742,354)(746,353)(750,352)(754,351)(758,350)(762,348)
\thinlines \path(762,348)(766,347)(770,345)(774,343)(778,341)(782,339)(786,337)(790,335)(794,332)(798,330)(802,327)(806,324)(810,321)(814,319)(818,316)(822,312)(826,309)(829,306)(833,303)(837,299)(841,296)(845,292)(849,288)(853,285)(857,281)(861,277)(865,273)(869,269)(873,265)(877,262)(881,258)(885,254)(889,249)(893,249)(897,249)(901,249)(905,249)(909,249)(913,249)(917,249)(921,249)(925,249)(929,249)(932,249)(936,249)(940,249)(944,249)(948,249)(952,249)(956,249)(960,249)
\thinlines \path(960,249)(964,249)(968,249)(972,249)(976,249)(980,249)(984,249)(988,249)(992,249)(996,249)(1000,249)(1004,249)(1008,249)(1012,249)(1016,249)(1020,249)(1024,249)(1028,249)(1032,249)(1035,249)(1039,249)(1043,249)(1047,249)(1051,249)(1055,249)(1059,249)(1063,249)(1067,249)(1071,249)(1075,249)(1079,249)(1083,249)(1087,249)(1091,249)(1095,249)(1099,249)(1103,249)(1107,249)(1111,249)(1115,249)(1119,249)(1123,249)(1127,249)(1131,249)(1134,249)(1138,249)(1142,249)(1146,249)(1150,249)(1154,249)(1158,249)
\thinlines \path(1158,249)(1162,249)(1166,249)(1170,249)(1174,249)(1178,249)(1182,249)(1186,249)(1190,249)(1194,249)(1198,249)(1202,249)(1206,249)(1210,249)(1214,249)(1218,249)(1222,249)(1226,249)(1230,249)(1234,249)(1237,249)(1241,249)(1245,249)(1249,249)(1253,249)(1257,249)(1261,249)(1265,249)(1269,249)(1273,252)(1277,257)(1281,261)(1285,265)(1289,269)(1293,273)(1297,277)(1301,281)(1305,285)(1309,289)(1313,293)(1317,296)(1321,300)(1325,304)(1329,307)(1333,311)(1337,314)(1340,318)(1344,321)(1348,324)(1352,327)(1356,331)
\thinlines \path(1356,331)(1360,335)(1364,339)(1368,343)(1372,347)(1376,350)(1380,354)(1384,357)(1388,361)(1392,364)(1396,367)(1400,369)(1404,372)(1408,375)(1412,377)(1416,379)(1420,382)(1424,384)(1428,386)(1432,387)(1436,389)(1440,390)(1443,392)(1447,393)(1451,394)(1455,395)(1459,395)(1463,396)(1467,396)(1471,397)(1475,397)(1479,397)(1483,397)(1487,397)(1491,396)(1495,396)(1499,395)(1503,394)(1507,393)(1511,392)(1515,391)(1519,390)(1523,388)(1527,387)(1531,385)(1535,383)(1539,381)(1542,379)(1546,377)(1550,375)(1554,373)
\thinlines \path(1554,373)(1558,370)(1562,368)(1566,365)(1570,363)(1574,360)(1578,357)(1582,355)(1586,352)(1590,349)(1594,346)(1598,343)(1602,339)(1606,336)(1610,333)(1614,330)(1618,327)(1622,323)(1626,320)(1630,317)(1634,313)(1638,310)(1642,307)(1645,303)(1649,300)(1653,297)(1657,294)(1661,290)(1665,287)(1669,284)(1673,281)(1677,277)(1681,274)(1685,271)(1689,268)(1693,265)(1697,262)(1701,260)(1705,257)(1709,254)(1713,252)(1717,249)(1721,249)(1725,249)(1729,249)(1733,249)(1737,249)(1741,249)(1745,249)(1748,249)(1752,249)
\thinlines \path(1752,249)(1756,249)(1760,249)(1764,249)(1768,249)(1772,249)(1776,249)(1780,249)(1784,249)(1788,249)(1792,249)(1796,249)
\thinlines \path(370,249)(370,249)(374,249)(378,249)(382,249)(386,249)(390,249)(394,249)(398,249)(402,249)(406,249)(410,249)(414,249)(418,249)(421,249)(425,249)(429,249)(433,249)(437,249)(441,249)(445,249)(449,249)(453,249)(457,249)(461,249)(465,249)(469,249)(473,249)(477,249)(481,249)(485,249)(489,249)(493,249)(497,249)(501,249)(505,249)(509,251)(513,257)(517,264)(521,271)(524,277)(528,284)(532,291)(536,298)(540,305)(544,312)(548,318)(552,325)(556,332)(560,339)(564,346)
\thinlines \path(564,346)(568,352)(572,359)(576,365)(580,372)(584,378)(588,384)(592,390)(596,396)(600,402)(604,408)(608,413)(612,419)(616,424)(620,429)(624,434)(627,439)(631,443)(635,448)(639,452)(643,456)(647,459)(651,463)(655,466)(659,469)(663,472)(667,474)(671,477)(675,479)(679,481)(683,482)(687,483)(691,486)(695,488)(699,490)(703,491)(707,492)(711,493)(715,494)(719,494)(723,494)(727,494)(730,493)(734,492)(738,491)(742,489)(746,487)(750,485)(754,483)(758,480)(762,477)
\thinlines \path(762,477)(766,474)(770,470)(774,466)(778,462)(782,458)(786,453)(790,448)(794,443)(798,437)(802,432)(806,426)(810,420)(814,413)(818,407)(822,400)(826,393)(829,386)(833,378)(837,371)(841,363)(845,355)(849,347)(853,339)(857,331)(861,322)(865,314)(869,305)(873,297)(877,288)(881,279)(885,270)(889,261)(893,252)(897,249)(901,249)(905,249)(909,249)(913,249)(917,249)(921,249)(925,249)(929,249)(932,249)(936,249)(940,249)(944,249)(948,249)(952,249)(956,249)(960,249)
\thinlines \path(960,249)(964,249)(968,249)(972,249)(976,249)(980,249)(984,249)(988,249)(992,249)(996,249)(1000,249)(1004,249)(1008,249)(1012,249)(1016,249)(1020,249)(1024,249)(1028,249)(1032,249)(1035,249)(1039,249)(1043,249)(1047,249)(1051,249)(1055,249)(1059,249)(1063,249)(1067,249)(1071,249)(1075,249)(1079,249)(1083,249)(1087,249)(1091,249)(1095,249)(1099,249)(1103,249)(1107,249)(1111,249)(1115,249)(1119,249)(1123,249)(1127,249)(1131,249)(1134,249)(1138,249)(1142,249)(1146,249)(1150,249)(1154,249)(1158,249)
\thinlines \path(1158,249)(1162,249)(1166,249)(1170,249)(1174,249)(1178,249)(1182,249)(1186,249)(1190,249)(1194,249)(1198,249)(1202,249)(1206,249)(1210,249)(1214,249)(1218,249)(1222,249)(1226,249)(1230,249)(1234,249)(1237,249)(1241,249)(1245,249)(1249,249)(1253,249)(1257,249)(1261,249)(1265,249)(1269,254)(1273,263)(1277,272)(1281,281)(1285,290)(1289,299)(1293,308)(1297,317)(1301,325)(1305,334)(1309,343)(1313,351)(1317,359)(1321,367)(1325,375)(1329,383)(1333,391)(1337,398)(1340,405)(1344,412)(1348,419)(1352,426)(1356,434)
\thinlines \path(1356,434)(1360,442)(1364,450)(1368,458)(1372,465)(1376,472)(1380,478)(1384,485)(1388,491)(1392,497)(1396,502)(1400,508)(1404,513)(1408,518)(1412,522)(1416,526)(1420,530)(1424,533)(1428,537)(1432,539)(1436,542)(1440,544)(1443,546)(1447,548)(1451,549)(1455,550)(1459,551)(1463,551)(1467,551)(1471,551)(1475,550)(1479,549)(1483,548)(1487,546)(1491,545)(1495,542)(1499,540)(1503,537)(1507,534)(1511,531)(1515,528)(1519,524)(1523,520)(1527,515)(1531,511)(1535,506)(1539,501)(1542,496)(1546,490)(1550,485)(1554,479)
\thinlines \path(1554,479)(1558,473)(1562,467)(1566,460)(1570,454)(1574,447)(1578,440)(1582,433)(1586,426)(1590,419)(1594,411)(1598,404)(1602,396)(1606,389)(1610,381)(1614,373)(1618,365)(1622,358)(1626,350)(1630,342)(1634,334)(1638,326)(1642,318)(1645,310)(1649,303)(1653,295)(1657,287)(1661,280)(1665,272)(1669,265)(1673,257)(1677,250)(1681,249)(1685,249)(1689,249)(1693,249)(1697,249)(1701,249)(1705,249)(1709,249)(1713,249)(1717,249)(1721,249)(1725,249)(1729,249)(1733,249)(1737,249)(1741,249)(1745,249)(1748,249)(1752,249)
\thinlines \path(1752,249)(1756,249)(1760,249)(1764,249)(1768,249)(1772,249)(1776,249)(1780,249)(1784,249)(1788,249)(1792,249)(1796,249)
\thinlines \path(370,249)(370,249)(374,249)(378,249)(382,249)(386,249)(390,249)(394,249)(398,249)(402,249)(406,249)(410,249)(414,249)(418,249)(421,249)(425,249)(429,249)(433,249)(437,249)(441,249)(445,249)(449,249)(453,249)(457,249)(461,249)(465,249)(469,249)(473,249)(477,249)(481,249)(485,249)(489,249)(493,249)(497,249)(501,249)(505,249)(509,249)(513,249)(517,249)(521,251)(524,264)(528,277)(532,290)(536,303)(540,316)(544,329)(548,342)(552,355)(556,367)(560,380)(564,393)
\thinlines \path(564,393)(568,406)(572,418)(576,431)(580,443)(584,455)(588,467)(592,479)(596,490)(600,502)(604,513)(608,524)(612,534)(616,544)(620,554)(624,564)(627,573)(631,582)(635,591)(639,599)(643,607)(647,614)(651,621)(655,628)(659,634)(663,640)(667,645)(671,650)(675,655)(679,659)(683,663)(687,666)(691,670)(695,674)(699,677)(703,680)(707,683)(711,684)(715,685)(719,686)(723,686)(727,686)(730,685)(734,684)(738,682)(742,679)(746,676)(750,672)(754,668)(758,664)(762,658)
\thinlines \path(762,658)(766,653)(770,646)(774,640)(778,633)(782,625)(786,617)(790,608)(794,599)(798,590)(802,580)(806,570)(810,559)(814,548)(818,536)(822,524)(826,512)(829,499)(833,487)(837,473)(841,460)(845,446)(849,432)(853,418)(857,403)(861,388)(865,373)(869,358)(873,343)(877,328)(881,312)(885,297)(889,281)(893,265)(897,249)(901,249)(905,249)(909,249)(913,249)(917,249)(921,249)(925,249)(929,249)(932,249)(936,249)(940,249)(944,249)(948,249)(952,249)(956,249)(960,249)
\thinlines \path(960,249)(964,249)(968,249)(972,249)(976,249)(980,249)(984,249)(988,249)(992,249)(996,249)(1000,249)(1004,249)(1008,249)(1012,249)(1016,249)(1020,249)(1024,249)(1028,249)(1032,249)(1035,249)(1039,249)(1043,249)(1047,249)(1051,249)(1055,249)(1059,249)(1063,249)(1067,249)(1071,249)(1075,249)(1079,249)(1083,249)(1087,249)(1091,249)(1095,249)(1099,249)(1103,249)(1107,249)(1111,249)(1115,249)(1119,249)(1123,249)(1127,249)(1131,249)(1134,249)(1138,249)(1142,249)(1146,249)(1150,249)(1154,249)(1158,249)
\thinlines \path(1158,249)(1162,249)(1166,249)(1170,249)(1174,249)(1178,249)(1182,249)(1186,249)(1190,249)(1194,249)(1198,249)(1202,249)(1206,249)(1210,249)(1214,249)(1218,249)(1222,249)(1226,249)(1230,249)(1234,249)(1237,249)(1241,249)(1245,249)(1249,249)(1253,249)(1257,249)(1261,249)(1265,249)(1269,263)(1273,279)(1277,295)(1281,311)(1285,327)(1289,343)(1293,358)(1297,374)(1301,389)(1305,404)(1309,419)(1313,433)(1317,448)(1321,462)(1325,476)(1329,490)(1333,503)(1337,516)(1340,529)(1344,541)(1348,553)(1352,565)(1356,579)
\thinlines \path(1356,579)(1360,592)(1364,604)(1368,617)(1372,629)(1376,640)(1380,651)(1384,661)(1388,671)(1392,681)(1396,690)(1400,698)(1404,706)(1408,714)(1412,721)(1416,727)(1420,733)(1424,738)(1428,743)(1432,747)(1436,750)(1440,753)(1443,756)(1447,758)(1451,759)(1455,760)(1459,760)(1463,760)(1467,759)(1471,758)(1475,756)(1479,753)(1483,750)(1487,747)(1491,743)(1495,738)(1499,733)(1503,727)(1507,721)(1511,715)(1515,708)(1519,700)(1523,692)(1527,684)(1531,675)(1535,666)(1539,656)(1542,646)(1546,636)(1550,625)(1554,614)
\thinlines \path(1554,614)(1558,603)(1562,591)(1566,579)(1570,567)(1574,554)(1578,542)(1582,529)(1586,515)(1590,502)(1594,488)(1598,475)(1602,461)(1606,447)(1610,433)(1614,418)(1618,404)(1622,390)(1626,376)(1630,361)(1634,347)(1638,333)(1642,318)(1645,304)(1649,290)(1653,276)(1657,262)(1661,249)(1665,249)(1669,249)(1673,249)(1677,249)(1681,249)(1685,249)(1689,249)(1693,249)(1697,249)(1701,249)(1705,249)(1709,249)(1713,249)(1717,249)(1721,249)(1725,249)(1729,249)(1733,249)(1737,249)(1741,249)(1745,249)(1748,249)(1752,249)
\thinlines \path(1752,249)(1756,249)(1760,249)(1764,249)(1768,249)(1772,249)(1776,249)(1780,249)(1784,249)(1788,249)(1792,249)(1796,249)
\thinlines \dashline[-20]{16}(370,984)(370,984)(384,984)(399,984)(413,984)(428,984)(442,984)(456,984)(471,984)(485,984)(500,984)(514,984)(528,984)(543,984)(557,984)(572,984)(586,984)(600,984)(615,984)(629,984)(644,984)(658,984)(672,984)(687,984)(701,984)(716,984)(730,984)(745,984)(759,984)(773,984)(788,984)(802,984)(817,984)(831,984)(845,984)(860,984)(874,984)(889,984)(903,984)(917,984)(932,984)(946,984)(961,984)(975,984)(989,984)(1004,984)(1018,984)(1033,984)(1047,984)(1061,984)(1076,984)
\thinlines \dashline[-20]{16}(1076,984)(1090,984)(1105,984)(1119,984)(1133,984)(1148,984)(1162,984)(1177,984)(1191,984)(1205,984)(1220,984)(1234,984)(1249,984)(1263,984)(1277,984)(1292,984)(1306,984)(1321,984)(1335,984)(1349,984)(1364,984)(1378,984)(1393,984)(1407,984)(1421,984)(1436,984)(1450,984)(1465,984)(1479,984)(1494,984)(1508,984)(1522,984)(1537,984)(1551,984)(1566,984)(1580,984)(1594,984)(1609,984)(1623,984)(1638,984)(1652,984)(1666,984)(1681,984)(1695,984)(1710,984)(1724,984)(1738,984)(1753,984)(1767,984)(1782,984)(1796,984)
\thinlines \dashline[-20]{16}(370,540)(370,540)(384,540)(399,540)(413,540)(428,540)(442,540)(456,540)(471,540)(485,540)(500,540)(514,540)(528,540)(543,540)(557,540)(572,540)(586,540)(600,540)(615,540)(629,540)(644,540)(658,540)(672,540)(687,540)(701,540)(716,540)(730,540)(745,540)(759,540)(773,540)(788,540)(802,540)(817,540)(831,540)(845,540)(860,540)(874,540)(889,540)(903,540)(917,540)(932,540)(946,540)(961,540)(975,540)(989,540)(1004,540)(1018,540)(1033,540)(1047,540)(1061,540)(1076,540)
\thinlines \dashline[-20]{16}(1076,540)(1090,540)(1105,540)(1119,540)(1133,540)(1148,540)(1162,540)(1177,540)(1191,540)(1205,540)(1220,540)(1234,540)(1249,540)(1263,540)(1277,540)(1292,540)(1306,540)(1321,540)(1335,540)(1349,540)(1364,540)(1378,540)(1393,540)(1407,540)(1421,540)(1436,540)(1450,540)(1465,540)(1479,540)(1494,540)(1508,540)(1522,540)(1537,540)(1551,540)(1566,540)(1580,540)(1594,540)(1609,540)(1623,540)(1638,540)(1652,540)(1666,540)(1681,540)(1695,540)(1710,540)(1724,540)(1738,540)(1753,540)(1767,540)(1782,540)(1796,540)
\thinlines \dashline[-20]{16}(370,1206)(370,1206)(384,1206)(399,1206)(413,1206)(428,1206)(442,1206)(456,1206)(471,1206)(485,1206)(500,1206)(514,1206)(528,1206)(543,1206)(557,1206)(572,1206)(586,1206)(600,1206)(615,1206)(629,1206)(644,1206)(658,1206)(672,1206)(687,1206)(701,1206)(716,1206)(730,1206)(745,1206)(759,1206)(773,1206)(788,1206)(802,1206)(817,1206)(831,1206)(845,1206)(860,1206)(874,1206)(889,1206)(903,1206)(917,1206)(932,1206)(946,1206)(961,1206)(975,1206)(989,1206)(1004,1206)(1018,1206)(1033,1206)(1047,1206)(1061,1206)(1076,1206)
\thinlines \dashline[-20]{11}(1076,1206)(1090,1206)(1105,1206)(1119,1206)(1133,1206)(1148,1206)(1162,1206)(1177,1206)(1191,1206)(1205,1206)(1220,1206)(1234,1206)(1249,1206)(1263,1206)(1277,1206)(1292,1206)(1306,1206)(1321,1206)(1335,1206)(1349,1206)(1364,1206)(1378,1206)(1393,1206)(1407,1206)(1421,1206)(1436,1206)(1450,1206)(1465,1206)(1479,1206)(1494,1206)(1508,1206)(1522,1206)(1537,1206)(1551,1206)(1566,1206)(1580,1206)(1594,1206)(1609,1206)(1623,1206)(1638,1206)(1652,1206)(1666,1206)(1681,1206)(1695,1206)(1710,1206)(1724,1206)(1738,1206)(1753,1206)(1767,1206)(1782,1206)(1796,1206)
\thinlines \dashline[-20]{11}(370,318)(370,318)(384,318)(399,318)(413,318)(428,318)(442,318)(456,318)(471,318)(485,318)(500,318)(514,318)(528,318)(543,318)(557,318)(572,318)(586,318)(600,318)(615,318)(629,318)(644,318)(658,318)(672,318)(687,318)(701,318)(716,318)(730,318)(745,318)(759,318)(773,318)(788,318)(802,318)(817,318)(831,318)(845,318)(860,318)(874,318)(889,318)(903,318)(917,318)(932,318)(946,318)(961,318)(975,318)(989,318)(1004,318)(1018,318)(1033,318)(1047,318)(1061,318)(1076,318)
\thinlines \dashline[-20]{11}(1076,318)(1090,318)(1105,318)(1119,318)(1133,318)(1148,318)(1162,318)(1177,318)(1191,318)(1205,318)(1220,318)(1234,318)(1249,318)(1263,318)(1277,318)(1292,318)(1306,318)(1321,318)(1335,318)(1349,318)(1364,318)(1378,318)(1393,318)(1407,318)(1421,318)(1436,318)(1450,318)(1465,318)(1479,318)(1494,318)(1508,318)(1522,318)(1537,318)(1551,318)(1566,318)(1580,318)(1594,318)(1609,318)(1623,318)(1638,318)(1652,318)(1666,318)(1681,318)(1695,318)(1710,318)(1724,318)(1738,318)(1753,318)(1767,318)(1782,318)(1796,318)
\end{picture}
\eec
\end{minipage} 
\caption{The lines show the maximum bound of $R_c-R_n$ for $r_{EW}$ and
    $\delta^{EW} $
    at $r_T = 0.2 $ and $40^\circ < \phi_3 < 80^\circ $ 
    under constraint $-0.123 < A_{CP}^{+-} < -0.067$ }
    \label{fig:4}
\end{center}
\end{figure}

In Eq.~(\ref{B+-B0+}), which leads to Fleischer-Mannel bound~\cite{FMB}, 
if $r_{EW}^C$ and $r_A$ are negligible, 
\bea 
R\, \equiv\, \frac{\tau^+}{\tau^0}\frac{\bar{B}_K^{+-}}{\bar{B}_K^{0+}}\, 
=\,
  1  
    - 2 r_{T} \cos\delta^{T} \cos\phi_3  
    + r_{T}^2\, =\, 0.90\pm0.07\;, 
\eea 
and to satisfy 
$R = 0.90\pm 0.07 < 1 $ we need positive $\cos\delta^T$ so that 
the range $\delta^T \sim 10^\circ-20^\circ $ is favored. Taking
account of these constraints for $\delta^T$ from $A_{CP}^{+-}$, we plot
the maximum bound of $R_c-R_n$ as the functions of  
$\delta^{EW}$ and $r_{EW}$ in Fig.~\ref{fig:4}, respectively.
Then the allowed regions for 
$\delta^{EW}-\delta^{T} $ around $0^\circ$ and $180^\circ$
disappear. $R_c-R_n$ seems to favor  
$45^\circ < |\delta^{EW}|<135^\circ$, but the constraint from
$A_{CP}^{+-}$ is strongly suggesting that the strong phase, $\delta^T$, 
should be around $15^\circ$. In consequence, $\delta^{EW}-\delta^{T} =
0$ as the theoretical prospect is disfavored.    

What we found from $B\rightarrow K\pi $ decays are that we need larger
$r_{EW} > 0.3$ and large strong phase differences, $\delta^{EW} - \delta^{T}$. 
The constraint, $\delta^T \sim 10^\circ-20^\circ$,  
from direct CP asymmetry is differ from the favored range of the
strong phase of EW penguin $\delta^{EW} > 45^\circ $ so that 
$\omega = \delta^{EW}-\delta^T\simeq 0 $, which is favored in
theoretically, will not be satisfied.  
What the quite large strong phase difference is requested may be a
serious problem in these modes. 
If SU(3) symmetry is good one,
these properties should also appear in $B\rightarrow \pi \pi $.

\section{SU(3) breaking effect in gluonic penguin }  
When we consider the ratios among the branching ratios 
for $B\rightarrow \pi \pi$ decays, 
\bea
\frac{2 \bar{B}_\pi^{00}}{\bar{B}_\pi^{+-}} &=& \frac{ 
   \tilde{r}_C^2 + \tilde{r}_P^2 (1 + r_{EW}^2 - 2 r_{EW}
   \cos\delta^{EW}) - 2 \tilde{r}_P \tilde{r}_C (\cos\delta^T  
        - {r}_{EW}\cos\omega ) \cos(\phi_1+\phi_3 ) }
   { 1 + \tilde{r}_P^2 + 2 \tilde{r}_P \cos\delta^T \cos(\phi_1+\phi_3
   ) }\;, \nn \\
 & & \\[2mm]
\frac{\tau^0}{\tau^+}\frac{2 \bar{B}_\pi^{+0}}{\bar{B}_\pi^{+-}} &=& \frac{ 
   1 + \tilde{r}_C^2 + 2 \tilde{r}_C 
     + \tilde{r}_P^2 {r}_{EW}^2  + 2 \tilde{r}_P {r}_{EW} (\cos\omega   
        + {r}_{C}\cos\omega ) \cos(\phi_1+\phi_3 ) }
   { 1 + \tilde{r}_P^2 + 2 \tilde{r}_P \cos\delta^T
     \cos(\phi_1+\phi_3)}\;, \nn \\
  & &    
\eea   
there is also discrepancy between theoretical expectation and
experimental data. In above equations, $\delta^C$ is taken to be equal
to $\delta^T$. 
From a rough estimation, $\tilde{r}_P \sim 0.3$, 
$\frac{\tau^0}{\tau^+}\frac{2 \bar{B}_\pi^{0+}}{\bar{B}_\pi^{+-}} \sim
1 $, $\frac{\bar{B}_\pi^{00}}{\bar{B}_\pi^{+-}} \sim
0.1 $, but the experimental data (\ref{Bpp}) are quite large
values and are not consistent with them. To explain the discrepancy, 
the denominator seems to be smaller
value so that $\cos\delta^T $ should be negative or $\phi_1+\phi_3 $
should be larger than $90^\circ $ to reduce the
denominator. The negative $\cos\delta^T $ case is inconsistent 
with the condition 
$R= \frac{\tau^+}{\tau^{-}}\frac{\bar{B}^{+-}}{\bar{B}^{00}} =
0.90\pm0.07 < 1 $. As the result, negative $\cos(\phi_1+\phi_3 )$ is
favored. However it is not enough to explain the differences and 
we will also have to take account of SU(3) breaking effect. 

The ratio of the direct CP asymmetries may show the SU(3) breaking
effect. We consider the following ratio between 
$B\rightarrow K^+\pi^- $ and $B\rightarrow \pi^+ \pi^- $:
\bea 
\frac{|\bar{A}_\pi^{+-}|^2 - |{A}_\pi^{+-}|^2}
        {|\bar{A}_K^{+-}|^2 - |{A}_K^{+-}|^2 } 
&=& - \frac{|T+E||P+P_{EW}^C| \sin\delta^T }
         {|T|~~~~|P+P_{EW}^C| \sin\delta^T }\, 
 \sim\, - \frac{f_\pi}{f_K}\frac{|P_\pi| 
               ~\sin\delta^T_\pi}{|P_K|\sin\delta^T_K}\nn \\[3mm]
&=&
 \frac{Br^{\pi^+\pi^-}~A_{CP}^{\pi^+\pi^-}}{Br^{K^+\pi^-}~A_{CP}^{K^+\pi^-}}\,
=\, \left\{ 
  \begin{array}{l}  -f_\pi^2/f_K^2 = - 0.66 ~~(\mbox{factorization}) \\
                            -1.54\pm0.66 ~~~(\mbox{Belle~\cite{CP2BELLE}}) \\
                            -0.50\pm0.55 ~~~(\mbox{BaBar~\cite{CP2BABAR}}) \\
                            -1.12\pm0.49 ~~~(\mbox{Average~\cite{HFAG}}) 
          \end{array} \right.
\label{BAp-BAk} 
\eea  
where each values in Eq.~(\ref{BAp-BAk}) corresponds to the each experimental
data of the direct CP asymmetry $A_{CP}^{\pi^+\pi^-}$ and the other
data was used the averaged values. The factors of the CKM matrix
elements are completely canceled. If SU(3) is exact symmetry, the
ratio must be $-1$. {}From the experimental data 
in Eq.~(\ref{BAp-BAk}), one can find
that a possibility of large SU(3) breaking in gluonic penguin contribution
is remaining. When we
assume there is no SU(3) breaking effect in tree type diagram except
for the difference of decay constants in sense of factorization because
it is good agreement in $\frac{B\rightarrow D\pi}{B\rightarrow DK}$, the
gluonic penguin contribution might have the SU(3) breaking effect 
like $P_\pi/P_K \sim 2$.  

\section{Large EW penguin contribution in $B\rightarrow \pi \pi $ } 
\begin{figure}[htbp]
\begin{center}
\begin{minipage}[c]{1.0\textwidth}
\includegraphics[scale=1.0,angle=0]{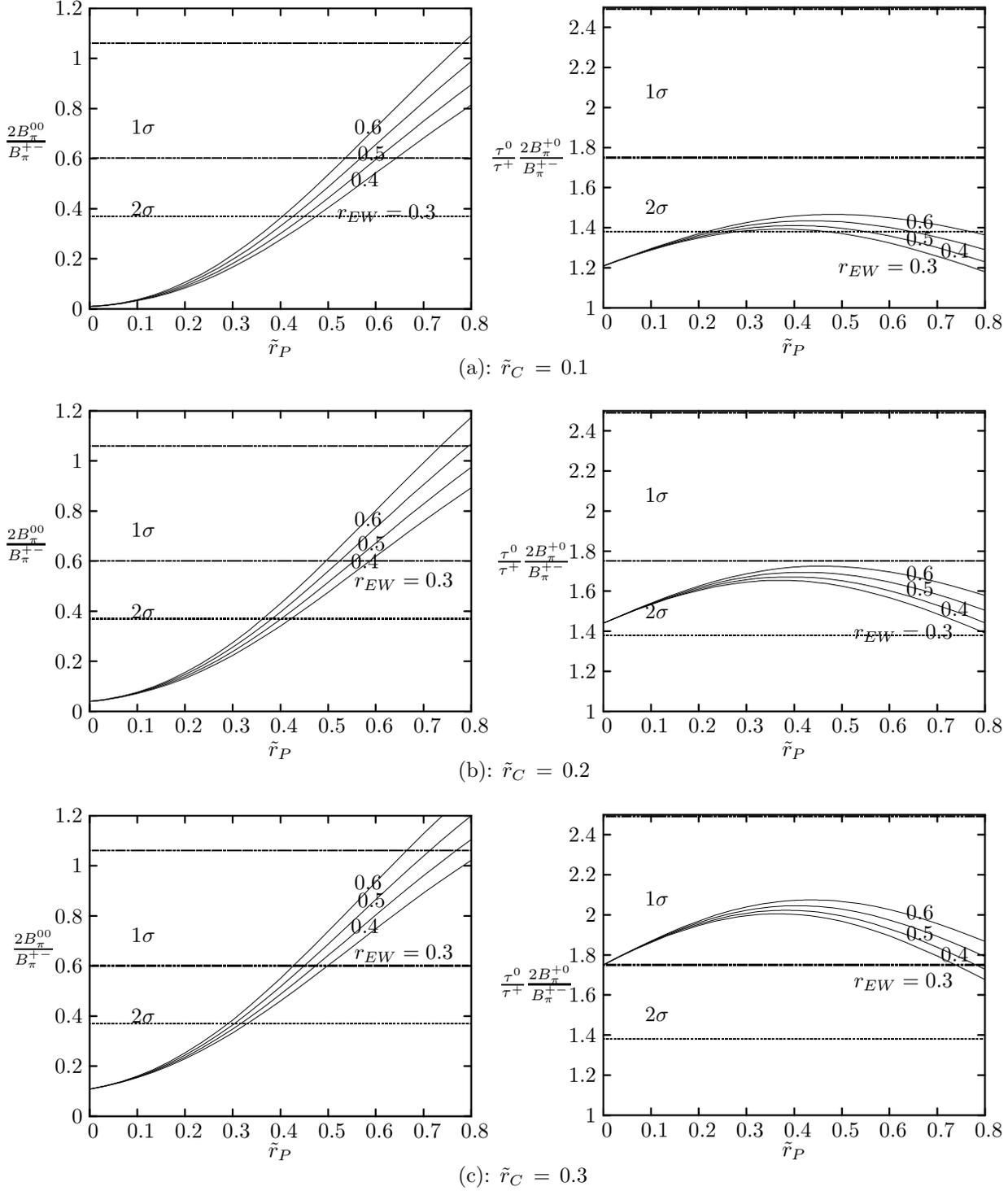}
\end{minipage}
\vspace*{-90mm}
\caption{ $\frac{2B^{00}_\pi }{B^{+-}_\pi }$ and 
          $ \frac{2B^{+0}_\pi }{B^{+-}_\pi }$ as the functions of
           $\tilde{r}_P $ at (a) $\tilde{r}_C=0.1$, (b) $\tilde{r}_C=0.2$
 and (c) $\tilde{r}_C=0.3$. }
    \label{fig:5}
\end{center}
\end{figure}

We discuss about a role of $r_{EW}$ in $B\rightarrow \pi \pi $ decays. To
explain the $B\rightarrow \pi \pi $ modes, we may need SU(3) breaking
effect in gluonic penguin as we discussed in previous section. In this
section, 
we assume that the strong phases almost satisfy the SU(3) symmetry
as an anzatz. 
To enhance the both
ratios $ \frac{2B^{00}_\pi }{B^{+-}_\pi } $ and 
       $ \frac{2B^{+0}_\pi }{B^{+-}_\pi } $,  
smaller denominator will be favored so that the cross term 
$2\tilde{r}_P \cos\delta^T \cos(\phi_1+\phi_3)$ should have negative
sign. Since $\cos \delta^T $ should be positive from the data of
$A_{CP}^{+-}$, $\cos(\phi_1+\phi_3)$ has to be negative value. 
As an example, we plot the
ratios as the function of $\tilde{r}_P$ and $r_{EW}$ in 
a case with $\delta^{EW} = 110^\circ $, $\delta^C = \delta^T =
10^\circ $ and $\phi_1+\phi_3 = 110^\circ$ for $\tilde{r}_C = 0.1,0.2
$ and $0.3$ in Figs.~\ref{fig:5}.  $\phi_1+\phi_3 =
110^\circ$ is almost maximum 
value of allowed region and $\delta^T = 10^\circ $ to satisfy  
$A_{CP}^{K^+\pi^-}$. 
{}From the figures, we can find that to explain 
the discrepancy between the theoretical estimation and the
experimental data, $b-d$ gluonic penguin contribution $P_\pi$
should be larger than $b-s$ gluonic penguin $P_K$ without the CKM
factor. It shows  
SU(3) breaking effect must appear in these decay modes. In addition,
large EW penguin contribution also help to enhance the ratios. 

{}From these figures, we find that the ratios are enhanced by
$\tilde{r}_P$, $\tilde{r}_C$ and $r_{EW}$. However $\tilde{r}_C = C/T$ 
is $0.1$ for the naive estimation by factorization and it will be
at largest up to $1/N_c \sim 0.3$ because it is the simple ratio of
two tree diagrams between color-allowed and color-suppressed types.
Large $\tilde{r}_P$ is an evidence to explain the discrepancies but it 
also has some constraints from $B\rightarrow KK$ decays which are pure
$b-d$ gluonic penguin, $(\sim P_\pi)$ processes. The upper bounds of 
$B\rightarrow KK$ decays \cite{HFAG} are 
\bea
Br(B^+\rightarrow K^+ \bar{K}^0 ) \propto |P~V_{tb}^*V_{td}|^2 
                                       &< & 2.5 \times 10^{-6}\;,  \\
Br(B^0\rightarrow K^0 \bar{K}^0 ) \propto |P~V_{tb}^*V_{td}|^2 
                                       &< & 1.5 \times 10^{-6}\;,  
\eea    
where $P$ is the gluonic penguin contribution without the CKM factor and $P
\sim P_\pi $ under SU(3) symmetry. The constraint to $P_\pi/P_K $
comes from 
\bea
\frac{Br(B^0\rightarrow K^0 \bar{K}^0 )}
     {Br(B^+\rightarrow{K}^0 \pi^+)}\, \sim\, 
\frac{|P_\pi~V_{tb}^*V_{td}|^2 }{|P_K~V_{tb}^*V_{ts}|^2 }\, <\, 7.3 \times
10^{-2}\;,  
\eea  
\bea
\frac{P_\pi}{P_K}\, <\, 1.5\;.   
\eea 
Thus $\tilde{r}_P$ may be allowed up to $0.3 \times 1.5 = 0.45
$. In addition, we will also need the help from $r_{EW} = P_{EW}/P$ to
enhance $B_\pi^{00}$. 

It is slightly difficult to get the values within the $1\sigma
$ region unless larger $r_C$ is allowed. However we feel that it may be
unnatural that such tree diagram obtains the larger contribution than
usual estimation. Therefore we consider the case keeping small $r_C$ and 
including some new effects in penguin contribution.

\section{New physics contribution} 
If the deviations come from new physics contribution, it
has to be included in the penguin like contribution with new weak
phases because it is very difficult to produce such large strong phase 
difference as $\omega = \delta^{EW}-\delta^T \sim 100^\circ $ within the
SM. 
$B\rightarrow K\pi $ decays need large EW penguin contributions so
that it may be including the new physics contribution with new
weak phase in the EW penguin. 
Besides, the effect must appear also in the direct CP asymmetries. 
For example, $A_{CP}^{K^0\pi^0} \propto 2 r_{EW} \sin\delta^{EW}
\sin\theta^{New} $, so that we will have to check carefully these
modes. 

We consider a possibility of new physics in the penguin 
contributions as follows:
\bea
P\, =\, P^{SM} + P^{New}\;, ~~~~ 
P_{EW}\, =\,  P^{SM}_{EW} + P^{New}_{EW}\;, 
\eea    
where $P^{New}$ and $P^{New}_{EW}$ are gluonic and EW penguin type
contributions coming from new physics,
respectively. For simplicity, we assume that the strong phase of the
penguin diagram within the SM is the same with one from new physics. Here 
we parameterize the phases as follows:
\bea
P\, V^*_{tb} V_{ts}\, &\equiv & -|P\, V^*_{tb} V_{ts}|\, e^{-i\theta^P }\;,\\
P_{EW} V^*_{tb}\, V_{ts} &\equiv & -|P_{EW} V^*_{tb} V_{ts}|\, 
                           e^{i\delta^{EW}}e^{i\theta^{EW} }\;,
\eea 
where $\theta^P$ and $\theta^{EW}$ are the weak phases coming 
from new physics contributions. The ratios among the parameters are
\bea 
\frac{T\, V_{ub}^*V_{us} }{P\, V_{tb}^* V_{ts} } &=& 
     r_T\, e^{i\delta^T} e^{i (\phi_3+\theta^P) }\;, \\
\frac{P_{EW} V_{tb}^*V_{ts} }{P\, V_{tb}^* V_{ts} } &=& 
     r_{EW}\, e^{i\delta^{EW}} e^{i (\theta^{EW}+ \theta^P ) }\;. 
\eea 
Using this parameterization, 
\bea
R_c - R_n  &=&
   2 r_{EW}^2 \left( 1 - 2 \cos^2\delta^{EW}
   \cos^2(\theta^{EW}+\theta^P) 
              \right)
    - 2 r_{EW}~r_T \cos(\delta^{EW}-\delta^{T})\cos(\phi_3 -
   \theta^{EW})
                \nn \\
& &  + 4 r_{EW} r_{T} 
       \cos\delta^{EW}\cos\delta^T \cos(\phi_3+\theta^P)
           \cos(\theta^{EW}+\theta^P ) , \\[3mm]
S ~~~ &=&  2 r_{EW}^2 
          - 2 r_{EW} r_{T}  
        \cos(\delta^{EW}-\delta^{T})\cos(\phi_3 - \theta^{EW}). 
\eea 
Because of the new weak phase $\theta^{EW}$ and $\theta^P$, the
constraints for the strong phases is fairly relaxed. To keep the first
term to positive value, $\cos^2\delta^{EW} \cos^2(\theta^{EW}+\theta^P)$
must be less than $1/2$ so that smaller $|\cos(\theta^{EW}+\theta^P)|$ 
will be favored. In the second term, 
if $\cos(\phi_3 - \theta^{EW})$ was negative, small $\omega 
= \delta^{EW}-\delta^T $ was not excluded in contrast with the SM case. 
The
constraint for $r_{EW}$ is almost same but one for the strong
phases is changed and almost region for the strong phase
$\delta^{EW}$ is allowed. Furthermore, small $\omega$ is also allowed in this
case. In other words, the constraint for $\delta^{EW}$ is
replaced to one for the new weak phase and their magnitude is not negligible
value.    
Therefore, to investigate the direct
CP asymmetries for $B\rightarrow K\pi $ 
will become more important to know the information
about the new CP phases. Especially, $A_{CP}^{00}$ will be
important mode. The CP asymmetries for $B\rightarrow K\pi $ decays are 
\bea
A_{CP}^{+-} &\propto & - 2 r_T \sin\delta^T \sin(\phi_3 + \theta^P ) 
                       - r_T^2 \sin2\delta^T \sin2(\phi_3+\theta^P)\;,
\\[5mm]  
A_{CP}^{00} &\propto & - 2 r_{EW} \sin{\delta^{EW}} 
                            \sin(\theta^{EW}+\theta^P) 
           - r_{EW}^2 \sin2\delta^{EW}\sin2(\theta^{EW}+\theta^P)\nn
                            \\
   & & + 2 r_C \sin\delta^C \sin(\phi_3 +\theta^P)\;,
                            \\[5mm]
A_{CP}^{+0} &\propto & 2 r_{EW} \sin{\delta^{EW}} 
                            \sin(\theta^{EW}+\theta^P) 
                       - 2 r_T \sin\delta^T \sin(\phi_3+\theta^P) \nn
                            \\
      & &   + 2 r_T r_{EW} \sin(\delta^{EW} + \delta^T) 
                        \sin(\theta^{EW}+\phi_3+2\theta^P ) \nn \\ 
      & &   - r_{EW}^2 \sin2\delta^{EW}\sin2(\theta^{EW}+\theta^P)
            - r_{T}^2 \sin2\delta^{T}\sin2(\phi_3+\theta^P)\;. 
\eea 
To explain the deviation of $R_c-R_n$, large $r_{EW}$ and large 
$\sin{\delta^{EW}} \sin(\theta^{EW}+\theta^P)$ are favored so that if 
there is some new physics contributions in $B\rightarrow K\pi $, 
sizable $A_{CP}^{K^0\pi^0}$ will appear in the $B$ factory experiment in the
near future. At present time, these experimental data have still
large uncertainties.  

\begin{figure}[htbp]
\begin{center}
\begin{minipage}[c]{1.0\textwidth}
\includegraphics[scale=1.0,angle=0]{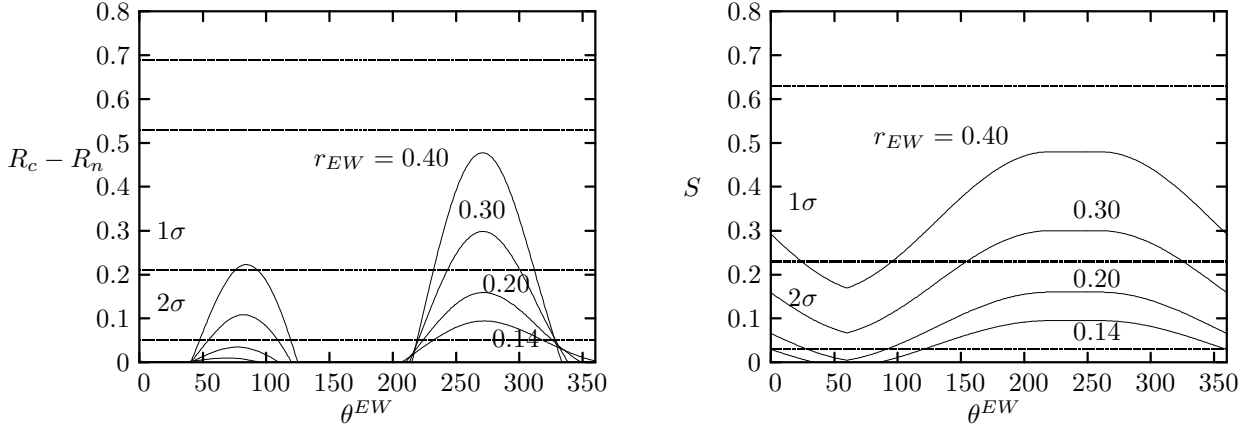}
\end{minipage}
\vspace*{-233mm}
\caption{The lines show the maximum bound of $R_c-R_n$ and $S$ for
         $\theta^{EW}$ at 
    $\omega = 0^\circ $ 
    and $r_T = 0.2 $  
    under constraint $-0.123 < A_{CP}^{+-} < -0.067$, $\theta^P = 0^\circ
    $ and $40^\circ <\phi_3<80^\circ $.}
    \label{fig:8}
\end{center}
\end{figure}

\begin{figure}[tbhp]
\begin{center}
\begin{minipage}[c]{3in} 
\begin{center}
\setlength{\unitlength}{0.100450pt}
\begin{picture}(2699,2069)(400,0)
\footnotesize
\thicklines \path(370,249)(411,249)
\thicklines \path(2576,249)(2535,249)
\put(329,249){\makebox(0,0)[r]{ 0}}
\thicklines \path(370,491)(411,491)
\thicklines \path(2576,491)(2535,491)
\put(329,491){\makebox(0,0)[r]{ 50}}
\thicklines \path(370,732)(411,732)
\thicklines \path(2576,732)(2535,732)
\put(329,732){\makebox(0,0)[r]{ 100}}
\thicklines \path(370,974)(411,974)
\thicklines \path(2576,974)(2535,974)
\put(329,974){\makebox(0,0)[r]{ 150}}
\thicklines \path(370,1215)(411,1215)
\thicklines \path(2576,1215)(2535,1215)
\put(329,1215){\makebox(0,0)[r]{ 200}}
\thicklines \path(370,1457)(411,1457)
\thicklines \path(2576,1457)(2535,1457)
\put(329,1457){\makebox(0,0)[r]{ 250}}
\thicklines \path(370,1698)(411,1698)
\thicklines \path(2576,1698)(2535,1698)
\put(329,1698){\makebox(0,0)[r]{ 300}}
\thicklines \path(370,1940)(411,1940)
\thicklines \path(2576,1940)(2535,1940)
\put(329,1940){\makebox(0,0)[r]{ 350}}
\thicklines \path(370,249)(370,290)
\thicklines \path(370,1988)(370,1947)
\put(370,166){\makebox(0,0){ 0}}
\thicklines \path(676,249)(676,290)
\thicklines \path(676,1988)(676,1947)
\put(676,166){\makebox(0,0){ 50}}
\thicklines \path(983,249)(983,290)
\thicklines \path(983,1988)(983,1947)
\put(983,166){\makebox(0,0){ 100}}
\thicklines \path(1289,249)(1289,290)
\thicklines \path(1289,1988)(1289,1947)
\put(1289,166){\makebox(0,0){ 150}}
\thicklines \path(1596,249)(1596,290)
\thicklines \path(1596,1988)(1596,1947)
\put(1596,166){\makebox(0,0){ 200}}
\thicklines \path(1902,249)(1902,290)
\thicklines \path(1902,1988)(1902,1947)
\put(1902,166){\makebox(0,0){ 250}}
\thicklines \path(2208,249)(2208,290)
\thicklines \path(2208,1988)(2208,1947)
\put(2208,166){\makebox(0,0){ 300}}
\thicklines \path(2515,249)(2515,290)
\thicklines \path(2515,1988)(2515,1947)
\put(2515,166){\makebox(0,0){ 350}}
\thicklines \path(370,249)(2576,249)(2576,1988)(370,1988)(370,249)
\put(22,1118){\makebox(0,0)[l]{{$\theta^{P}$}}}
\put(1473,42){\makebox(0,0){$ 	\theta^{EW} $ }}
\thinlines
    \path(370,626)(502,626)(502,626)(502,635)(502,645)(502,655)(502,664)(502,674)(502,684)
(502,693)(480,703)(480,713)(458,722)(436,732)(436,742)(414,751)(414,761)(392,771)(502,780)(502,790)
(502,800)(502,809)(502,819)(480,829)(480,838)(458,848)(458,858)(436,867)(436,877)(414,887)(414,896)
(392,906)(370,916)(370,925)
\thinlines
\path(370,1495)(502,1495)(502,1495)(502,1505)(502,1515)(502,1524)(502,1534)(502,1544)
(502,1553)(502,1563)(480,1573)(480,1582)(458,1592)(436,1602)(436,1611)(414,1621)(414,1631)(392,1640)
(502,1650)(502,1660)(502,1669)(502,1679)(502,1689)(480,1698)(480,1708)(458,1717)(458,1727)(436,1737)
(436,1746)(414,1756)(414,1766)(392,1775)(370,1785)(370,1795)
\thinlines
\path(370,626)(370,626)(370,635)(370,645)(370,655)(370,664)(370,674)(370,684)(370,693)
(370,703)(370,713)(370,722)(370,732)(370,742)(370,751)(370,761)(370,771)(370,780)(370,790)(370,800)
(370,809)(370,819)(370,829)(370,838)(370,848)(370,858)(370,867)(370,877)(370,887)(370,896)(370,906)
(370,916)(370,925)
\thinlines
\path(370,1495)(370,1495)(370,1505)(370,1515)(370,1524)(370,1534)(370,1544)(370,1553)
(370,1563)(370,1573)(370,1582)(370,1592)(370,1602)(370,1611)(370,1621)(370,1631)(370,1640)(370,1650)
(370,1660)(370,1669)(370,1679)(370,1689)(370,1698)(370,1708)(370,1717)(370,1727)(370,1737)(370,1746)
(370,1756)(370,1766)(370,1775)(370,1785)(370,1795)
\thinlines
\path(2267,249)(2267,249)(2267,259)(2245,268)(2223,278)(2223,288)(2201,297)(2201,307)
(2179,317)(2179,326)(2157,336)(2135,346)(2135,355)(2113,365)(2113,375)(2091,384)(2069,394)(2069,404)
(2047,413)(2047,423)(2025,433)(2002,442)(2002,452)(1980,462)(1980,471)(1958,481)(1936,491)(1936,500)
(1914,510)(1892,520)(1892,529)(1870,539)(1870,548)(1848,558)(1826,568)(1826,577)(1804,587)(1782,597)
(1782,606)(1760,616)(2576,626)(2576,635)(2576,645)(2576,655)(2576,664)(2576,674)(2576,684)(2576,693)
(2576,703)(2576,713)(2576,722)
\thinlines
\path(2576,722)(2576,732)(2576,742)(2576,751)(2576,761)(2576,771)(2576,780)(2576,790)
(2576,800)(2576,809)(2576,819)(2576,829)(2576,838)(2576,848)(2576,858)(2576,867)(2576,877)(2576,887)
(2576,896)(2576,906)(2576,916)(2576,925)(2554,935)(2554,945)(2532,954)(2532,964)(2510,974)(2510,983)
(2488,993)(2466,1003)(2422,1012)(2400,1022)(2377,1032)(2377,1041)(2355,1051)(2333,1061)(2333,1070)
(2311,1080)(2311,1090)(2289,1099)(2289,1109)(2267,1119)(2267,1128)(2245,1138)(2223,1147)(2223,1157)
(2201,1167)(2201,1176)(2179,1186)(2179,1196)(2157,1205)
\thinlines
\path(2157,1205)(2135,1215)(2135,1225)(2113,1234)(2113,1244)(2091,1254)(2069,1263)(2069,1273)
(2047,1283)(2047,1292)(2025,1302)(2002,1312)(2002,1321)(1980,1331)(1980,1341)(1958,1350)(1936,1360)
(1936,1370)(1914,1379)(1892,1389)(1892,1399)(1870,1408)(1870,1418)(1848,1428)(1826,1437)(1826,1447)
(1804,1457)(1782,1466)(1782,1476)(1760,1486)(2576,1495)(2576,1505)(2576,1515)(2576,1524)(2576,1534)
(2576,1544)(2576,1553)(2576,1563)(2576,1573)(2576,1582)(2576,1592)(2576,1602)(2576,1611)(2576,1621)
(2576,1631)(2576,1640)(2576,1650)(2576,1660)(2576,1669)(2576,1679)(2576,1689)
\thinlines
\path(2576,1689)(2576,1698)(2576,1708)(2576,1717)(2576,1727)(2576,1737)(2576,1746)(2576,1756)
(2576,1766)(2576,1775)(2576,1785)(2576,1795)(2554,1804)(2554,1814)(2532,1824)(2532,1833)(2510,1843)
(2510,1853)(2488,1862)(2466,1872)(2422,1882)(2400,1891)(2377,1901)(2377,1911)(2355,1920)(2333,1930)
(2333,1940)(2311,1949)(2311,1959)(2289,1969)(2289,1978)(2267,1988)
\thinlines
\path(1804,249)(1804,249)(1782,259)(1782,268)(1760,278)(1760,288)(1738,297)(1716,307)
(1716,317)(1694,326)(1672,336)(1672,346)(1649,355)(1649,365)(1627,375)(1605,384)(1605,394)(1583,404)
(1583,413)(1561,423)(1561,433)(1539,442)(1517,452)(1517,462)(1495,471)(1495,481)(1473,491)(1451,500)
(1451,510)(1429,520)(1429,529)(1407,539)(1407,548)(1385,558)(1385,568)(1363,577)(1341,587)(1341,597)
(1319,606)(1319,616)(1297,626)(1274,635)(1252,645)(1208,655)(1186,664)(1186,674)(1164,684)(1164,693)
(1142,703)(1142,713)(1120,722)
\thinlines
\path(1120,722)(1098,732)(1098,742)(1076,751)(1076,761)(1054,771)(1054,780)(1032,790)
(1032,800)(1010,809)(988,819)(988,829)(966,838)(966,848)(966,858)(966,867)(966,877)(1076,887)(1076,896)
(1054,906)(1032,916)(1032,925)(1010,935)(1010,945)(988,954)(988,964)(966,974)(966,983)(966,993)(966,1003)
(966,1012)(966,1022)(966,1032)(1914,1041)(1914,1051)(1892,1061)(1870,1070)(1870,1080)(1848,1090)
(1826,1099)(1826,1109)(1804,1119)(1782,1128)(1782,1138)(1760,1147)(1760,1157)(1738,1167)(1716,1176)
(1716,1186)(1694,1196)(1672,1205)
\thinlines
\path(1672,1205)(1672,1215)(1649,1225)(1649,1234)(1627,1244)(1605,1254)(1605,1263)(1583,1273)
(1583,1283)(1561,1292)(1561,1302)(1539,1312)(1517,1321)(1517,1331)(1495,1341)(1495,1350)(1473,1360)
(1451,1370)(1451,1379)(1429,1389)(1429,1399)(1407,1408)(1407,1418)(1385,1428)(1385,1437)(1363,1447)
(1341,1457)(1341,1466)(1319,1476)(1319,1486)(1297,1495)(1274,1505)(1252,1515)(1208,1524)(1186,1534)
(1186,1544)(1164,1553)(1164,1563)(1142,1573)(1142,1582)(1120,1592)(1098,1602)(1098,1611)(1076,1621)
(1076,1631)(1054,1640)(1054,1650)(1032,1660)(1032,1669)(1010,1679)(988,1689)
\thinlines
\path(988,1689)(988,1698)(966,1708)(966,1717)(966,1727)(966,1737)(966,1746)(1076,1756)
(1076,1766)(1054,1775)(1032,1785)(1032,1795)(1010,1804)(1010,1814)(988,1824)(988,1833)(966,1843)
(966,1853)(966,1862)(966,1872)(966,1882)(966,1891)(966,1901)(1914,1911)(1914,1920)(1892,1930)(1870,1940)
(1870,1949)(1848,1959)(1826,1969)(1826,1978)(1804,1988)
\thinlines\dashline[-20]{11}(2201,249)(2201,249)(2179,259)(2179,268)(2157,278)(2157,288)(2135,297)
(2135,307)(2113,317)(2091,326)(2091,336)(2069,346)(2069,355)(2047,365)(2047,375)(2025,384)(2025,394)
(2002,404)(1980,413)(1980,423)(1958,433)(1958,442)(1936,452)(1914,462)(1914,471)(1892,481)(1870,491)
(1870,500)(1848,510)(1848,520)(1826,529)(1804,539)(1804,548)(1782,558)(1760,568)(1760,577)(1738,587)
(1716,597)(1694,606)(1694,616)(1672,626)(1649,635)(1649,645)(2355,655)(2355,664)(2355,674)(2355,684)
(2333,693)(2333,703)(2311,713)(2311,722)
\thinlines\dashline[-20]{11}(2311,722)(2289,732)(2289,742)(2267,751)(2245,761)(2245,771)(2355,780)
(2355,790)(2355,800)(2355,809)(2355,819)(2355,829)(2355,838)(2333,848)(2333,858)(2311,867)(2311,877)
(2289,887)(2289,896)(2267,906)(2245,916)(2355,925)(2355,935)(2355,945)(2355,954)(2355,964)(2355,974)
(2355,983)(2355,993)(2355,1003)(2333,1012)(2311,1022)(2311,1032)(2289,1041)(2289,1051)(2267,1061)
(2267,1070)(2245,1080)(2245,1090)(2223,1099)(2201,1109)(2201,1119)(2179,1128)(2179,1138)(2157,1147)
(2157,1157)(2135,1167)(2135,1176)(2113,1186)(2091,1196)(2091,1205)
\thinlines\dashline[-20]{11}(2091,1205)(2069,1215)(2069,1225)(2047,1234)(2047,1244)(2025,1254)
(2025,1263)(2002,1273)(1980,1283)(1980,1292)(1958,1302)(1958,1312)(1936,1321)(1914,1331)(1914,1341)
(1892,1350)(1870,1360)(1870,1370)(1848,1379)(1848,1389)(1826,1399)(1804,1408)(1804,1418)(1782,1428)
(1760,1437)(1760,1447)(1738,1457)(1716,1466)(1694,1476)(1694,1486)(1672,1495)(1649,1505)(1649,1515)
(2355,1524)(2355,1534)(2355,1544)(2355,1553)(2333,1563)(2333,1573)(2311,1582)(2311,1592)(2289,1602)
(2289,1611)(2267,1621)(2245,1631)(2245,1640)(2355,1650)(2355,1660)(2355,1669)(2355,1679)(2355,1689)
\thinlines\dashline[-20]{11}(2355,1689)(2355,1698)(2355,1708)(2333,1717)(2333,1727)(2311,1737)
(2311,1746)(2289,1756)(2289,1766)(2267,1775)(2245,1785)(2355,1795)(2355,1804)(2355,1814)(2355,1824)
(2355,1833)(2355,1843)(2355,1853)(2355,1862)(2355,1872)(2333,1882)(2311,1891)(2311,1901)(2289,1911)
(2289,1920)(2267,1930)(2267,1940)(2245,1949)(2245,1959)(2223,1969)(2201,1978)(2201,1988)
\thinlines\dashline[-20]{8}(1870,249)(1870,249)(1848,259)(1848,268)(1826,278)(1826,288)(1804,297)
(1782,307)(1782,317)(1760,326)(1738,336)(1738,346)(1716,355)(1716,365)(1694,375)(1672,384)(1672,394)
(1649,404)(1649,413)(1627,423)(1627,433)(1605,442)(1583,452)(1583,462)(1561,471)(1561,481)(1539,491)
(1539,500)(1517,510)(1517,520)(1495,529)(1473,539)(1473,548)(1451,558)(1451,568)(1429,577)(1429,587)
(1407,597)(1407,606)(1385,616)(1363,626)(1363,635)(1341,645)(1341,655)(1319,664)(1319,674)(1319,684)
(1319,693)(1319,703)(1319,713)(1319,722)
\thinlines\dashline[-20]{11}(1319,722)(1319,732)(1429,742)(1407,751)(1407,761)(1385,771)(1385,780)
(1363,790)(1363,800)(1341,809)(1319,819)(1319,829)(1319,838)(1319,848)(1319,858)(1319,867)(1319,877)
(1429,887)(1429,896)(1407,906)(1407,916)(1385,925)(1385,935)(1363,945)(1363,954)(1341,964)(1319,974)
(1319,983)(1319,993)(1319,1003)(2025,1012)(2025,1022)(2002,1032)(2002,1041)(1980,1051)(1958,1061)
(1936,1070)(1936,1080)(1914,1090)(1892,1099)(1892,1109)(1870,1119)(1848,1128)(1848,1138)(1826,1147)
(1826,1157)(1804,1167)(1782,1176)(1782,1186)(1760,1196)(1738,1205)
\thinlines\dashline[-20]{11}(1738,1205)(1738,1215)(1716,1225)(1716,1234)(1694,1244)(1672,1254)(1672,1263)
(1649,1273)(1649,1283)(1627,1292)(1627,1302)(1605,1312)(1583,1321)(1583,1331)(1561,1341)(1561,1350)
(1539,1360)(1539,1370)(1517,1379)(1517,1389)(1495,1399)(1473,1408)(1473,1418)(1451,1428)(1451,1437)
(1429,1447)(1429,1457)(1407,1466)(1407,1476)(1385,1486)(1363,1495)(1363,1505)(1341,1515)(1341,1524)
(1319,1534)(1319,1544)(1319,1553)(1319,1563)(1319,1573)(1319,1582)(1319,1592)(1319,1602)(1429,1611)
(1407,1621)(1407,1631)(1385,1640)(1385,1650)(1363,1660)(1363,1669)(1341,1679)(1319,1689)
\thinlines\dashline[-20]{11}(1319,1689)(1319,1698)(1319,1708)(1319,1717)(1319,1727)(1319,1737)(1319,1746)
(1429,1756)(1429,1766)(1407,1775)(1407,1785)(1385,1795)(1385,1804)(1363,1814)(1363,1824)(1341,1833)
(1319,1843)(1319,1853)(1319,1862)(1319,1872)(2025,1882)(2025,1891)(2002,1901)(2002,1911)(1980,1920)
(1958,1930)(1936,1940)(1936,1949)(1914,1959)(1892,1969)(1892,1978)(1870,1988)
\end{picture}
\end{center}
\end{minipage} 
\caption{The lines show the allowed region for $\theta^{EW}$ and
         $\theta^P $ satisfying the data of $R_c-R_n $ and $S$ a
         $1\sigma $ level. The solid line shows that the case of $r_{EW}
         = 0.40 $ and the dashed lines is $r_{EW} = 0.30 $ at 
    $\omega =0^\circ $ and $r_T = 0.2 $  
    under constraint $-0.123 < A_{CP}^{+-} < -0.067$ 
    and $40^\circ <\phi_3<80^\circ $.}
    \label{fig:9}
\eec
\end{figure}
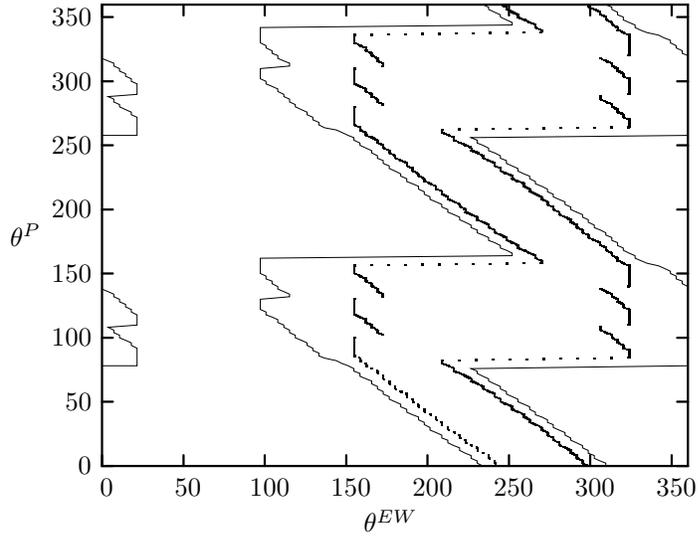

As we used $A_{CP}^{+-}$ before, it seems good accurate one so that 
it is good examples to plot the maximum bound of $R_c-R_n$ and $S$
as the function of $\theta^{EW}$ for each $r_{EW}$
at $\omega =0^\circ$ and $r_T=0.2$ under constraint of $A_{CP}^{+-} $. 
Here we take $\theta^P =0 $, for simplicity. Figure~\ref{fig:8} shows that  
the allowed region for both constraints of  
$R_c-R_n$ and $S$ exists even if $\omega = \delta^{EW} - \delta^T = 0^\circ$ 
and it is about $240^\circ  < \theta^{EW} <  300^\circ $. 
If the $\theta^P $ is non zero values, the allowed region should be
wider.   
In Fig.~\ref{fig:9}, the allowed region on the plane of $\theta^{EW}$ and
$\theta^{P}$ are plotted.  

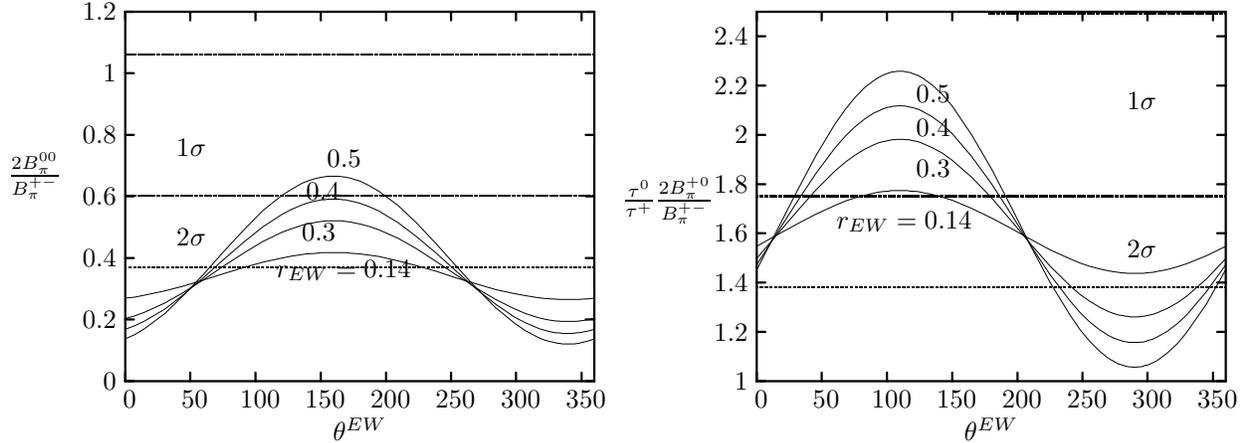
\begin{figure}[tb]
\bec
\hspace{5mm}
\begin{minipage}[l]{3.0in}
\begin{center}
\setlength{\unitlength}{0.080450pt}
\begin{picture}(2699,2069)(0,0)
\footnotesize
\thicklines \path(370,249)(411,249)
\thicklines \path(2576,249)(2535,249)
\put(329,249){\makebox(0,0)[r]{ 0}}
\thicklines \path(370,539)(411,539)
\thicklines \path(2576,539)(2535,539)
\put(329,539){\makebox(0,0)[r]{ 0.2}}
\thicklines \path(370,829)(411,829)
\thicklines \path(2576,829)(2535,829)
\put(329,829){\makebox(0,0)[r]{ 0.4}}
\thicklines \path(370,1119)(411,1119)
\thicklines \path(2576,1119)(2535,1119)
\put(329,1119){\makebox(0,0)[r]{ 0.6}}
\thicklines \path(370,1408)(411,1408)
\thicklines \path(2576,1408)(2535,1408)
\put(329,1408){\makebox(0,0)[r]{ 0.8}}
\thicklines \path(370,1698)(411,1698)
\thicklines \path(2576,1698)(2535,1698)
\put(329,1698){\makebox(0,0)[r]{ 1}}
\thicklines \path(370,1988)(411,1988)
\thicklines \path(2576,1988)(2535,1988)
\put(329,1988){\makebox(0,0)[r]{ 1.2}}
\thicklines \path(370,249)(370,290)
\thicklines \path(370,1988)(370,1947)
\put(370,166){\makebox(0,0){ 0}}
\thicklines \path(676,249)(676,290)
\thicklines \path(676,1988)(676,1947)
\put(676,166){\makebox(0,0){ 50}}
\thicklines \path(983,249)(983,290)
\thicklines \path(983,1988)(983,1947)
\put(983,166){\makebox(0,0){ 100}}
\thicklines \path(1289,249)(1289,290)
\thicklines \path(1289,1988)(1289,1947)
\put(1289,166){\makebox(0,0){ 150}}
\thicklines \path(1596,249)(1596,290)
\thicklines \path(1596,1988)(1596,1947)
\put(1596,166){\makebox(0,0){ 200}}
\thicklines \path(1902,249)(1902,290)
\thicklines \path(1902,1988)(1902,1947)
\put(1902,166){\makebox(0,0){ 250}}
\thicklines \path(2208,249)(2208,290)
\thicklines \path(2208,1988)(2208,1947)
\put(2208,166){\makebox(0,0){ 300}}
\thicklines \path(2515,249)(2515,290)
\thicklines \path(2515,1988)(2515,1947)
\put(2515,166){\makebox(0,0){ 350}}
\thicklines \path(370,249)(2576,249)(2576,1988)(370,1988)(370,249)
\put(-180,1218){\makebox(0,0)[l]{$\frac{2 B^{00}_\pi}{B^{+-}_\pi}$}}
\put(1080,770){\makebox(0,0)[l]{$r_{EW} = 0.14$}}
\put(1473,22){\makebox(0,0){$ \theta^{EW} $}}
\put(1300,1140){\makebox(0,0){$ 0.4 $}}
\put(1400,1300){\makebox(0,0){$ 0.5 $}}
\put(1280,950){\makebox(0,0){$ 0.3 $}}
\put(680,1350){\makebox(0,0){$ 1\sigma $}}
\put(680,930){\makebox(0,0){$ 2\sigma $}}
\thinlines
\path(370,544)(370,544)(414,556)(459,572)(557,617)(645,667)(739,727)(829,788)
(925,851)(1018,906)(1107,950)(1152,967)(1202,983)(1224,989)(1249,994)(1272,998)(1293,1001)
(1304,1002)(1315,1003)(1321,1003)(1327,1003)(1331,1004)(1337,1004)(1340,1004)(1342,1004)
(1344,1004)(1346,1004)(1347,1004)(1349,1004)(1350,1004)(1352,1004)(1355,1004)(1357,1004)
(1358,1004)(1362,1004)(1364,1004)(1367,1004)(1373,1003)(1385,1003)(1396,1002)(1407,1001)
(1430,998)(1452,994)(1473,989)(1519,977)(1562,962)(1653,921)(1741,871)(1834,811)(1924,751)(2020,688)
\thinlines
\path(2020,688)(2112,633)(2201,588)(2250,568)(2295,553)(2317,547)(2342,541)(2364,537)
(2386,534)(2397,533)(2408,532)(2419,531)(2426,530)(2432,530)(2438,530)(2441,530)(2443,530)
(2444,530)(2447,530)(2450,530)(2453,530)(2455,530)(2458,530)(2460,530)(2462,530)(2465,530)
(2468,530)(2474,530)(2481,530)(2492,531)(2503,532)(2527,535)(2550,539)(2576,544)(2576,544)
\thinlines
\path(370,493)(370,493)(414,509)(459,530)(557,589)(645,656)(739,737)(829,818)(925,902)
(1018,975)(1107,1033)(1152,1057)(1202,1078)(1224,1086)(1249,1093)(1272,1098)(1293,1102)(1304,1103)
(1315,1104)(1321,1105)(1327,1105)(1331,1105)(1337,1106)(1340,1106)(1342,1106)(1344,1106)(1346,1106)
(1347,1106)(1349,1106)(1350,1106)(1352,1106)(1355,1106)(1357,1106)(1358,1106)(1362,1106)(1364,1105)
(1367,1105)(1373,1105)(1385,1104)(1396,1103)(1407,1102)(1430,1097)(1452,1092)(1473,1086)(1519,1070)
(1562,1050)(1653,995)(1741,929)(1834,849)(1924,768)(2020,684)
\thinlines
\path(2020,684)(2112,611)(2201,551)(2250,525)(2295,505)(2317,497)(2342,489)(2364,483)
(2386,479)(2397,477)(2408,476)(2419,475)(2426,474)(2432,474)(2438,474)(2441,474)(2443,474)(2444,474)
(2447,474)(2450,474)(2453,474)(2455,474)(2458,474)(2460,474)(2462,474)(2465,474)(2468,474)(2474,474)
(2481,475)(2492,476)(2503,477)(2527,481)(2550,486)(2576,493)(2576,493)
\thinlines
\path(370,448)(370,448)(414,468)(459,494)(557,569)(645,653)(739,753)(829,854)(925,959)
(1018,1051)(1107,1124)(1152,1153)(1202,1180)(1224,1189)(1249,1198)(1272,1205)(1293,1209)(1304,1211)
(1315,1212)(1321,1213)(1327,1213)(1331,1214)(1337,1214)(1340,1214)(1342,1214)(1344,1214)(1346,1214)
(1347,1214)(1349,1214)(1350,1214)(1352,1214)(1355,1214)(1357,1214)(1358,1214)(1362,1214)(1364,1214)
(1367,1214)(1373,1213)(1385,1212)(1396,1211)(1407,1209)(1430,1204)(1452,1197)(1473,1190)(1519,1169)
(1562,1144)(1653,1075)(1741,993)(1834,893)(1924,793)(2020,687)
\thinlines
\path(2020,687)(2112,595)(2201,521)(2250,488)(2295,463)(2317,453)(2342,443)(2364,436)
(2386,431)(2397,429)(2408,427)(2419,426)(2426,425)(2432,425)(2438,424)(2441,424)(2443,424)(2444,424)
(2447,424)(2450,424)(2453,424)(2455,424)(2458,424)(2460,424)(2462,424)(2465,424)(2468,424)(2474,425)
(2481,425)(2492,426)(2503,428)(2527,433)(2550,439)(2576,448)(2576,448)
\thinlines
\path(370,640)(370,640)(414,645)(459,653)(557,674)(645,697)(739,725)(829,754)(925,783)
(1018,809)(1107,829)(1152,837)(1202,845)(1224,848)(1249,850)(1272,852)(1293,853)(1304,854)(1315,854)
(1321,854)(1327,854)(1331,854)(1337,854)(1340,854)(1342,854)(1344,854)(1346,854)(1347,854)(1349,854)
(1350,854)(1352,854)(1355,854)(1357,854)(1358,854)(1362,854)(1364,854)(1367,854)(1373,854)(1385,854)
(1396,854)(1407,853)(1430,852)(1452,850)(1473,848)(1519,842)(1562,835)(1653,816)(1741,792)(1834,765)
(1924,736)(2020,707)
\thinlines
\path(2020,707)(2112,681)(2201,660)(2250,651)(2295,644)(2317,641)(2342,639)(2364,637)
(2386,635)(2397,635)(2408,634)(2419,634)(2426,634)(2432,633)(2438,633)(2441,633)(2443,633)(2444,633)
(2447,633)(2450,633)(2453,633)(2455,633)(2458,633)(2460,633)(2462,633)(2465,633)(2468,633)(2474,633)
(2481,634)(2492,634)(2503,634)(2527,636)(2550,637)(2576,640)(2576,640)
\thinlines\dashline[-20]{18}(370,1785)(370,1785)(392,1785)(415,1785)(437,1785)(459,1785)(481,1785)
(504,1785)(526,1785)(548,1785)(571,1785)(593,1785)(615,1785)(637,1785)(660,1785)(682,1785)(704,1785)
(727,1785)(749,1785)(771,1785)(793,1785)(816,1785)(838,1785)(860,1785)(883,1785)(905,1785)(927,1785)
(949,1785)(972,1785)(994,1785)(1016,1785)(1038,1785)(1061,1785)(1083,1785)(1105,1785)(1128,1785)
(1150,1785)(1172,1785)(1194,1785)(1217,1785)(1239,1785)(1261,1785)(1284,1785)(1306,1785)(1328,1785)
(1350,1785)(1373,1785)(1395,1785)(1417,1785)(1440,1785)(1462,1785)
\thinlines\dashline[-20]{18}(1462,1785)(1484,1785)(1506,1785)(1529,1785)(1551,1785)(1573,1785)
(1596,1785)(1618,1785)(1640,1785)(1662,1785)(1685,1785)(1707,1785)(1729,1785)(1752,1785)(1774,1785)
(1796,1785)(1818,1785)(1841,1785)(1863,1785)(1885,1785)(1908,1785)(1930,1785)(1952,1785)(1974,1785)
(1997,1785)(2019,1785)(2041,1785)(2063,1785)(2086,1785)(2108,1785)(2130,1785)(2153,1785)(2175,1785)
(2197,1785)(2219,1785)(2242,1785)(2264,1785)(2286,1785)(2309,1785)(2331,1785)(2353,1785)(2375,1785)
(2398,1785)(2420,1785)(2442,1785)(2465,1785)(2487,1785)(2509,1785)(2531,1785)(2554,1785)(2576,1785)
\thinlines\dashline[-20]{18}(370,1119)(370,1119)(392,1119)(415,1119)(437,1119)(459,1119)(481,1119)
(504,1119)(526,1119)(548,1119)(571,1119)(593,1119)(615,1119)(637,1119)(660,1119)(682,1119)(704,1119)
(727,1119)(749,1119)(771,1119)(793,1119)(816,1119)(838,1119)(860,1119)(883,1119)(905,1119)(927,1119)
(949,1119)(972,1119)(994,1119)(1016,1119)(1038,1119)(1061,1119)(1083,1119)(1105,1119)(1128,1119)
(1150,1119)(1172,1119)(1194,1119)(1217,1119)(1239,1119)(1261,1119)(1284,1119)(1306,1119)(1328,1119)
(1350,1119)(1373,1119)(1395,1119)(1417,1119)(1440,1119)(1462,1119)
\thinlines\dashline[-20]{18}(1462,1119)(1484,1119)(1506,1119)(1529,1119)(1551,1119)(1573,1119)
(1596,1119)(1618,1119)(1640,1119)(1662,1119)(1685,1119)(1707,1119)(1729,1119)(1752,1119)(1774,1119)
(1796,1119)(1818,1119)(1841,1119)(1863,1119)(1885,1119)(1908,1119)(1930,1119)(1952,1119)(1974,1119)
(1997,1119)(2019,1119)(2041,1119)(2063,1119)(2086,1119)(2108,1119)(2130,1119)(2153,1119)(2175,1119)
(2197,1119)(2219,1119)(2242,1119)(2264,1119)(2286,1119)(2309,1119)(2331,1119)(2353,1119)(2375,1119)
(2398,1119)(2420,1119)(2442,1119)(2465,1119)(2487,1119)(2509,1119)(2531,1119)(2554,1119)(2576,1119)
\thinlines\dashline[-20]{11}(370,785)(370,785)(392,785)(415,785)(437,785)(459,785)(481,785)(504,785)
(526,785)(548,785)(571,785)(593,785)(615,785)(637,785)(660,785)(682,785)(704,785)(727,785)(749,785)
(771,785)(793,785)(816,785)(838,785)(860,785)(883,785)(905,785)(927,785)(949,785)(972,785)(994,785)
(1016,785)(1038,785)(1061,785)(1083,785)(1105,785)(1128,785)(1150,785)(1172,785)(1194,785)(1217,785)
(1239,785)(1261,785)(1284,785)(1306,785)(1328,785)(1350,785)(1373,785)(1395,785)(1417,785)(1440,785)
(1462,785)
\thinlines\dashline[-20]{11}(1462,785)(1484,785)(1506,785)(1529,785)(1551,785)(1573,785)(1596,785)
(1618,785)(1640,785)(1662,785)(1685,785)(1707,785)(1729,785)(1752,785)(1774,785)(1796,785)(1818,785)
(1841,785)(1863,785)(1885,785)(1908,785)(1930,785)(1952,785)(1974,785)(1997,785)(2019,785)(2041,785)
(2063,785)(2086,785)(2108,785)(2130,785)(2153,785)(2175,785)(2197,785)(2219,785)(2242,785)(2264,785)
(2286,785)(2309,785)(2331,785)(2353,785)(2375,785)(2398,785)(2420,785)(2442,785)(2465,785)(2487,785)
(2509,785)(2531,785)(2554,785)(2576,785)
\end{picture}
\eec
\end{minipage}
\hspace*{5mm}
\begin{minipage}[r]{3in}
\bec
\setlength{\unitlength}{0.080450pt}
\begin{picture}(2699,2069)(0,0)
\footnotesize
\thicklines \path(370,249)(411,249)
\thicklines \path(2576,249)(2535,249)
\put(329,249){\makebox(0,0)[r]{ 1}}
\thicklines \path(370,481)(411,481)
\thicklines \path(2576,481)(2535,481)
\put(329,481){\makebox(0,0)[r]{ 1.2}}
\thicklines \path(370,713)(411,713)
\thicklines \path(2576,713)(2535,713)
\put(329,713){\makebox(0,0)[r]{ 1.4}}
\thicklines \path(370,945)(411,945)
\thicklines \path(2576,945)(2535,945)
\put(329,945){\makebox(0,0)[r]{ 1.6}}
\thicklines \path(370,1176)(411,1176)
\thicklines \path(2576,1176)(2535,1176)
\put(329,1176){\makebox(0,0)[r]{ 1.8}}
\thicklines \path(370,1408)(411,1408)
\thicklines \path(2576,1408)(2535,1408)
\put(329,1408){\makebox(0,0)[r]{ 2}}
\thicklines \path(370,1640)(411,1640)
\thicklines \path(2576,1640)(2535,1640)
\put(329,1640){\makebox(0,0)[r]{ 2.2}}
\thicklines \path(370,1872)(411,1872)
\thicklines \path(2576,1872)(2535,1872)
\put(329,1872){\makebox(0,0)[r]{ 2.4}}
\thicklines \path(370,249)(370,290)
\thicklines \path(370,1988)(370,1947)
\put(370,166){\makebox(0,0){ 0}}
\thicklines \path(676,249)(676,290)
\thicklines \path(676,1988)(676,1947)
\put(676,166){\makebox(0,0){ 50}}
\thicklines \path(983,249)(983,290)
\thicklines \path(983,1988)(983,1947)
\put(983,166){\makebox(0,0){ 100}}
\thicklines \path(1289,249)(1289,290)
\thicklines \path(1289,1988)(1289,1947)
\put(1289,166){\makebox(0,0){ 150}}
\thicklines \path(1596,249)(1596,290)
\thicklines \path(1596,1988)(1596,1947)
\put(1596,166){\makebox(0,0){ 200}}
\thicklines \path(1902,249)(1902,290)
\thicklines \path(1902,1988)(1902,1947)
\put(1902,166){\makebox(0,0){ 250}}
\thicklines \path(2208,249)(2208,290)
\thicklines \path(2208,1988)(2208,1947)
\put(2208,166){\makebox(0,0){ 300}}
\thicklines \path(2515,249)(2515,290)
\thicklines \path(2515,1988)(2515,1947)
\put(2515,166){\makebox(0,0){ 350}}
\thicklines \path(370,249)(2576,249)(2576,1988)(370,1988)(370,249)
\put(-262,1100){\makebox(0,0)[l]{{$\frac{\tau^0}{\tau^+}
                                    \frac{2 B^{+0}_\pi}{B^{+-}_\pi} $}}}
\put(1473,22){\makebox(0,0){$ \theta^{EW} $}}
\put(1380,1000){\makebox(0,0)[r]{$r_{EW} = 0.14$}}
\put(1200,1440){\makebox(0,0){$ 0.4 $}}
\put(1200,1600){\makebox(0,0){$ 0.5 $}}
\put(1200,1250){\makebox(0,0){$ 0.3 $}}
\put(2180,1570){\makebox(0,0){$ 1\sigma $}}
\put(2180,870){\makebox(0,0){$ 2\sigma $}}
\thinlines \path(370,827)(370,827)(459,930)(557,1046)(649,1149)
(737,1237)(786,1279)(831,1312)(878,1341)(921,1362)(942,1370)(965,1377)
(989,1382)(1002,1384)(1008,1385)(1014,1385)(1020,1386)(1025,1386)
(1031,1387)(1034,1387)(1036,1387)(1038,1387)(1040,1387)(1043,1387)
(1046,1387)(1049,1387)(1052,1387)(1054,1387)(1057,1387)(1060,1387)
(1066,1386)(1071,1386)(1081,1385)(1092,1383)(1103,1381)(1127,1376)
(1148,1369)(1196,1348)(1240,1324)(1288,1290)(1382,1208)(1473,1113)
(1569,1001)(1661,892)(1750,791)(1845,697)(1892,657)(1937,625)(1982,597)
(2024,577)(2047,569)
\thinlines \path(2047,569)(2071,561)(2082,559)(2094,557)(2104,555)
(2114,554)(2120,553)(2127,552)(2130,552)(2133,552)(2135,552)(2138,552)
(2140,552)(2141,552)(2143,552)(2145,552)(2147,552)(2148,552)(2150,552)
(2152,552)(2154,552)(2156,552)(2158,552)(2161,552)(2164,552)(2170,553)
(2176,553)(2187,554)(2199,556)(2211,559)(2233,564)(2254,571)(2302,592)
(2349,619)(2392,649)(2484,730)(2573,823)(2576,827)
\thinlines \path(370,797)(370,797)(459,935)(557,1089)(649,1227)(737,1345)
(786,1401)(831,1445)(878,1483)(921,1511)(942,1521)(965,1531)(989,1538)
(1002,1540)(1008,1542)(1014,1542)(1020,1543)(1025,1544)(1031,1544)
(1034,1544)(1036,1544)(1038,1544)(1040,1545)(1043,1545)(1046,1545)
(1049,1545)(1052,1544)(1054,1544)(1057,1544)(1060,1544)(1066,1544)
(1071,1543)(1081,1541)(1092,1539)(1103,1537)(1127,1529)(1148,1520)
(1196,1493)(1240,1460)(1288,1415)(1382,1306)(1473,1179)(1569,1030)
(1661,884)(1750,750)(1845,624)(1892,571)(1937,528)(1982,492)(2024,465)
(2047,453)
\thinlines \path(2047,453)(2071,444)(2082,440)(2094,437)(2104,435)
(2114,433)(2120,432)(2127,432)(2130,432)(2133,431)(2135,431)(2138,431)
(2140,431)(2141,431)(2143,431)(2145,431)(2147,431)(2148,431)(2150,431)
(2152,431)(2154,431)(2156,431)(2158,431)(2161,431)(2164,431)(2170,432)
(2176,433)(2187,434)(2199,437)(2211,440)(2233,447)(2254,457)(2302,484)
(2349,521)(2392,561)(2484,669)(2573,793)(2576,797)
\thinlines
\path(370,773)(370,773)(459,946)(557,1138)(649,1311)(737,1457)
(786,1528)(831,1583)(878,1631)(921,1665)(942,1678)(965,1690)(989,1699)
(1002,1702)(1008,1704)(1014,1705)(1020,1706)(1025,1706)(1031,1707)
(1034,1707)(1036,1707)(1038,1707)(1040,1707)(1043,1707)(1046,1707)
(1049,1707)(1052,1707)(1054,1707)(1057,1707)(1060,1707)(1066,1706)
(1071,1705)(1081,1703)(1092,1701)(1103,1697)(1127,1688)(1148,1677)
(1196,1643)(1240,1601)(1288,1546)(1382,1409)(1473,1250)(1569,1064)
(1661,881)(1750,714)(1845,557)(1892,490)(1937,437)(1982,391)(2024,357)(2047,343)
\thinlines \path(2047,343)(2071,331)(2082,327)(2094,323)(2104,320)
(2114,318)(2120,317)(2127,316)(2130,316)(2133,316)(2135,315)(2138,315)
(2140,315)(2141,315)(2143,315)(2145,315)(2147,315)(2148,315)(2150,315)
(2152,315)(2154,315)(2156,315)(2158,315)(2161,316)(2164,316)(2170,317)
(2176,317)(2187,319)(2199,323)(2211,326)(2233,336)(2254,347)(2302,382)
(2349,427)(2392,478)(2484,612)(2573,768)(2576,773)
\thinlines
\path(370,884)(370,884)(459,933)(557,987)(649,1035)(737,1076)
(786,1096)(831,1111)(878,1124)(921,1134)(942,1138)(965,1141)(989,1144)
(1002,1145)(1008,1145)(1014,1145)(1020,1145)(1025,1146)(1031,1146)
(1034,1146)(1036,1146)(1038,1146)(1040,1146)(1043,1146)(1046,1146)
(1049,1146)(1052,1146)(1054,1146)(1057,1146)(1060,1146)(1066,1146)
(1071,1145)(1081,1145)(1092,1144)(1103,1143)(1127,1141)(1148,1137)
(1196,1128)(1240,1116)(1288,1101)(1382,1062)(1473,1018)(1569,966)
(1661,915)(1750,868)(1845,824)(1892,805)(1937,790)(1982,777)(2024,768)
(2047,764)
\thinlines \path(2047,764)(2071,761)(2082,759)(2094,758)(2104,758)
(2114,757)(2120,757)(2127,756)(2130,756)(2133,756)(2135,756)(2138,756)
(2140,756)(2141,756)(2143,756)(2145,756)(2147,756)(2148,756)(2150,756)
(2152,756)(2154,756)(2156,756)(2158,756)(2161,756)(2164,756)(2170,757)
(2176,757)(2187,757)(2199,758)(2211,759)(2233,762)(2254,765)(2302,775)
(2349,788)(2392,802)(2484,839)(2573,883)(2576,884)
\thinlines\dashline[-20]{18}(1462,1976)(1484,1976)(1506,1976)(1529,1976)
(1551,1976)(1573,1976)(1596,1976)(1618,1976)(1640,1976)(1662,1976)
(1685,1976)(1707,1976)(1729,1976)(1752,1976)(1774,1976)(1796,1976)
(1818,1976)(1841,1976)(1863,1976)(1885,1976)(1908,1976)(1930,1976)(1952,1976)
(1974,1976)(1997,1976)(2019,1976)(2041,1976)(2063,1976)(2086,1976)(2108,1976)
(2130,1976)(2153,1976)(2175,1976)(2197,1976)(2219,1976)(2242,1976)(2264,1976)
(2286,1976)(2309,1976)(2331,1976)(2353,1976)(2375,1976)(2398,1976)(2420,1976)
(2442,1976)(2465,1976)(2487,1976)(2509,1976)(2531,1976)(2554,1976)(2576,1976)
\thinlines\dashline[-20]{18}(370,1118)(370,1118)(392,1118)(415,1118)(437,1118)
(459,1118)(481,1118)(504,1118)(526,1118)(548,1118)(571,1118)(593,1118)
(615,1118)(637,1118)(660,1118)(682,1118)(704,1118)(727,1118)(749,1118)
(771,1118)(793,1118)(816,1118)(838,1118)(860,1118)(883,1118)(905,1118)
(927,1118)(949,1118)(972,1118)(994,1118)(1016,1118)(1038,1118)(1061,1118)
(1083,1118)(1105,1118)(1128,1118)(1150,1118)(1172,1118)(1194,1118)(1217,1118)
(1239,1118)(1261,1118)(1284,1118)(1306,1118)(1328,1118)(1350,1118)(1373,1118)
(1395,1118)(1417,1118)(1440,1118)(1462,1118)
\thinlines\dashline[-20]{18}(1462,1118)(1484,1118)(1506,1118)(1529,1118)
(1551,1118)(1573,1118)(1596,1118)(1618,1118)(1640,1118)(1662,1118)(1685,1118)
(1707,1118)(1729,1118)(1752,1118)(1774,1118)(1796,1118)(1818,1118)(1841,1118)
(1863,1118)(1885,1118)(1908,1118)(1930,1118)(1952,1118)(1974,1118)(1997,1118)
(2019,1118)(2041,1118)(2063,1118)(2086,1118)(2108,1118)(2130,1118)(2153,1118)
(2175,1118)(2197,1118)(2219,1118)(2242,1118)(2264,1118)(2286,1118)(2309,1118)
(2331,1118)(2353,1118)(2375,1118)(2398,1118)(2420,1118)(2442,1118)(2465,1118)
(2487,1118)(2509,1118)(2531,1118)(2554,1118)(2576,1118)
\thinlines\dashline[-20]{11}(370,690)(370,690)(392,690)(415,690)(437,690)
(459,690)(481,690)(504,690)(526,690)(548,690)(571,690)(593,690)(615,690)
(637,690)(660,690)(682,690)(704,690)(727,690)(749,690)(771,690)(793,690)
(816,690)(838,690)(860,690)(883,690)(905,690)(927,690)(949,690)(972,690)
(994,690)(1016,690)(1038,690)(1061,690)(1083,690)(1105,690)(1128,690)
(1150,690)(1172,690)(1194,690)(1217,690)(1239,690)(1261,690)(1284,690)
(1306,690)(1328,690)(1350,690)(1373,690)(1395,690)(1417,690)(1440,690)
(1462,690)
\thinlines\dashline[-20]{11}(1462,690)(1484,690)(1506,690)(1529,690)
(1551,690)(1573,690)(1596,690)(1618,690)(1640,690)(1662,690)(1685,690)
(1707,690)(1729,690)(1752,690)(1774,690)(1796,690)(1818,690)(1841,690)
(1863,690)(1885,690)(1908,690)(1930,690)(1952,690)(1974,690)(1997,690)
(2019,690)(2041,690)(2063,690)(2086,690)(2108,690)(2130,690)(2153,690)
(2175,690)(2197,690)(2219,690)(2242,690)(2264,690)(2286,690)(2309,690)
(2331,690)(2353,690)(2375,690)(2398,690)(2420,690)(2442,690)(2465,690)
(2487,690)(2509,690)(2531,690)(2554,690)(2576,690)
\end{picture}
\eec
\end{minipage}
\caption{$\frac{2B^{00}_\pi}{B^{+-}_\pi}$ and 
         $\frac{\tau^0}{\tau^+} \frac{2B^{+0}_\pi}{B^{+-}_\pi}$ 
         as the function of $\theta^{EW}$ at $\tilde{r}_C=0.2, 
         \omega = 0^\circ $ and $\theta^P=0^\circ $. 
         The gluonic penguin contribution
         $\tilde{r}_P$ is $0.45$, which is almost $1.5$ times as large as the
         usual estimation in the SM. }
    \label{fig:10}
\end{center}
\end{figure}

As well as $K\pi$, we have to reconsider $B\rightarrow \pi \pi
$ modes in the case with new physics contributions. Under the same
conditions which are $\omega=0^\circ $ and $\theta^{P} = 0^\circ $ but
$\tilde{r}_C = 0.2 $ which is slightly larger than usual estimation, 
the ratios among the branching fractions are plotted in
Fig.~\ref{fig:10}. The allowed region for $\theta^{EW}$, however, seems
to be slightly different between $K\pi $ and $\pi\pi $. 
Moreover, $\tilde{r}_P=0.45$, which is almost $1.5$ times as large as the SM
expectation, and larger $r_{EW}$ will be requested to
satisfy $\frac{B^{00}_\pi }{B^{+-}_\pi }$. 
In the allowed region of
$\theta^{EW} $ for $B\rightarrow K\pi $, there is no region satisfying 
the data of $B\rightarrow \pi\pi $ at $1\sigma$ level even if $\tilde{r}_C =
0.2 $. 
It may suggest that the other angles as the parameters of new physics
are needed. Hence to find the allowed region to explain
both discrepancies between $B\rightarrow K\pi$ and $B\rightarrow \pi
\pi $ may be slightly difficult without considering the SU(3)
breaking effects and assuming that these parameters should be
independent each other between $K \pi$ and $\pi
\pi $ even if $\theta^P$ is non-zero value.  

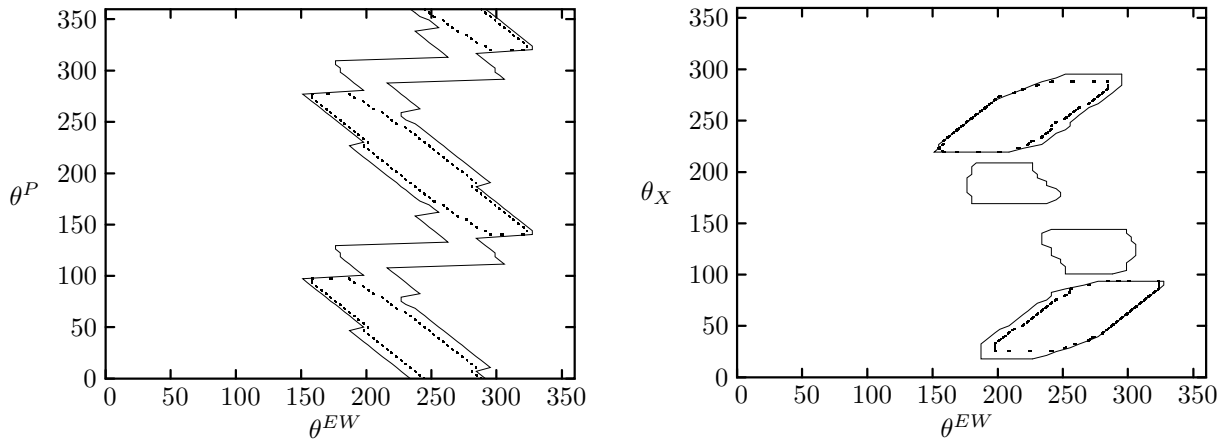
\begin{figure}[thb]
\bec
\hspace*{3mm}
\begin{minipage}[l]{3.0in}
\begin{center}
\setlength{\unitlength}{0.080450pt}
\begin{picture}(2699,2069)(0,0)
\footnotesize
\thicklines \path(370,249)(411,249)
\thicklines \path(2576,249)(2535,249)
\put(329,249){\makebox(0,0)[r]{ 0}}
\thicklines \path(370,491)(411,491)
\thicklines \path(2576,491)(2535,491)
\put(329,491){\makebox(0,0)[r]{ 50}}
\thicklines \path(370,732)(411,732)
\thicklines \path(2576,732)(2535,732)
\put(329,732){\makebox(0,0)[r]{ 100}}
\thicklines \path(370,974)(411,974)
\thicklines \path(2576,974)(2535,974)
\put(329,974){\makebox(0,0)[r]{ 150}}
\thicklines \path(370,1215)(411,1215)
\thicklines \path(2576,1215)(2535,1215)
\put(329,1215){\makebox(0,0)[r]{ 200}}
\thicklines \path(370,1457)(411,1457)
\thicklines \path(2576,1457)(2535,1457)
\put(329,1457){\makebox(0,0)[r]{ 250}}
\thicklines \path(370,1698)(411,1698)
\thicklines \path(2576,1698)(2535,1698)
\put(329,1698){\makebox(0,0)[r]{ 300}}
\thicklines \path(370,1940)(411,1940)
\thicklines \path(2576,1940)(2535,1940)
\put(329,1940){\makebox(0,0)[r]{ 350}}
\thicklines \path(370,249)(370,290)
\thicklines \path(370,1988)(370,1947)
\put(370,166){\makebox(0,0){ 0}}
\thicklines \path(676,249)(676,290)
\thicklines \path(676,1988)(676,1947)
\put(676,166){\makebox(0,0){ 50}}
\thicklines \path(983,249)(983,290)
\thicklines \path(983,1988)(983,1947)
\put(983,166){\makebox(0,0){ 100}}
\thicklines \path(1289,249)(1289,290)
\thicklines \path(1289,1988)(1289,1947)
\put(1289,166){\makebox(0,0){ 150}}
\thicklines \path(1596,249)(1596,290)
\thicklines \path(1596,1988)(1596,1947)
\put(1596,166){\makebox(0,0){ 200}}
\thicklines \path(1902,249)(1902,290)
\thicklines \path(1902,1988)(1902,1947)
\put(1902,166){\makebox(0,0){ 250}}
\thicklines \path(2208,249)(2208,290)
\thicklines \path(2208,1988)(2208,1947)
\put(2208,166){\makebox(0,0){ 300}}
\thicklines \path(2515,249)(2515,290)
\thicklines \path(2515,1988)(2515,1947)
\put(2515,166){\makebox(0,0){ 350}}
\thicklines \path(370,249)(2576,249)(2576,1988)(370,1988)(370,249)
\put(-82,1118){\makebox(0,0)[l]{{$\theta^{P}$}}}
\put(1473,22){\makebox(0,0){$\theta^{EW} $ }}
\thinlines
 \path(2157,249)(2157,249)(2135,266)(2113,284)(2179,301)(2157,319)(2135,336)
(2113,353)(2091,371)(2069,388)(2047,406)(2025,423)(2002,440)(1980,458)
(1958,475)(1936,492)(1914,510)(1892,527)(1870,545)(1848,562)(1826,579)
(1782,597)(1760,614)(1760,632)(1848,649)
(1826,666)(1804,684)(1782,701)(1760,719)(1738,736)(1716,753)
(1694,771)(2245,788)(2223,805)
(2201,823)(2201,840)(2179,858)(2157,875)(2135,892)(2113,910)
(2377,927)(2377,945)(2355,962)
(2333,979)(2311,997)(2289,1014)(2267,1032)(2245,1049)(2223,1066)
(2201,1084)(2179,1101)
\thinlines
\path(2179,1101)(2157,1119)(2135,1136)(2113,1153)(2179,1171)(2157,1188)(2135,1205)
(2113,1223)(2091,1240)(2069,1258)(2047,1275)(2025,1292)(2002,1310)(1980,1327)(1958,1345)
(1936,1362)(1914,1379)(1892,1397)(1870,1414)(1848,1432)(1826,1449)(1782,1466)(1760,1484)
(1760,1501)(1848,1518)(1826,1536)(1804,1553)(1782,1571)(1760,1588)(1738,1605)(1716,1623)
(1694,1640)(2245,1658)(2223,1675)(2201,1692)(2201,1710)(2179,1727)(2157,1745)(2135,1762)
(2113,1779)(2377,1797)(2377,1814)(2355,1831)(2333,1849)(2311,1866)(2289,1884)(2267,1901)
(2245,1918)(2223,1936)(2201,1953)(2179,1971)
\thinlines \path(2179,1971)(2157,1988)
\thinlines \path(1804,249)(1804,249)(1782,266)(1760,284)(1738,301)(1716,319)(1694,336)
(1672,353)(1649,371)(1627,388)(1605,406)(1583,423)(1561,440)(1539,458)(1517,475)(1583,492)
(1561,510)(1539,527)(1517,545)(1495,562)(1473,579)(1451,597)(1429,614)(1407,632)(1385,649)
(1363,666)(1341,684)(1319,701)(1297,719)(1583,736)(1561,753)(1539,771)(1517,788)(1495,805)
(1473,823)(1473,840)(1451,858)(1451,875)(1980,892)(1958,910)(1936,927)(1914,945)(1892,962)
(1870,979)(1848,997)(1826,1014)(1936,1032)(1914,1049)(1892,1066)(1848,1084)(1826,1101)
\thinlines
\path(1826,1101)(1804,1119)(1782,1136)(1760,1153)(1738,1171)(1716,1188)(1694,1205)
(1672,1223)(1649,1240)(1627,1258)(1605,1275)(1583,1292)(1561,1310)(1539,1327)(1517,1345)
(1583,1362)(1561,1379)(1539,1397)(1517,1414)(1495,1432)(1473,1449)(1451,1466)(1429,1484)
(1407,1501)(1385,1518)(1363,1536)(1341,1553)(1319,1571)(1297,1588)(1583,1605)(1561,1623)
(1539,1640)(1517,1658)(1495,1675)(1473,1692)(1473,1710)(1451,1727)(1451,1745)(1980,1762)
(1958,1779)(1936,1797)(1914,1814)(1892,1831)(1870,1849)(1848,1866)(1826,1884)(1936,1901)
(1914,1918)(1892,1936)(1848,1953)(1826,1971)
\thinlines \path(1826,1971)(1804,1988)
\thinlines\dashline[-20]{11}(2135,249)(2135,249)(2113,266)(2091,284)(2113,301)(2091,319)
(2069,336)(2047,353)(2025,371)(2002,388)(1980,406)(1958,423)(1936,440)(1914,458)(1892,475)
(1870,492)(1848,510)(1826,527)(1804,545)(1782,562)(1738,579)(1716,597)(1694,614)(1672,632)
(1627,649)(1605,666)(1583,684)(1539,701)(1517,719)
\thinlines\dashline[-20]{11}(2333,927)(2333,927)(2355,945)(2333,962)(2311,979)(2289,997)
(2267,1014)(2245,1032)(2223,1049)(2201,1066)(2179,1084)(2157,1101)(2135,1119)(2113,1136)
(2091,1153)(2113,1171)(2091,1188)(2069,1205)(2047,1223)(2025,1240)(2002,1258)(1980,1275)
(1958,1292)(1936,1310)(1914,1327)(1892,1345)(1870,1362)(1848,1379)(1826,1397)(1804,1414)
(1782,1432)(1738,1449)(1716,1466)(1694,1484)(1672,1501)(1627,1518)(1605,1536)(1583,1553)
(1539,1571)(1517,1588)
\thinlines\dashline[-20]{11}(2179,1797)(2279,1797)(2333,1797)(2333,1797)(2355,1814)(2333,1831)
(2311,1849)(2289,1866)(2267,1884)(2245,1901)(2223,1918)(2201,1936)(2179,1953)(2157,1971)(2135,1988)
\thinlines\dashline[-20]{11}(1870,249)(1870,249)(1848,266)(1826,284)(1804,301)(1782,319)
(1760,336)(1738,353)(1716,371)(1694,388)(1672,406)(1649,423)(1627,440)(1605,458)(1583,475)
(1605,492)(1583,510)(1561,527)(1539,545)(1517,562)(1495,579)(1473,597)(1451,614)(1429,632)
(1407,649)(1385,666)(1363,684)(1341,701)(1341,719)(1430,719)(1517,719)
\thinlines\dashline[-20]{11}(2333,927)(2233,927)(2179,927)(2179,927)(2157,945)(2113,962)
(2091,979)(2069,997)(2025,1014)(2002,1032)(1980,1049)(1958,1066)(1914,1084)(1892,1101)
(1870,1119)(1848,1136)(1826,1153)(1804,1171)(1782,1188)(1760,1205)(1738,1223)(1716,1240)
(1694,1258)(1672,1275)(1649,1292)(1627,1310)(1605,1327)(1583,1345)(1605,1362)(1583,1379)
(1561,1397)(1539,1414)(1517,1432)(1495,1449)(1473,1466)(1451,1484)(1429,1501)(1407,1518)
(1385,1536)(1363,1553)(1341,1571)(1341,1588)(1417,1588)(1517,1588)
\thinlines\dashline[-20]{11}(2179,1797)(2179,1797)(2157,1814)(2113,1831)
(2091,1849)(2069,1866)(2025,1884)(2002,1901)(1980,1918)(1958,1936)(1914,1953)
(1892,1971)(1870,1988)
\end{picture}
\eec
\end{minipage}
\hspace*{5mm}
\begin{minipage}[r]{3in}
\setlength{\unitlength}{0.080450pt}
\begin{picture}(2699,2069)(0,0)
\footnotesize
\thicklines \path(370,249)(411,249)
\thicklines \path(2576,249)(2535,249)
\put(329,249){\makebox(0,0)[r]{ 0}}
\thicklines \path(370,491)(411,491)
\thicklines \path(2576,491)(2535,491)
\put(329,491){\makebox(0,0)[r]{ 50}}
\thicklines \path(370,732)(411,732)
\thicklines \path(2576,732)(2535,732)
\put(329,732){\makebox(0,0)[r]{ 100}}
\thicklines \path(370,974)(411,974)
\thicklines \path(2576,974)(2535,974)
\put(329,974){\makebox(0,0)[r]{ 150}}
\thicklines \path(370,1215)(411,1215)
\thicklines \path(2576,1215)(2535,1215)
\put(329,1215){\makebox(0,0)[r]{ 200}}
\thicklines \path(370,1457)(411,1457)
\thicklines \path(2576,1457)(2535,1457)
\put(329,1457){\makebox(0,0)[r]{ 250}}
\thicklines \path(370,1698)(411,1698)
\thicklines \path(2576,1698)(2535,1698)
\put(329,1698){\makebox(0,0)[r]{ 300}}
\thicklines \path(370,1940)(411,1940)
\thicklines \path(2576,1940)(2535,1940)
\put(329,1940){\makebox(0,0)[r]{ 350}}
\thicklines \path(370,249)(370,290)
\thicklines \path(370,1988)(370,1947)
\put(370,166){\makebox(0,0){ 0}}
\thicklines \path(676,249)(676,290)
\thicklines \path(676,1988)(676,1947)
\put(676,166){\makebox(0,0){ 50}}
\thicklines \path(983,249)(983,290)
\thicklines \path(983,1988)(983,1947)
\put(983,166){\makebox(0,0){ 100}}
\thicklines \path(1289,249)(1289,290)
\thicklines \path(1289,1988)(1289,1947)
\put(1289,166){\makebox(0,0){ 150}}
\thicklines \path(1596,249)(1596,290)
\thicklines \path(1596,1988)(1596,1947)
\put(1596,166){\makebox(0,0){ 200}}
\thicklines \path(1902,249)(1902,290)
\thicklines \path(1902,1988)(1902,1947)
\put(1902,166){\makebox(0,0){ 250}}
\thicklines \path(2208,249)(2208,290)
\thicklines \path(2208,1988)(2208,1947)
\put(2208,166){\makebox(0,0){ 300}}
\thicklines \path(2515,249)(2515,290)
\thicklines \path(2515,1988)(2515,1947)
\put(2515,166){\makebox(0,0){ 350}}
\thicklines \path(370,249)(2576,249)(2576,1988)(370,1988)(370,249)
\put(-82,1118){\makebox(0,0)[l]{{$\theta_X $}}}
\put(1473,22){\makebox(0,0){$ \theta^{EW} $ }}
\thinlines\path(1760,336)(1760,336)(1826,353)(1870,371)(1936,388)(1980,406)
(2025,423)(2069,440)(2091,458)(2113,475)(2135,492)(2157,510)(2179,527)(2201,545)
(2223,562)(2245,579)(2267,597)(2289,614)(2311,632)(2333,649)(2355,666)(2377,684)(2377,701)
\thinlines\path(2135,736)(2135,736)(2201,753)(2201,771)(2201,788)(2223,805)
(2245,823)(2245,840)(2245,858)(2245,875)(2223,892)(2223,910)(2201,927)(2201,945)
\thinlines\path(1473,1066)(1826,1066)(1826,1066)(1870,1084)(1892,1101)(1892,1119)
(1870,1136)(1826,1153)(1826,1171)(1782,1188)(1782,1205)(1760,1223)(1760,1240)(1760,1258)
\thinlines\path(1649,1310)(1649,1310)(1716,1327)(1804,1345)(1826,1362)(1848,1379)
(1870,1397)(1914,1414)(1936,1432)(1936,1449)(1958,1466)(1980,1484)(2002,1501)(2025,1518)
(2069,1536)(2091,1553)(2113,1571)(2135,1588)(2157,1605)(2179,1623)(2179,1640)
(2179,1658)(2179,1675)
\thinlines\path(1760,336)(1517,336)(1517,336)(1517,353)(1517,371)(1517,388)(1517,406)
(1539,423)(1561,440)(1583,458)(1605,475)(1649,492)(1672,510)(1694,527)(1716,545)(1738,562)
(1760,579)(1782,597)(1826,614)(1848,632)(1848,649)(1914,666)(2002,684)(2069,701)(2377,701)
\thinlines\path(2135,736)(1914,736)(1914,736)(1914,753)(1914,771)(1892,788)(1892,805)
(1848,823)(1848,840)(1848,858)(1804,875)(1804,892)(1804,910)(1804,927)(1848,945)
(2201,945)
\thinlines\path(1473,1066)(1473,1066)(1473,1084)(1473,1101)(1451,1119)(1451,1136)
(1451,1153)(1451,1171)(1451,1188)(1473,1205)(1473,1223)(1473,1240)(1495,1258)
(1760,1258)
\thinlines\path(1649,1310)(1297,1310)(1297,1310)(1319,1327)(1319,1345)(1341,1362)
(1363,1379)(1385,1397)(1407,1414)(1429,1432)(1451,1449)(1473,1466)(1495,1484)
(1517,1501)(1539,1518)(1561,1536)(1583,1553)(1627,1571)(1672,1588)(1716,1605)
(1760,1623)(1804,1640)(1870,1658)(1914,1675)(2179,1675)
\thinlines\dashline[-20]{13}(1848,371)(1848,371)(1914,388)(1958,406)(2025,423)
(2047,440)(2091,458)(2113,475)(2135,492)(2157,510)(2179,527)(2201,545)(2223,562)
(2245,579)(2267,597)(2289,614)(2311,632)(2333,649)(2355,666)(2355,684)(2355,701)
(2213,701)(2163,701)(2113,701)
\thinlines\dashline[-20]{13}(1605,1310)(1605,1310)(1694,1327)(1738,1345)(1760,1362)
(1804,1379)(1826,1397)(1848,1414)(1848,1432)(1870,1449)(1914,1466)(1936,1484)
(1958,1501)(1980,1518)(2025,1536)(2047,1553)(2069,1571)(2091,1588)(2113,1605)
(2113,1623)(2113,1640)(2048,1640)(1948,1640)(1848,1640)
\thinlines\dashline[-20]{13}(1848,371)(1748,371)(1648,371)(1583,371)(1583,371)
(1583,388)(1583,406)(1605,423)(1627,440)(1649,458)(1672,475)(1716,492)(1738,510)
(1760,527)(1782,545)(1804,562)(1826,579)(1848,597)(1870,614)(1914,632)(1936,649)
(1936,666)(2025,684)(2113,701)
\thinlines\dashline[-20]{13}(1605,1310)(1505,1310)(1405,1310)(1341,1310)(1341,1310)
(1319,1327)(1341,1345)(1341,1362)(1363,1379)(1385,1397)(1407,1414)(1429,1432)(1451,1449)
(1473,1466)(1495,1484)(1517,1501)(1539,1518)(1561,1536)(1583,1553)(1605,1571)(1672,1588)
(1716,1605)(1782,1623)(1848,1640)
\end{picture}
\end{minipage}
\caption{The allowed region for the new weak phases
         $\theta^{EW},\theta^P$ of each penguin
         diagrams and $\theta_X$ which is the phase difference 
         between $\pi\pi $ and $K\pi$ final state modes, 
         under a assumption that the tree diagrams do not
         include any new physics and $\tilde{r}_C=0.1$ and the 
         all strong phase differences do not also so that 
         $\omega=\delta^{EW}-\delta^T=0^\circ $. The regions are
         satisfying $R_c-R_n = 0.37\pm0.16, S = 0.43\pm 0.20, 
         A_{CP}^{K^+\pi^-} = -0.095\pm 0.028,
         A_{CP}^{K^0\pi^0}=0.11\pm0.23, 
         \frac{2B^{00}_\pi}{B^{+-}_\pi} = 0.83\pm0.23 $ and
         $\frac{\tau^0}{\tau^+}\frac{2B^{+0}_\pi}{B^{+-}_\pi}=2.08\pm0.37$. 
         And at $2\sigma $ level $R$ is used to put a constraint
         because we neglected some small terms to get the bound.  
         Here the KM weak phases are used $\phi_1 = 23.7^\circ $ and 
         $ 40^\circ < \phi_3 < 80^\circ $.   
         The gluonic penguin contribution in $B\rightarrow \pi \pi $, 
         $\tilde{r}_P$ is 0.45 which is almost 1.5 times of the
         usual estimation. In both figures, the solid line show
         the case of $r_{EW}=0.4$ and the dashed line is $r_{EW}=0.3$. }  
    \label{fig:11}
\end{center}
\end{figure}

In order to solve them, we introduce one more parameter as a phase 
difference of the new weak phases between $K\pi $ and  
$\pi \pi $. 
The phase difference is defined as 
\bea
\theta_X\, \equiv\, \theta_\pi^P - \theta^P\, =\, \theta^{EW} -
\theta_\pi^{EW}\;, 
\eea    
which is a phase difference of penguin diagrams between  
$B\rightarrow K\pi $ and $B\rightarrow \pi \pi $ modes except for the
KM phase, where 
\bea
P_\pi V_{tb}^*V_{td}\, \equiv\, |P_\pi V_{tb}^*V_{td}|\,e^{-i\phi_1}
                                                  e^{-i\theta_\pi^P}\;, 
~~~ P_{EW\pi} V_{tb}^*V_{td}\, 
   \equiv\, |P_{EW\pi } V_{tb}^*V_{td}|\,e^{-i\phi_1}e^{i\delta^{EW}}
                                     e^{i\theta_\pi^{EW}}\;. 
\eea 
We assume that the SU(3) breaking effect in the strong phase is not
large because the final state is same even if the modes are including 
some new physics effects. Using this parameterization the branching
 ratios are  
\bea 
2 \bar{B}_\pi^{00} &\propto&  \tilde{r}_C^2 + \tilde{r}_P^2 \left( 1 +
              r_{EW}^2 - 2 r_{EW} \cos\delta^{EW}
              \cos(\theta^{EW}+\theta^P) \right)  \\
      & & ~~~ - 2 \tilde{r}_C \tilde{r}_P
             \left( \cos\delta^T
              \cos(\phi_1+\phi_3+\theta^P+\theta_X) 
             - r_{EW} \cos\omega
              \cos(\phi_1+\phi_3-\theta^{EW}+\theta_X)
               \right)\;,\nn  \\[2mm]
\bar{B}_\pi^{+-} &\propto& 1 + \tilde{r}_P^2 + 2 \tilde{r}_P 
                    \cos\delta^T 
                   \cos(\phi_1+\phi_3+\theta^P+\theta_X)\;, \\[2mm]
2 \bar{B}_\pi^{+0} &\propto& 1+\tilde{r}_C^2 +\tilde{r}_P^2 r_{EW}^2 
                               + 2 \tilde{r}_C 
                    + 2 \tilde{r}_P r_{EW} 
           ( 1 + \tilde{r}_C )\cos \omega 
                  \cos(\phi_1+\phi_3-\theta^{EW}+\theta_X )\;.
\eea 
Considering several constraints from the experimental values,
$R_c-R_n = 0.37\pm0.16$, $A_{CP}^{+-} = -0.095\pm 0.028$, $A_{CP}^{00}=
0.11 \pm 0.23$, $R= 0.90 \pm 0.14$, 
$\frac{2B^{00}_\pi}{B^{+-}_\pi} = 0.83\pm0.23$ and 
$\frac{\tau^0}{\tau^+}\frac{2B^{+0}_\pi}{B^{+-}_\pi}=2.08\pm0.37$, 
the allowed region for the three new phases, $\theta^P$, $\theta^{EW}$ 
and $\theta_X$ at $\omega = 0^\circ$, $r_T= 0.2$, $\tilde{r}_C= 0.1$ 
and $\tilde{r}_P = 0.45 $ in Fig.~\ref{fig:11}, where $\tilde{r}_P$
with some new physics contribution may be taken to be almost maximum value
from constraints by $B\rightarrow KK$ and $\tilde{r}_C$, which are used
in usual estimation. To enhance the ratios, the denominator,
 $\bar{B}_\pi^{+-}$, should be reduced so that the cross term is
important and negative $\cos\delta^T
\cos(\phi_1+\phi_3+\theta^P-\theta_X) $ will
be favored. $\phi_1+\phi_3~$ is about $60^\circ \sim 100^\circ$ so that
they will be strongly enhanced around $\theta^P - \theta_X \sim
100^\circ$. However, if $\theta^P$ is a large angle, we have to note that 
the region for $\delta^T$ is also changed since the constraint from 
$A_{CP}^{K^+\pi^-}$ will be relaxed by $\theta^P$. 
Here $A_{CP}^{K^0\pi^0}$ was taken account and it give a constraint 
to $\theta^{EW}$. Hence it may be slightly complicated to understand 
the allowed region.           

{}From Fig.~\ref{fig:11}, we can find that satisfying the several
 experimental data for  
$B\rightarrow \pi\pi $ and $B\rightarrow K\pi $ at once does not only
require the large $r_{EW}$ but also large new weak
phases, $\theta^{EW}$ and a phase difference 
$\theta_X = \theta^{EW} - \theta^{EW}_\pi $, which may suggest 
SU(3) breaking effects for the penguin diagrams. The right figure shows
that there is no solution for $\theta_X = 0^\circ $ if we take the
current experimental data seriously.   

In this analysis, we assumed that the tree processes do not have
any new physics contributions and there are not so large strong phase 
differences which satisfy also SU(3) symmetry. If some new physics
exists, it is including in the penguin-type diagrams and they should
appear as the large EW penguin contribution with large new weak
phase and it may cause some SU(3) breaking effects. As a result we can find 
several allowed regions such like Fig.~\ref{fig:11} for the new weak
phases of penguin contributions. In $B\rightarrow \pi\pi $ mode, the role 
of EW penguin is not so important within the SM because its magnitude
much smaller than that of tree. However if we consider some new physics 
contribution to explain both $B\rightarrow K\pi $ and $B\rightarrow \pi\pi
$, then the role of EW penguin contribution with the new weak phase 
will be more important so that it should not be neglected even if we
consider new physics effects in $\pi\pi$.

\section{Conclusion}
In this paper, we discussed a possibility of large EW penguin
contribution in $B\rightarrow K \pi$ and $B\rightarrow \pi \pi$ 
from recent experimental data. To satisfy several relations among the
branching ratios of $B\rightarrow K \pi$, the larger EW penguin
contribution with non-negligible strong phase differences. It seems to
be difficult to explain them in the SM. 
If the EW penguin estimated from experimental data is quite large compared 
with the theoretical estimation, which is usually smaller 
than tree contributions~\cite{NEU,FM,BBNS}, then it may be including 
some new physics effects. 
In addition, to avoid the large strong phase difference, the EW penguin
must have new weak phase. 

When we respect the allowed region for 
the parameters in $B\rightarrow K \pi$, then 
they could not satisfy $B\rightarrow \pi \pi$ modes under the SU(3) 
symmetry. To explain the
both modes at once, SU(3) breaking effects in gluonic and EW penguin
diagrams with new phase will be strongly requested. 
In consequence, the role of the EW penguin contribution
will be more important even in $B\rightarrow \pi \pi$ modes. 
In several recent works discussing about the branching ratios 
in the $B\rightarrow \pi \pi$ modes, EW penguin contribution is 
usually neglected because of the smallness and suggest the other
large contribution such like color-suppressed tree to explain the
deviations from experimental data. However it is unnatural that 
the color-suppressed tree diagram is including such large
new contributions so that in this paper we discussed about the
explanation by the penguin-type diagrams. 

If there is any new physics and the effects appear 
through the loop effect in these modes, 
$B\rightarrow K \pi$ and $B\rightarrow \pi \pi$, will be helpful
modes to examine and search for the evidence of new physics. 
At the present situation, the deviation from the SM in $B\rightarrow K
\pi$ is still within the 
$2 \sigma$ level if large strong phase difference is allowed. Thus 
we need more accurate experimental data. 
In near future, we can use these modes to test the SM~\cite{S-U} 
or the several new models~\cite{AH,DILPSS,KMXY,EKOU,NEWWORK}. 
For this purpose, the project the $B$ factories are upgrade\cite{LOI} 
is helpful
and important.

\section*{Acknowledgments}
We would like to thank Profs. A.I.~Sanda, A.~Ali, J.L.~Rosner,
C.S.~Kim, H.-n.~Li, Y.-Y.~Charng, Y.-Y.~Keum, K.~Terasaki,
T.~Morozumi, T.~Kurimoto and E.~Kou for many useful comments and discussions.  
The work of S.M. was supported by the Grant-in-Aid for Scientific Research
in Priority Areas from the Ministry of Education, Culture, Science,
Sports and Technology of Japan (No.14046201). The work of T.Y. was  
supported by 21st Century COE Program of Nagoya University provided 
by JSPS (15COEG01).

\end{document}